\documentclass[a4paper,11pt]{article}

\title{Modulation theory and resonant regimes for dispersive shock waves in nematic liquid crystals}
\author{Saleh Baqer, \\
School of Mathematics, University of Edinburgh,\\
Edinburgh, Scotland, EH9 3FD, U.K. \and 
Noel F. Smyth, \\
School of Mathematics, University of Edinburgh,\\
Edinburgh, Scotland, EH9 3FD, U.K. and \\
School of Mathematics and Applied Statistics, University of Wollongong,\\
Northfields Avenue, Wollongong, N.S.W., Australia, 2522.}
\date{}
\usepackage{a4wide}
\usepackage{graphicx}
\def\sech{\mathop{\rm sech}\nolimits}
\def\cn{\mathop{\rm cn}\nolimits}

\begin{document}



 
 
 
 
 

\maketitle

\begin{abstract}
 A full analysis of all regimes for optical dispersive shock wave (DSW) propagation in nematic 
 liquid crystals is undertaken.  These dispersive shock waves are generated from step initial
 conditions for the optical field and are resonant in that linear diffractive waves are in resonance
 with the DSW, resulting in a resonant linear wavetrain propagating ahead of it.  It is found that there are
 six regimes, which are distinct and require different solution methods.  In previous studies, the same solution
 method was used for all regimes, which does not yield solutions in full agreement with numerical solutions.  Indeed, 
 the standard DSW structure disappears for sufficiently large initial jumps.  Asymptotic theory, approximate methods or 
 Whitham modulation theory are used to find solutions for these resonant dispersive shock waves in a given regime.  
 The solutions are found to be in excellent agreement with numerical solutions of the nematic equations in all regimes.
 It is found that for small initial jumps, the resonant wavetrain is unstable, but that it stabilises above
 a critical jump height.  It is additionally found that the DSW is unstable, except for small jump heights
 for which there is no resonance and large jump heights for which there is no standard DSW structure.
\end{abstract}

\section{Introduction}

The generic solutions of nonlinear, dispersive wave equations, such as the Korteweg-de Vries (KdV) and nonlinear 
Schr\"odinger (NLS) equations, are the solitary wave, soliton for integrable equations, and the dispersive shock wave 
(DSW), also termed an undular bore in fluid mechanics applications.  Of these, the solitary wave (soliton) is the most 
widely studied \cite{whitham,newell} as it is a steady solution, so that its solution is given by an ordinary differential 
equation, rather than the original full, dispersive, nonlinear partial differential equation.  In addition, for integrable 
equations, solitons naturally arise from the point spectrum of the associated scattering problem \cite{whitham,newell}.  In 
general, a DSW is a modulated wavetrain which links two distinct states in a smooth manner, in contrast to a compressible flow 
shock, which is a discontinuous jump between two flow states \cite{whitham}.  It should be noted that in fluid mechanics, 
bores come in two general forms, viscous bores and undular bores \cite{elreview}.  A viscous bore is a steady wave form due to 
a balance between viscosity and dispersion/nonlinearity, so that viscous bores dissipate energy \cite{whitham,elreview,johnson70}.  
In contrast, for a DSW, an undular bore in fluid mechanics, there is no loss and it is an unsteady, continually expanding, 
modulated wave for which dispersion stops nonlinear breaking, as occurs for a gas dynamic shock.  In its standard form, a 
DSW is a modulated periodic wavetrain, with solitary waves at one edge and linear dispersive waves at the other 
\cite{elreview}, the classic example being the KdV DSW.  DSWs arise in a wide range of applications, including meteorology 
\cite{christie,clarke,anne}, oceanography \cite{nwshelf,baines}, water waves \cite{baines, esler}, geophysics 
\cite{scott1,scott2,lowman,magma}, nonlinear optics \cite{fleischer2,fleischer,elopt,trillohump}, Bose-Einstein condensates 
\cite{bose}, Fermionic fluids \cite{hoefer1} and colloidal media \cite{colloid}, see \cite{elreview} for a summary and 
appreciation of these applications.

The unsteady nature of a DSW meant that the study of these wave forms fell behind that of solitary waves (solitons).  
The breakthrough in their study came from the development of Whitham modulation theory \cite{whitham,mod1,modproc}, also 
termed the method of averaged Lagrangians.  Whitham modulation theory is a powerful method to analyse slowly varying linear, 
and importantly, nonlinear, periodic dispersive waves.  It is related to the general method of multiple scales \cite{whitham}.  
Whitham modulation theory generates so-called modulation equations for the parameters , such as amplitude, wavenumber and mean 
height, describing a modulated (slowly) varying wavetrain.  If these modulation equations are hyperbolic, the underlying nonlinear 
periodic wave is stable and if they are elliptic, the wave is unstable.  Whitham used modulation theory to give a theoretical 
explanation of the Benjamin-Feir instability of gravity water waves \cite{whitham,modbenjamin}, showing that the modulation 
equations for these waves are elliptic if the product $kh$ of the depth of the water $h$ and the wavenumber $k$ is above a 
critical value.  The breakthrough for the theory of DSWs was the realisation that if the modulation equations for a nonlinear, 
dispersive wave equation form a hyperbolic system, then a simple wave solution of these equations corresponds to a DSW, the 
first DSW solution being that for the KdV equation \cite{gur} based on its previously derived modulation equations 
\cite{whitham,modproc}.  The original derivation of the KdV modulation equations \cite{whitham,modproc} relied on extensive 
manipulation of what are essentially elliptic integrals as the periodic wave solution of the KdV equation is given
in terms of the Jacobian elliptic cosine function $\cn z$.  The KdV equation is the standard example of an equation which is 
exactly integrable via the method of inverse scattering \cite{whitham,newell}.  Using techniques from functional analysis, 
it was found that this spectral solution could be used to immediately determine the modulation equations for the KdV equation 
\cite{flash}, with the methods applicable to any integrable nonlinear, dispersive wave equation.  

In principle, the DSW solution for any nonlinear, dispersive wave equation with a modulationally stable periodic wave solution 
can then be found.  However, most nonlinear, dispersive wave equations are non-integrable.  While it is still difficult 
to determine the full DSW structure for non-integrable equations, a general method exists to find its solitary wave and 
linear dispersive wave edges if the DSW is of ``KdV-type,'' that is it consists of a monotonic modulated periodic wavetrain 
with solitary waves at one edge and linear dispersive waves at the other \cite{elreview,chaos}.  The reason that this can 
be done is that for a modulationally stable, nonlinear, periodic dispersive wave, its modulation equations are degenerate at 
these two edges and have a standard structure, which can be determined without detailed knowledge of the full modulation 
equations \cite{elreview,chaos}.  

This work deals with a type of non-standard DSW, that for which there is a resonance between the waves of the DSW and 
linear, diffractive radiation.  This resonance has a major effect on the DSW structure as the resonant radiation leaks mass 
and energy from it.  The standard example equations governing such resonant DSWs are the Kahawara equation \cite{kaw}
\begin{equation}
 \frac{\partial w}{\partial t} + 6 w\frac{\partial w}{\partial x} + \mu \frac{\partial^{3} w}{\partial x^{3}} +
 \frac{\partial^{5} w}{\partial x^{5}} = 0
 \label{e:kdv5}
\end{equation}
for $\mu \ge 0$, which is the KdV equation with the next higher order, fifth order, dispersion \cite{patkdv} and the equivalent 
higher order NLS equation 
\begin{equation}
 i\frac{\partial u}{\partial t} + \frac{1}{2} \frac{\partial^{2} u}{\partial x^{2}} - |u|^{2}u + 
 i\mu \frac{\partial^{3} u}{\partial x^{3}} = 0
 \label{e:resnls}
\end{equation}
with the next higher order, third order, dispersion \cite{trillores,trilloresfour,trilloresnature,trilloreslossbore}.  
The Kahawara equation arises for surface water waves when the effect of surface tension is included and the NLS equation 
with third order dispersion arises in nonlinear fibre optics, with third order dispersion being one of the higher order 
effects included for femtosecond pulses \cite{kivshar}.  The effect of resonant radiation on the DSW structure will be 
discussed in detail below, but it is sufficient to state here that if the effect of fifth order dispersion for the 
Kahawara equation is large enough, $\mu \ge 0$ below a critical value, then the classical KdV DSW structure disappears 
\cite{patkdv,pat}, with a strong resonant wavetrain terminated by a Whitham shock, a modulation shock wave in the modulated wave
variables, rather than by a KdV-type DSW \cite{pat,patjump}, as illustrated in Figure \ref{f:types}(d).  For DSW bearing 
nonlinear, dispersive wave equations, the corresponding modulation equations are hyperbolic and so, in theory, possess sharp 
gas dynamic type shock waves in the wave modulation parameters.  When he originally derived modulation theory 
\cite{whitham,modproc,modbenjamin}, Whitham speculated on the applicability of these modulation shock solutions, which seem to 
contradict the slowly varying assumption behind modulation theory.  Recent work \cite{patjump,gav} has shown that these 
modulation jump conditions are applicable and can be used to derive a wide variety of solutions for nonlinear, dispersive wave 
equations which have been derived in the past using more involved methods, or not at all.  Of interest here, they can be used 
to derive non-KdV type DSW solutions for which there is resonance with radiation, which have been derived in the past using more 
ad-hoc methods \cite{patkdv,pat}.  

This work will derive the full range of solutions for the six DSW types which arise as solutions of the equations governing 
optical beam propagation in the nonlinear optical medium of a nematic liquid crystal \cite{Khoo,PR,Wiley}.  The equations 
governing optical beam propagation in a nematic liquid crystal consist of an NLS-type equation for the electric field of 
the optical beam and an elliptic equation for the nematic response.  Previous studies \cite{nembore,nemgennady} have found 
that linear radiation can be in resonance with the waves of the DSW if the initial beam has a strong enough intensity 
discontinuity, with a resonant wavetrain propagating ahead of the main DSW.  Again, if the intensity jump is large enough, 
a DSW as such does not occur, with a resonant wavetrain dominating the solution, as for the Kawahara equation (\ref{e:kdv5}) 
\cite{patkdv}.  These previous studies did not give satisfactory agreement with numerical solutions of the nematic equations 
over the full range of intensity jump strengths.  The only good agreement was obtained for the so-called dam break case for which 
the optical intensity rises from $0$ and the solution consists of an expansion wave with no DSW structure \cite{nembore}.  The 
reason that good agreement between theoretical and numerical solutions was not obtained for the other cases is that the first 
study \cite{nembore} assumed that the solution was of KdV DSW-type for all the other five regimes, while the second study 
\cite{nemgennady} assumed that the solution always has a gas dynamic shock structure for all these five regimes.  It is found here 
that these two assumptions do not hold over all solution regimes.  The DSW is only of (perturbed) KdV-type for small initial 
intensity jumps.  As the initial optical intensity jump increases, the solution structure changes to the previously mentioned 
strong resonant wavetrain terminated by a Whitham shock.  This is not the same as the shock structure assumed by 
\cite{nemgennady} as this work assumed that the shock was a non-dispersive shallow water equation type shock.  It is found 
here that this approximation is not satisfactory as the full Whitham shock, including dispersion, has to be used 
to construct the solution in this regime.  In addition to these two limiting cases for small and large initial intensity jumps, 
the solution for the transition regime between these two limits will be constructed.  

In the next section, the six solution regimes for the resonant nematic DSW will be described.  The basic method used to 
derive the five non-trivial nematic DSW structures and understand their stability is Whitham modulation theory for the nematic 
equations, which is detailed in Section \ref{s:modtheory}.  In the subsequent sections, the solutions for the five non-trivial 
regimes (excluding the known dam break solution) will be derived using this and a combination of perturbed and approximate 
DSW theory.  As the solutions for each regime are developed they are compared with full numerical solutions of the nematic 
equations.  In all cases, excellent comparisons are obtained.

\section{Nematic Equations and DSW types}
\label{s:types}

Let us consider the propagation of a linearly polarised beam of coherent light through the 
nonlinear optical medium of a cell filled with a nematic liquid crystal \cite{PR,Wiley}.  The 
light is assumed to propagate in the direction $z$ down the cell with the beam polarised in
the $y$ direction.  The $x$ coordinate then completes the coordinate triad.  Typical physical
cell dimensions are $200\mu\mbox{m}\times30\mu\mbox{m}\times1\mbox{mm}$ \cite{PR}.  Nematic molecules are biaxial \cite{PR} 
with a tensor refractive index.  There are then two eigendirections for optical beams, termed the ordinary and extraordinary 
polarisations \cite{whitham}.  Only the extarordinary polarisation is dispersive and able to support solitary waves 
\cite{whitham}.  Hence, the optical beam will be assumed to be extraodinarily polarised.  An added complication 
of the nematic medium is that if the nematic molecules are initially aligned with 
their axis, termed the director, orthogonal to the electric field, the optical 
Fre\'edericksz threshold exists so that a minimum electric field strength is required 
to overcome the elastic forces of the nematic medium before the molecules can rotate 
\cite{Khoo,PR}.  To overcome this threshold so that milliwatt power optical beams can rotate the nematic 
molecules (and so change the refractive index of the medium), an external static electric field is applied to 
pre-tilt the nematic molecules at an angle $\theta_{0}$ \cite{assantokhoo}.  Low power optical beams are necessary 
so that there is minimal heating of the nematic as excessive heating can cause the medium to change phase out of 
the nematic state.  A nematic is typically a focusing medium, so that rotation of the nematic molecules by an 
optical beam increases the refractive index, leading to self-focusing \cite{PR,Wiley}.  However, the formation of 
a DSW requires the medium to be defocusing \cite{elreview}.  It has been found that the addition of small amounts 
of azo dye to the nematic changes the medium response, so that the nematic becomes defocusing \cite{gaetanodark}.  The 
addition of the dye increases scattering losses for the optical beam, but these will be neglected in the present work.
With this background and these assumptions, in the paraxial, slowly varying envelope approximation \cite{whitham}, 
the non-dimensional equations governing the propagation of an optical beam in a defocusing nematic are 
\cite{PR,Wiley,gaetanodark,conti2}
\begin{eqnarray}
 i \frac{\partial u}{\partial z} + \frac{1}{2}\frac{\partial^{2} u}{\partial x^{2}} - 2\theta u & = & 0 ,
\label{e:eeqn} \\
\nu \frac{\partial^{2}\theta}{\partial x^{2}} - 2q\theta & = & - 2|u|^{2} . \label{e:direqn}
\end{eqnarray}
Here, $u$ is the complex-valued slowly varying envelope of the electric field of the optical beam and $\theta$ is the 
extra optically induced rotation of the director beyond the imposed pre-tilt $\theta_{0}$.  The parameter $q$ is 
proportional to the square of the imposed pre-tilting electric field and $\nu$ is the non-dimensional elasticity 
of the nematic.  The electric field equation (\ref{e:eeqn}) is a defocusing NLS-type equation and the director equation
(\ref{e:direqn}) is forced by the intensity of the electric field.  Note that the nematic equations describe bulk waves,
including bulk solitary waves, as $z$ is a spatial coordinate, even though it is ``time-like'' in the context of the 
NLS-type equation (\ref{e:eeqn}).  For typical experimental conditions $\nu = O(100)$ 
\cite{waveguide,curvenature,curvejosab,curvepra}.  In this limit the nematic response is termed non-local in the sense that 
the response of the nematic to the optical forcing extends far beyond the waist of the optical beam \cite{PR,Wiley,conti2}.  
This nonlocal response and large value of $\nu$ will play a dominating role in the variety of structures possible for the 
DSW solution of the nematic equations (\ref{e:eeqn}) and (\ref{e:direqn}).  We note that in the so-called local limit 
$\nu \to 0$, the nematic equations (\ref{e:eeqn}) and (\ref{e:direqn}) reduce to the defocusing NLS equation, whose DSW 
solution is known \cite{gennady}.  While the nematic equations are of NLS-type, the nematic DSW more resembles a KdV DSW 
\cite{nembore,nemgennady} than an NLS DSW \cite{gennady} due to the large nonlocality $\nu$.  

While the system of equations (\ref{e:eeqn}) and (\ref{e:direqn}) has been presented in the context of the nonlinear optics 
of nematic liquid crystals, they are more general than this.  The same system of equations arises in the optics of nonlinear 
thermal optical media \cite{kuz,trillo6}, for example lead glasses \cite{lead,segev,segev2}, and certain photorefractive crystals 
\cite{photo}.  A similar system of equations also arises in simplified models of fluid turbulence \cite{alpha1} and quantum 
gravity \cite{grav}.   

To generate a DSW we shall use the step initial condition at $z=0$
\begin{equation}
 u = \left\{ \begin{array}{cc}
              u_{-}e^{iv_{-}x}, & x < 0 \\
              u_{+}e^{iv_{+}x}, & x > 0
             \end{array}
     \right. , \qquad
     \theta = \left\{ \begin{array}{cc}
              \frac{u_{-}^{2}}{q}, & x < 0 \\
              \frac{u_{+}^{2}}{q}, & x > 0,
             \end{array}
     \right. 
\label{e:ic}
\end{equation}
which is a jump in the amplitude of the optical field and its wavenumber.  

The simplest manner in which to analyse the DSW solution of the nematic equations (\ref{e:eeqn}) and (\ref{e:direqn}) 
is to use the polar coordinate transform (Madelung transformation) \cite{elreview,nembore,nemgennady}
\begin{equation}
 u = \sqrt{\rho} e^{i\phi}, \quad v = \phi_{x} .
\label{e:med}
\end{equation}
The nematic equations then become the system
\begin{eqnarray}
 \frac{\partial \rho}{\partial z} + \frac{\partial }{\partial x} \left( \rho v \right) & = & 0, \label{e:mass} \\
 \frac{\partial v}{\partial z} + v \frac{\partial v}{\partial x} + 2\frac{\partial \theta}{\partial x} 
- \frac{\partial}{\partial x} \left( \frac{\rho_{xx}}{4\rho} - \frac{\rho_{x}^{2}}{8\rho^{2}} \right) & = & 0,
\label{e:mom} \\
\nu \frac{\partial^{2}\theta}{\partial x^{2}} - 2q\theta & = & -2\rho . \label{e:thm} 
\end{eqnarray}
The importance of this polar form is that in the non-dispersive limit they reduce to the shallow water equations 
\cite{whitham} with $\rho$ playing the role of the fluid depth and $v$ playing the role of the fluid velocity.  In 
addition, in the non-dispersive limit these equations are also those for ideal compressible gas 
flow \cite{whitham}.  The non-dispersive nematic equations can be set in the Riemann invariant form
\begin{eqnarray}
 v + \frac{2\sqrt{2}}{\sqrt{q}} \sqrt{\rho} = R_{+} = \mbox{constant} & \mbox{on} & C_{+}:  
\frac{dx}{dz} = V_{+} = v + \frac{\sqrt{2}}{\sqrt{q}} \sqrt{\rho}
\label{e:cp} \\
v - \frac{2\sqrt{2}}{\sqrt{q}} \sqrt{\rho}= R_{-} = \mbox{constant} & \mbox{on} & C_{-}:  
\frac{dx}{dz} = V_{-} = v - \frac{\sqrt{2}}{\sqrt{q}} \sqrt{\rho} ,
\label{e:cm} 
\end{eqnarray}
with $\theta = \rho/q$.
These non-dispersive equations are an important component of finding DSW solutions as they apply outside 
the DSW itself and its associated resonant wavetrain, if they exist \cite{elreview,nembore,nemgennady}.

\begin{figure}
\centering
 \includegraphics[width=0.3\textwidth,angle=270]{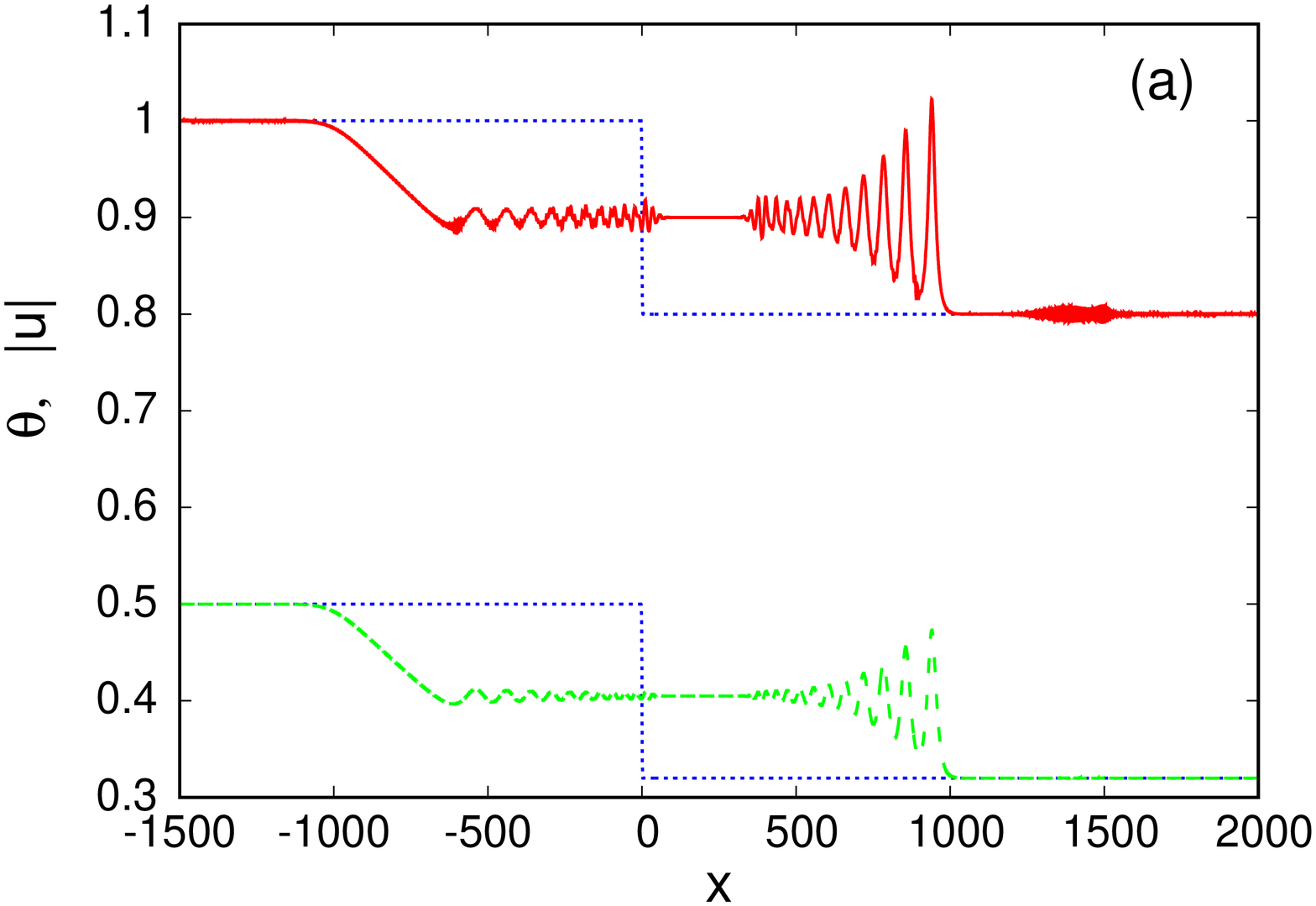}
 \includegraphics[width=0.3\textwidth,angle=270]{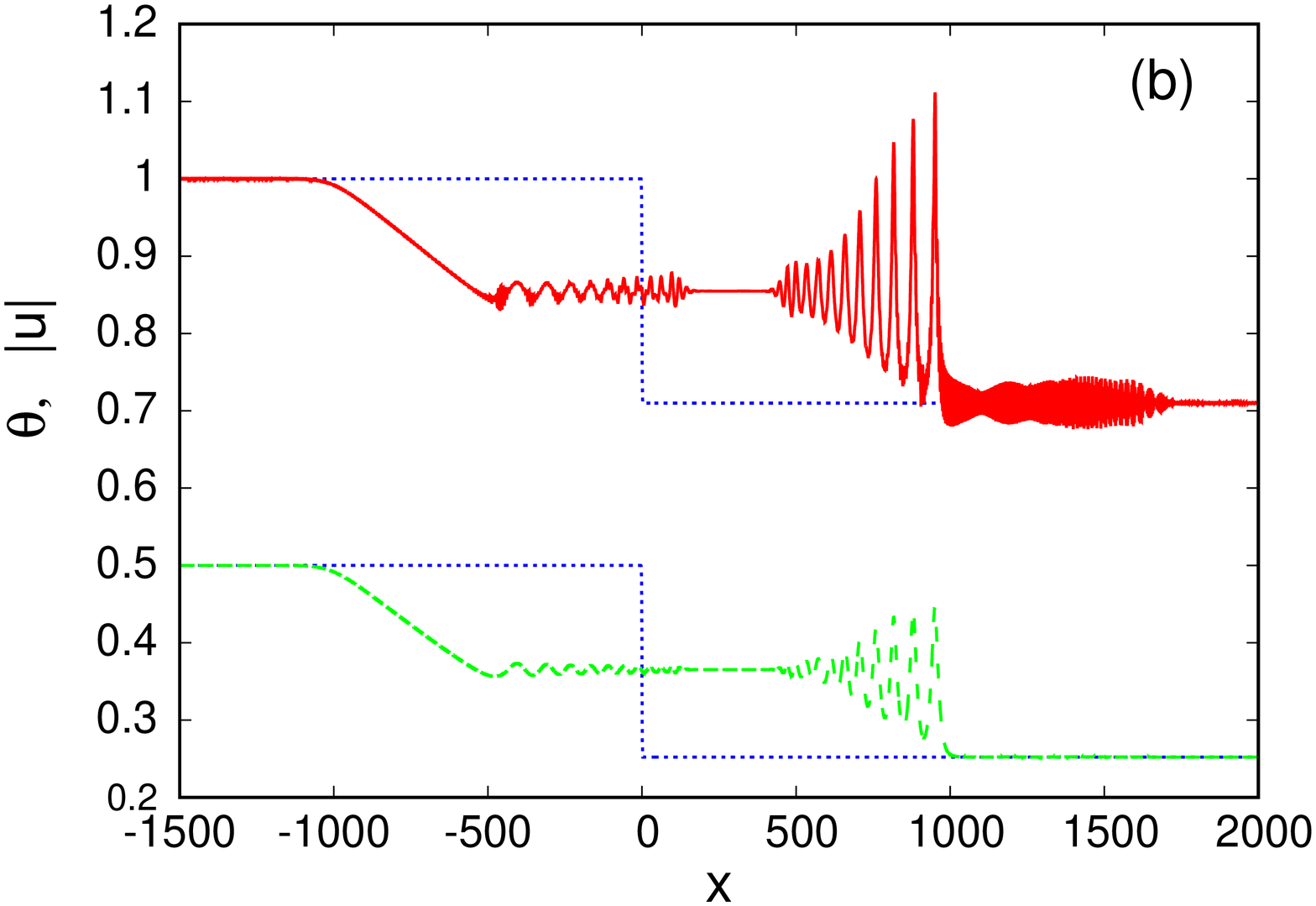}
 
 \includegraphics[width=0.3\textwidth,angle=270]{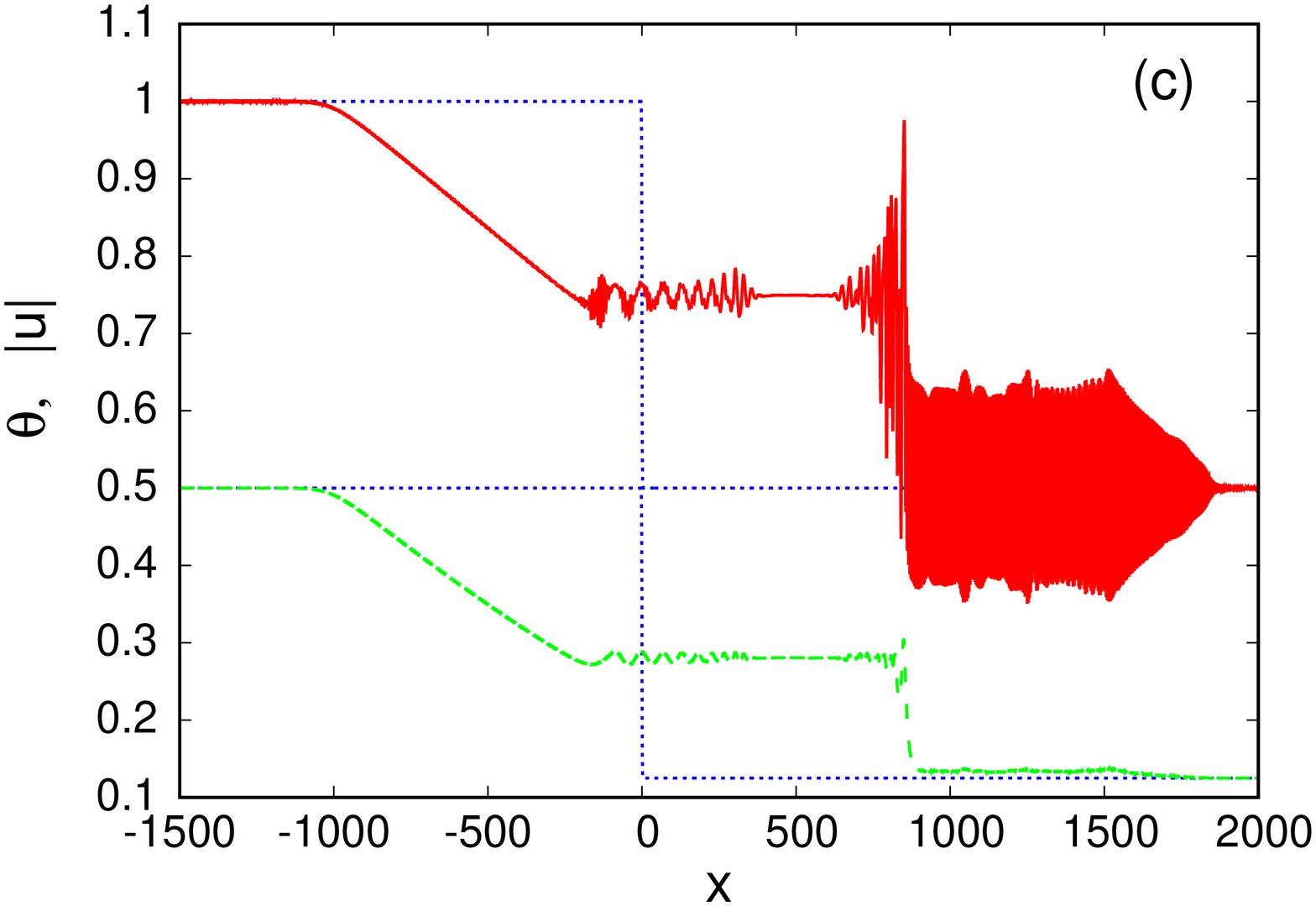}
 \includegraphics[width=0.3\textwidth,angle=270]{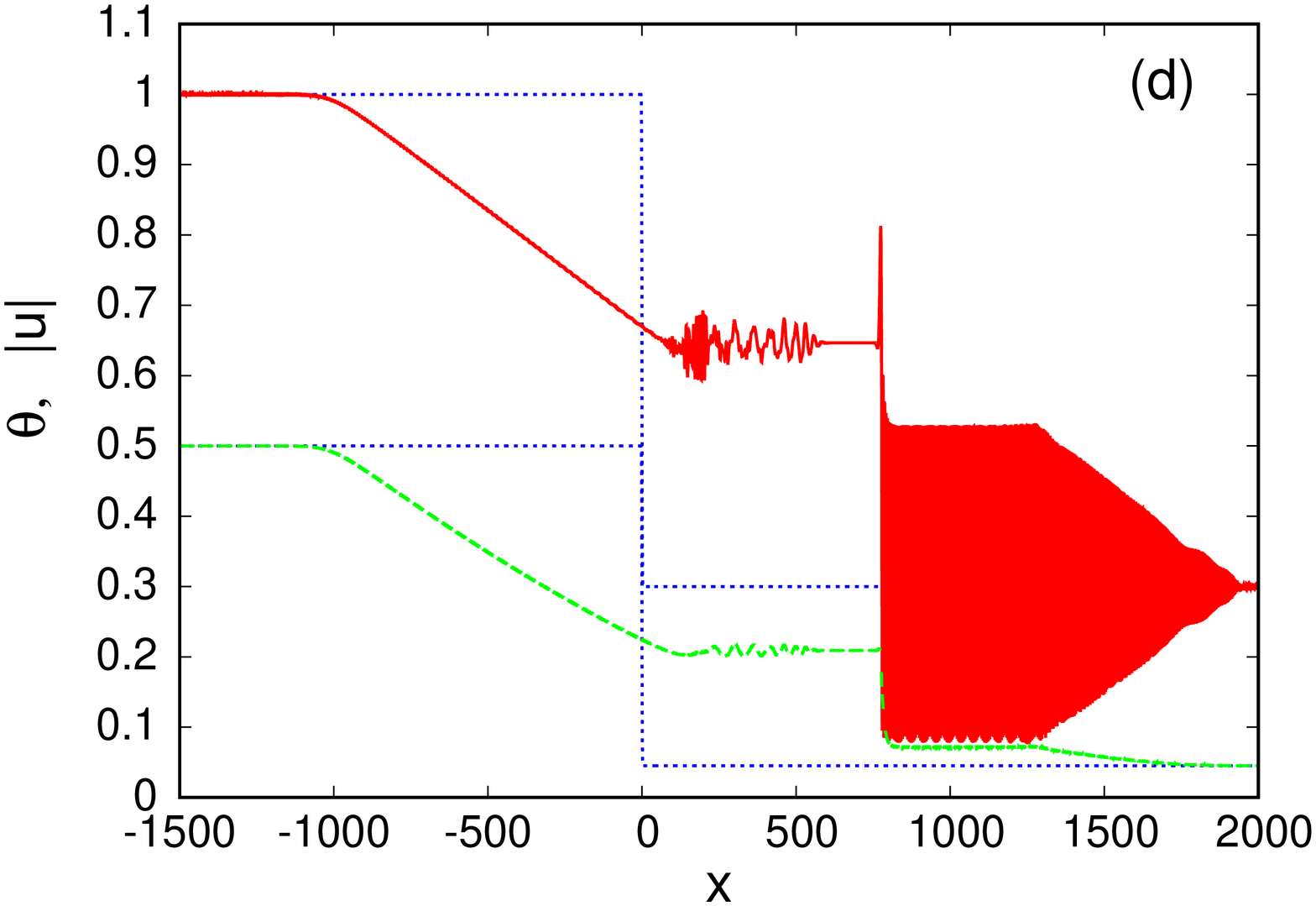}
 
 \includegraphics[width=0.3\textwidth,angle=270]{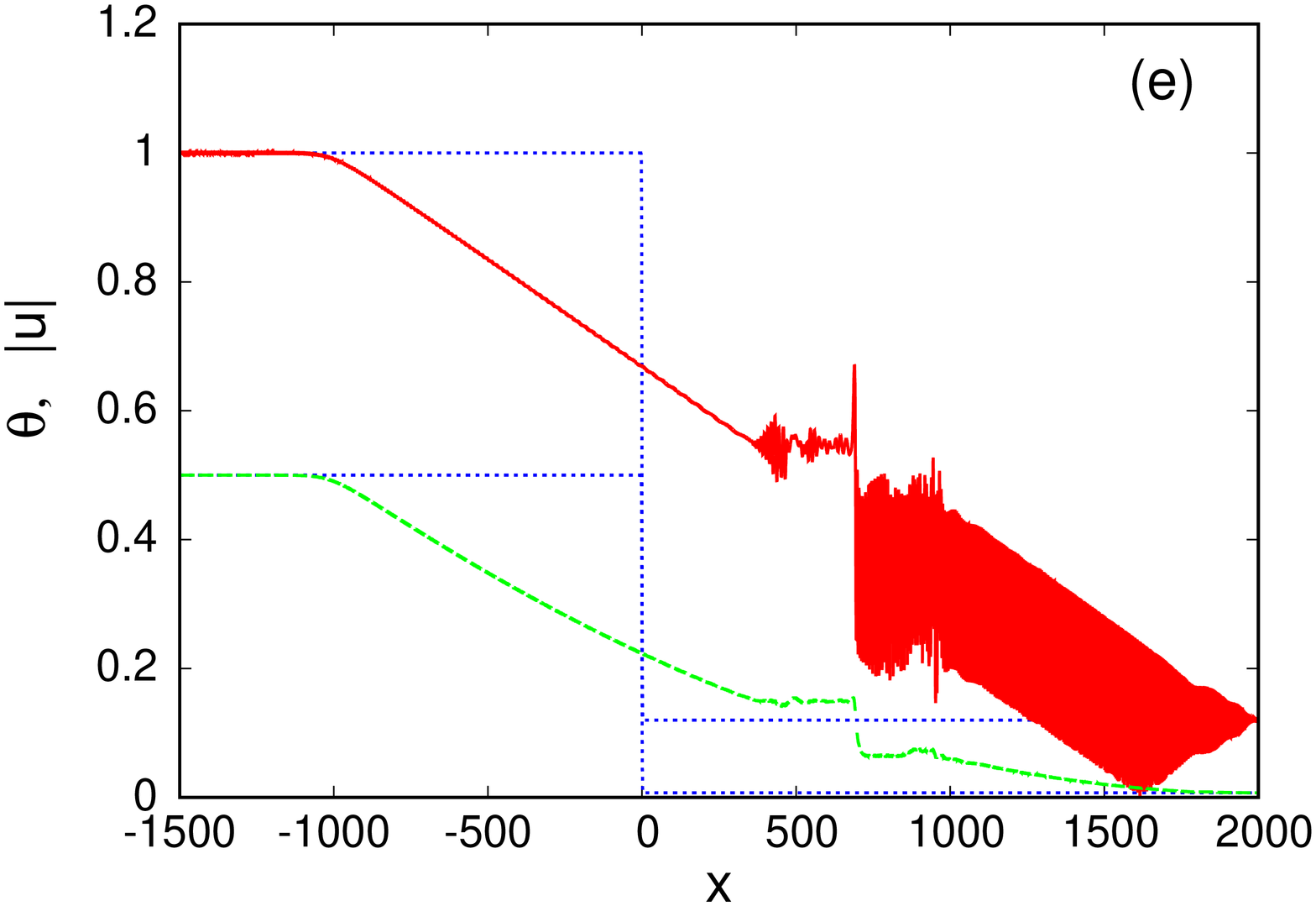}
 \includegraphics[width=0.3\textwidth,angle=270]{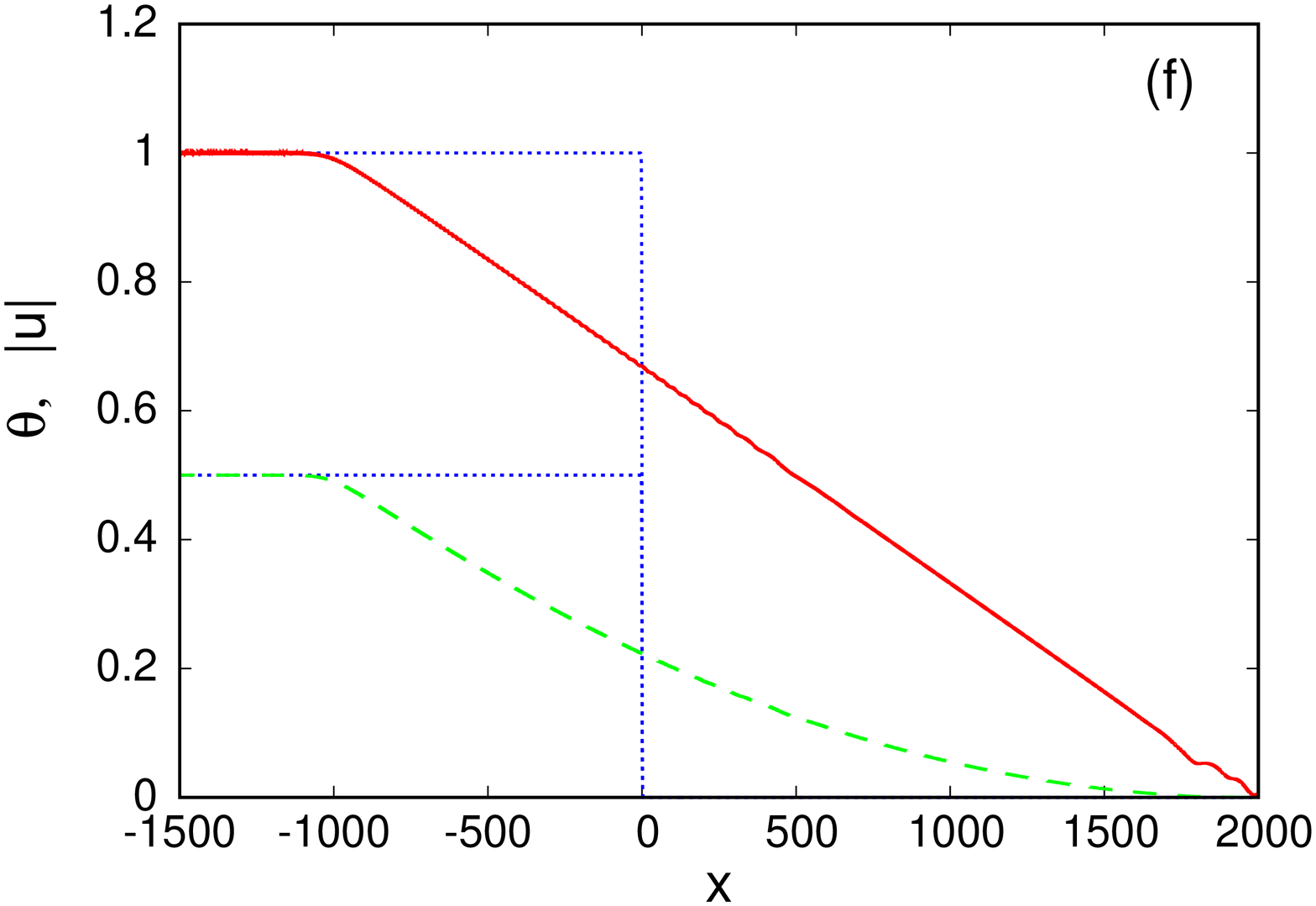}
\caption{Numerical solutions of nematic equations (\ref{e:eeqn}) and (\ref{e:direqn}) for
initial condition (\ref{e:ic}).  Red (solid) lines:  $|u|$ at $z=1000$; green (dashed)
lines $\theta$ at $z=1000$; blue (dotted) lines:  $|u|$ at $z=0$ (upper) and $\theta$ at $z=0$ (lower).
(a) PDSW with $u_{+}= 0.8$, (b) RDSW with $u_{+} = 0.71$, (c) CDSW with $u_{+} = 0.5$, (d) TDSW with
$u_{+} = 0.3$, (e) VDSW with $u_{+} = 0.12$, (f) dam break solution with $u_{+} = 0$.  Here $u_{-} = 1$, $\nu = 200$ 
and $q =2$.} 
\label{f:types}
\end{figure}

Before deriving the solutions for the various DSW regimes, same basic properties of solutions of the nematic equations will be 
presented.  The dispersion relation for linear waves will play a key role in the derivation of the nematic DSW 
solutions.  This is found by linearising about the background states $\bar{\rho}$ in $\rho$ and $\bar{v}$ in $v$, with
\begin{equation}
 \rho = \bar{\rho} + \tilde{\rho}e^{i(kx - \omega z)}, \quad v = \bar{v} + \tilde{v}e^{i(kx - \omega z)}, 
 \quad \theta = \frac{\bar{\rho}}{q} + \tilde{\theta}e^{i(kx - \omega z)},
\label{e:rho1}
\end{equation}
assuming that $|\tilde{\rho}| \ll \bar{\rho}$, $|\tilde{v}| \ll |\bar{v}|$ and $|\tilde{\theta}| \ll \bar{\rho}/q$.
The linear dispersion can then be found as \cite{nembore,nemgennady}
\begin{equation}
 \omega = k\bar{v} + \frac{\sqrt{\bar{\rho}k}}{\sqrt{\nu k^{2} + 2q}} \left[ 
\frac{\nu k^{2} + 2q}{4\bar{\rho}} k^{3} + 4k \right]^{1/2} .
\label{e:disp}
\end{equation}
A fundamental result of this work is that the DSW for the nematic equations (\ref{e:eeqn}) and (\ref{e:direqn}) is
itself on long $x$ and $z$ scales, while the resonant wavetrain it sheds are short waves with wavenumber $k$ satisfying
$\nu k^{2} \gg 1$.  In this short wave limit, the dispersion relation (\ref{e:disp}) becomes
\begin{equation} 
 \omega = k\bar{v} + \frac{1}{2} k^{2} + \ldots 
\label{e:displargenu}
\end{equation}

Let us first consider the solution of the nematic equations (\ref{e:eeqn}) and (\ref{e:direqn}) with the step initial 
condition (\ref{e:ic}) for no wavenumber jump, so that $v_{-} = v_{+} = 0$.  The solution
depends strongly on the height $u_{-} - u_{+}$ of the jump, with six distinct DSW types being identified.  Figure
\ref{f:types} shows typical examples of these six DSW types.  Four of these
DSW types are similar to those for the Kawahara equation (\ref{e:kdv5}).  The two remaining relate to the existence 
of vacuum cases with $\rho = 0$ at some point.  The details of these six DSW types are as follows.

\begin{enumerate}

 \item  Type 1, perturbed dispersive shock wave (PDSW):  
 This nematic DSW type resembles the standard KdV 
 DSW \cite{gur,bengt}, with a monotonic modulated wavetrain consisting of bright solitary waves at the leading
 edge and linear waves at the trailing edge.  There is no resonant wavetrain ahead of the DSW as there is
 no resonance between the phase velocity of a possible resonant wavetrain, determined by (\ref{e:displargenu}),
 with the DSW \cite{nembore,nemgennady}.  As the nematic equations (\ref{e:eeqn}) and (\ref{e:direqn}) are 
 bi-directional, there is also a backwards propagating expansion wave, as for the compressible flow shock
 tube problem \cite{whitham}, which links the level behind $u_{-}$ and an intermediate shelf of height $|u| = u_{i}$
 behind the DSW.  The DSW takes the place of the shock in the shock tube problem.  The solution for this DSW type 
 will be found as a perturbation of the KdV DSW. 
 
 \item  Type 2, radiating dispersive shock wave (RDSW):  When the jump height reaches a critical value \cite{nembore}, which 
 will be discussed in detail in Section \ref{s:pdsw}, resonance can exist between the DSW and diffractive waves, so that a 
 resonant wavetrain ahead of the DSW is generated, as shown in Figure \ref{f:types}(b).  The DSW itself resembles a KdV
 DSW, but with an attached resonant wavetrain.  This case resembles the equivalent RDSW type for the Kawahara
 equation (\ref{e:kdv5}) \cite{patkdv}.  The DSW itself will be found to be a perturbed KdV DSW, as for the PDSW.
 
 \item  Type 3, crossover dispersive shock wave (CDSW):  With increasing jump height, the DSW becomes unstable, as does
 its resonant wavetrain, as seen in Figure \ref{f:types}(c).  The DSW loses its rank ordered structure and the resonant 
 wavetrain has a highly modulated amplitude.  Again, this DSW structure is similar to its equivalent for the Kawahara equation 
 \cite{patkdv}.   
 
 \item  Type 4, travelling dispersive shock wave (TDSW):  At a critical jump height, the DSW itself disappears, leaving just 
 a resonant wavetrain, as shown in Figure \ref{f:types}(d).  This solution form is similar to the travelling DSW (TDSW) found 
 for the Kawahara equation (\ref{e:kdv5}) \cite{patkdv}.  A remnant of the DSW is left in the form of a negative polarity 
 solitary wave which links the resonant wavetrain to the intermediate level.  It is further seen that the resonant wavetrain 
 is of higher amplitude than in the CDSW case and has stabilised.
 
 \item  Type 5, vacuum dispersive shock wave (VDSW):  As $u_{+}$ decreases, $u_{-} - u_{+}$ increases, and the amplitude of 
 the resonant wavetrain grows, the minimum of its oscillation eventually hits the vacuum point $u=0$, at which point there 
 is a phase singularity and the DSW solution changes form, as seen in Figure \ref{f:types}(e).  A constant amplitude resonant 
 wavetrain now propagates on a varying mean level.  Behind this, there is a resonant wavetrain on a constant mean level, which 
 is linked to the intermediate level by a negative polarity wave, as for the TDSW.  A similar vacuum point solution occurs for 
 the defocusing NLS equation \cite{gennady}, although without the accompanying resonant wavetrain. 
 
 \item  Type 6 (dam break solution):  When the level ahead $u_{+}=0$, the solution becomes essentially non-diffractive, as seen 
 in Figure \ref{f:types}(f).  This solution is referred to as the dam break solution as it is the simple wave solution of 
 the shallow water equations (\ref{e:cp}) and (\ref{e:cm}) on the characteristic $C_{-}$ and arises as a solution for the 
 flow generated by a breaking dam \cite{whitham}.  As this dam break solution was considered in detail in previous work 
 \cite{nembore} and was found to be in near perfect agreement with numerical solutions of the nematic equations (\ref{e:eeqn}) 
 and (\ref{e:direqn}), it will not be considered in detail here.
 
\end{enumerate}

The solution types shown in Figure \ref{f:types} have the general form that outside the DSW and the resonant wavetrain
(if they exist) the solution is non-diffractive and governed by the non-dispersive equations (\ref{e:cp}) and (\ref{e:cm}).
There is an expansion fan which links the level $u_{-}$ to an intermediate level $u_{i}$, with the DSW (and resonant wavetrain)
then taking the solution from $u_{i} = \sqrt{\rho_{i}}$ to the level $u_{+}$ ahead.  The expansion wave is given by the 
simple wave solution on the characteristic $C_{-}$ of the Riemann invariant form (\ref{e:cp}) and (\ref{e:cm}) of the 
non-dispersive equations with the Riemann invariant on $C_{+}$ constant.  If we set the velocity of the trailing
edge of the DSW (or Whitham shock, depending on the regime) to be $s_{i}$, then this expansion wave solution is 
\cite{nembore,nemgennady}
\begin{equation}
|u| = \sqrt{\rho} = \left\{ \begin{array}{cc}
                        u_{-}, & \frac{x}{z} < -\frac{\sqrt{2}}{\sqrt{q}} u_{-}\\
                        \frac{\sqrt{q}}{3\sqrt{2}} \left[ \frac{2\sqrt{2}}{\sqrt{q}}u_{-} - \frac{x}{z} \right], &
-\frac{\sqrt{2}}{\sqrt{q}} u_{-}\le \frac{x}{z} \le \frac{\sqrt{2}}{\sqrt{q}} \left( 2u_{-} - 3\sqrt{\rho_{i}}
\right) \\
\sqrt{\rho_{i}}, & \frac{\sqrt{2}}{\sqrt{q}} \left( 2u_{-} - 3\sqrt{\rho_{i}} \right) < \frac{x}{z} \le s_{i} ,
\end{array}
\right.
\label{e:midsolnu}
\end{equation}
with $v = 2\sqrt{2}(u_{-} - \sqrt{\rho})/\sqrt{q}$.
It is then clear that
\begin{equation}
 v = v_{i} = \frac{2\sqrt{2}}{\sqrt{q}} \left( u_{-} - \sqrt{\rho_{i}} \right) 
\label{e:v2expr}
\end{equation}
on the intermediate shelf.  Previous work has shown that this expansion wave solution is in excellent agreement with full 
numerical solutions of the nematic equations (\ref{e:eeqn}) and (\ref{e:direqn}) \cite{nembore,nemgennady}, so it will not 
be compared with numerical solutions in this work.  To determine the height $u_{i} = \sqrt{\rho_{i}}$ of the intermediate
shelf and the velocity $s_{i}$ of its leading edge, the DSW needs to be determined.  The dam break problem corresponds to 
$\rho_{i}=0$, in which case $v_{i} = 2u_{-} \sqrt{2/q}$ \cite{nembore}.

In the small amplitude limit, the nematic equations (\ref{e:eeqn})--(\ref{e:direqn}) can be reduced to a 
KdV equation \cite{nembore,nemgennady,horikis} on using the expansions
\begin{eqnarray}
 |u| = \sqrt{\rho} & = & u_{0} + \epsilon^{2} u_{1}(\xi,\eta) + \epsilon^{4}u_{2}(\xi,\eta) + \ldots, \label{e:ukdv} \\
 v & = & \epsilon^{2}V_{1}(\xi,\eta) + \epsilon^{4} V_{2}(\xi,\eta)
+ \epsilon^{6}V_{3}(\xi,\eta) + \ldots, \label{e:phikdv} \\
 \theta & = & \frac{u_{0}^{2}}{q} + \epsilon^{2}\theta_{1}(\xi,\eta) + \epsilon^{4}\theta_{2}(\xi,\eta) 
+ \epsilon^{6}\theta_{3}(\xi,\eta) + \ldots \label{e:thetakdv}
\end{eqnarray}
for small deviations from the level $u_{0}$, with $|\epsilon| \ll 1$ being a measure of this 
deviation.  In the moving, scaled coordinates $\xi = \epsilon \left( x - U z \right)$ and 
$\eta = \epsilon^{3} z$, with $U^{2} = (2/q) u_{0}^{2}$, the correction $u_{1}$ satisfies a KdV equation with 
fifth order dispersion, or Kawahara equation, \cite{kaw,nemgennady}
\begin{equation}
\frac{\partial u_{1}}{\partial \eta} + 3\sqrt{\frac{2}{q}}u_{1}\frac{\partial u_{1}}{\partial \xi}
+ \sqrt{\frac{2}{q}}\frac{u_{0}}{4}\left( \frac{\nu }{q} - \frac{q}{4u_{0}^{2}} \right) \frac{\partial^{3}u_{1}}{\partial \xi^{3}} 
+ \sqrt{\frac{2}{q}}\frac{3\epsilon^{2}\nu^{2} u_{0}^{2}}{16q^{2}} \frac{\partial^{5} u_{1}}{\partial \xi^{5}} = 0.
\label{e:kdv5nem}
\end{equation}
Note that in the highly nonlocal limit $\nu \gg 1$ the coefficient of the third derivative term is positive, so that 
solitary waves are waves of elevation, which explains why the DSWs illustrated in Figure \ref{f:types} are similar
to KdV-type DSWs of elevation with solitary waves at the leading edge and linear diffractive waves at the 
trailing edge, not NLS-type DSWs, which consist of waves of depression for which the positions of the solitary waves
and linear diffractive waves are reversed \cite{gennady}.  The fifth derivative term is nominally of higher order,
but due to the high nonlocality $\nu$ the combination $\epsilon \nu$ can be $O(1)$ \cite{nemgennady}.  This reduction 
of the nematic equations to the Kawahara equation explains the presence of the resonant wavetrains seen in 
Figure \ref{f:types} as it has been shown that for the Kawahara equation (\ref{e:kdv5}) linear diffractive waves 
resonate with the DSW if $\mu > 0$.  In this case, the Kawahara DSW can be a RDSW, CDSW or TDSW, depending on the 
value of $\mu$.  The vacuum cases of Figures \ref{f:types}(d) and (e) cannot occur for a KdV DSW.  They arise
for the nematic DSW and NLS-type DSWs due to the requirement $|u| \ge 0$.  The Kawahara equation reduction of the full 
nematic equations will be used in Sections \ref{s:pdsw}--\ref{s:cross} to derive the DSW solution as the cases studied 
there correspond to a small jump height $u_{-} - u_{+}$.  

The analytical solutions for the DSW in the five regimes, excluding the previously studied dam break limit,
will be compared with full numerical solutions of the nematic equations.
The nematic equations (\ref{e:eeqn}) and (\ref{e:direqn}) were solved numerically using a 
pseudo-spectral method, with the $x$ derivatives calculated in Fourier space and the solution 
advanced in $z$ in Fourier space using the fourth order Runge-Kutta scheme.  This pseudo-spectral scheme is
based on the classic scheme of Fornberg and Whitham \cite{bengt}, but extended to improve
its stability and accuracy \cite{chan,tref}.  The propagation in $z$ using the Runge-Kutta 
scheme is done in Fourier space with the dispersion $u_{xx}$ propagated using an integrating 
factor, which improves stability at high wavenumbers, see \cite{chan}.  The step initial condition
(\ref{e:ic}) was smoothed using a hyperbolic tangent with a width $w$.  In addition, as a Fourier
method is used, periodicity at the ends of the computational domain is ensured by transforming the step 
into a ``top hat.''  The smoothed initial condition used in the numerical solutions was then
\begin{equation}
 u(x,0) = \left[ \frac{1}{2} \left( u_{-} - u_{+} \right) \left( \tanh \frac{x+D}{W} - \tanh \frac{x}{W} \right) + u_{+}
 \right] e^{i\phi }, 
 \label{e:smooth}
\end{equation}
with $v = \phi_{x}$ smoothed in a similar fashion with $u_{-}$ replaced by $v_{-}$ and $u_{+}$ replaced by $v_{+}$.
The width $W$ was chosen to be large enough to stop instabilities, but small enough to well approximate a step.
In practice, $W=1$ was found suitable.  The distance $D$ at which the initial condition goes back to the level $u_{+}$
was chosen large enough so that the waves generated from the step at $x=0$ do not interact with the step down,
with $D=-10000$ found suitable.  Finally, a suitable number of Fourier modes was $N = 131072$, with a $z$ step
$dz = 0.002$.  All the comparisons with numerical solutions will be for the parameter choices $q=2$ and $\nu = 200$.  In 
particular, the value of the nonlocality $\nu$ depends on the beam power and wavelength, but $\nu = 200$ is typical for 
near-infrared beams of milliwatt powers \cite{waveguide,curvenature,curvejosab,curvepra}.


\section{Modulation Theory}
\label{s:modtheory}

The key to understanding the behaviours of and for deriving solutions for the nematic DSW and its associated
resonant wavetrain will be a solution for the resonant wavetrain and a knowledge of its stability.  To study this stability, 
Whitham modulation theory will be used to generate modulation equations for the periodic wave solution of the nematic 
equations \cite{whitham}.  Unfortunately, there are no known general solitary wave or periodic wave solutions of the 
nematic equations.  However, the resonant wavetrain has small amplitude, so that the periodic wave solution will be found 
as a Stokes' expansion, as was done for the Kahawara equation \cite{patkdv} and the fifth order KdV equation, the Kahawara 
equation (\ref{e:kdv5}) with $\mu = 0$ \cite{pat}.  We then seek a Stokes' expansion solution of the nematic equations as
\begin{eqnarray}
 \rho & = & \bar{\rho} + a \cos \varphi + a^{2} \rho_{2} \cos 2\varphi + a^{3} \rho_{3} \cos 3\varphi + \ldots, \label{e:rhostokes} \\
 v & = & \bar{v} + av_{1}\cos \varphi + a^{2}v_{2} \cos 2\varphi + a^{3} v_{3} \cos 3\varphi + \ldots, \label{e:vstokes} \\
 \theta & = & \frac{\bar{\rho}}{q} + a\theta_{1} \cos \varphi + a^{2} \theta_{2} \cos 2\varphi 
 + a^{3} \theta_{3} \cos 3\varphi + \ldots , \label{e:thetastokes} \\
 \omega & = & \omega_{0} + a\omega_{1} + a^{2} \omega_{2} + \ldots, \label{e:omstokes}
\end{eqnarray}
where the phase is $\varphi = kx - \omega z$.  As is standard, we now substitute the expansions 
(\ref{e:rhostokes})--(\ref{e:omstokes}) into the nematic equations (\ref{e:eeqn})--(\ref{e:direqn}) and solve the 
equations at each order of $a$, eliminating secular terms by appropriately choosing the $\omega_{j}$.  At $O(a)$ we find the 
linear dispersion relation
\begin{equation}
 \omega_{0} = k\bar{v} + \frac{\sqrt{\bar{\rho}k}}{\sqrt{\nu k^{2} + 2q}} \left[ 
 \frac{\nu k^{2} + 2q}{4\bar{\rho}} k^{3} + 4k \right]^{1/2} ,
\label{e:order0}
\end{equation}
in accordance with (\ref{e:disp}).  At $O(a^{2})$ we find there are no secular terms and
\begin{equation}
 \omega_{1} = 0, \quad v_{1}=\frac {1}{\sqrt{\bar{\rho}\,k \left( \nu\,{k}^{2}+2\,q \right)}}
\left[ 4k+\frac { \left( \nu\,{k}^{2}+2\,q \right) {k}^{3}}{4\bar{\rho}}
\right]^{1/2}, \quad \theta_{1} = \frac{2}{\nu\,{k}^{2}+2\,q} .
 \label{e:order1}
\end{equation}
Finally, at $O(a^{3})$ to eliminate secular terms we find the nonlinear dispersion relation correction
\begin{equation}
\omega_{2}=-\left[ {\frac {{k}^{4}\nu+2\,{k}^{2}q+16\,\bar{\rho}}{\nu\,{k}^{2}+2\,q}}\right]^{1/2}
{\frac {{k}^{6}{\nu}^{2}+5{k}^{4}\nu\,q/2+12\,{k}^{2}\nu\,\bar{\rho}
+{k}^{2}{q}^{2}+8\,q\bar{\rho}}{8{\bar{\rho}}^{2}k \left( {k}^{4}{\nu}^{2}+5{
k}^{2}\nu\,q/2+{q}^{2}-4\,\nu\,\bar{\rho} \right) }} .
\label{e:omega2nu}
\end{equation}
The next order terms in the expansions $\rho_{2}$, $v_{2}$ and $\theta_{2}$ are not needed and so are not given here.

This Stokes' expansion will be used to calculate the Whitham modulation equations for the Stokes' wave.  The expressions 
for $v_{1}$ and $\omega_{2}$ are involved and not really suitable to calculate amenable modulation equations.  Fortunately, 
the resonant wavetrain is a short wave relative to the DSW and the nonlocal limit of $\nu$ large can be taken, as in calculating 
the short wave linear dispersion relation (\ref{e:displargenu}).  We also note that this is the physical limit in experiments 
\cite{waveguide,curvenature,curvejosab,curvepra}.  Taking this nonlocal limit, we have
\begin{equation}
 v_{1} = {\frac {k}{2\bar{\rho}}}+{\frac {4}{\nu\,{k}^{3}}}- \frac{8qk^{2} + 16 \bar{\rho}}{\nu^{2} k^{7}} + \ldots, 
 \quad \theta_{1} = {\frac {2}{\nu\,{k}^{2}}}-{\frac {4q}{{k}^{4}{\nu}^{2}}} + \ldots
\label{e:v1expand}
\end{equation}
and the dispersion relation amplitude correction
\begin{equation}
 \omega_{2} = -{\frac {{k}^{2}}{8\bar{\rho}^2}}-{\frac {3}{{k}^{2}\nu\,\bar{\rho}}}-
 \frac{20\bar{\rho} - 6qk^{2}}{\nu^{2}k^{6}\bar{\rho}}
 + \ldots 
\label{e:om2expand}
\end{equation}

                 
To calculate Whitham modulation equations, either a Lagrangian for the governing equations, or conservation
laws for these equations can be used, or a mixture of both \cite{whitham,mod1,modproc}.  The use of Lagrangians has resulted 
in this method of analysing slowly varying modulated wavetrains also being termed the method of averaged Lagrangians.  As for 
the determination of the modulation equations for the KdV equation, it was found that a combination of 
an averaged Lagrangian and conservation equations is optimal \cite{whitham,modproc}.  The nematic equations in shallow
water form (\ref{e:mass})--(\ref{e:thm}) have the Lagrangian
\begin{equation}
 L = -2\rho \phi_{z} - \frac{1}{4} \frac{\rho_{x}^{2}}{\rho} - \rho \phi_{x}^{2} - 4\rho \theta 
 + \nu \theta_{x}^{2} + 2q\theta^{2}.
 \label{e:lag}
\end{equation}
The nematic equations (\ref{e:mass})--(\ref{e:thm}) also have the ``mass'' conservation equation
\begin{equation}
 \frac{\partial}{\partial z} \rho + \frac{\partial}{\partial x} \rho v = 0,
 \label{e:masscons}
\end{equation}
the ``momentum'' conservation equation
\begin{equation}
 \frac{\partial}{\partial z} \left( \rho v \right) + \frac{\partial}{\partial x} \left[ \rho v^{2} - \frac{1}{4} \rho_{xx}
 + \frac{\rho_{x}^{2}}{4\rho} - \frac{1}{2} \nu \theta_{x}^{2} + q\theta^{2} \right] = 0
 \label{e:momcons}
\end{equation}
and the ``energy'' conservation equation
\begin{equation}
 \frac{\partial}{\partial z} \left[ \rho v^{2} + \frac{\rho_{x}^{2}}{4\rho} + 4\rho \theta - \nu \theta_{x}^{2}
 - 2q\theta^{2} \right] + \frac{\partial}{\partial x} \left[ \rho v^{3} + \frac{1}{2}\rho_{x}v_{x} 
 + \frac{3v\rho_{x}^{2}}{4\rho} - \frac{1}{2}v\rho_{xx} + 4\rho v \theta + 2\nu \theta_{x}\theta_{z} \right] = 0.
 \label{e:energycons}
\end{equation}
The mass and momentum conservation equations follow directly from (\ref{e:mass})--(\ref{e:thm}).  The easiest
method to determine the energy conservation equation is to use N\"other's Theorem on the Lagrangian (\ref{e:lag})
based on invariances with respect to $z$ \cite{gelfand}.  The terms mass, momentum and energy equations are used
in the sense of invariances of the Lagrangian (\ref{e:lag}) and how they arise in water wave theory.  While these
terms will be used here, in optics they have different interpretations.  For instance, the mass conservation 
equation (\ref{e:masscons}) is conservation of optical power.  Finally, there is the wave conservation equation
\begin{equation}
 \frac{\partial k}{\partial z} + \frac{\partial \omega}{\partial x} = 0.
 \label{e:wavecon}
\end{equation}
This is based on the definition of the slowly varying wavenumber $k = \varphi_{x}$ and frequency $\omega = -\varphi_{z}$
in terms of the (slowly varying) phase $\varphi$.  

The averaged Lagrangian for the Stokes' wave (\ref{e:rhostokes})--(\ref{e:thetastokes}) is calculated
by substituting this weakly nonlinear wave into the Lagrangian (\ref{e:lag}) and averaging by integrating over 
a period, that is integrating in $\varphi$ from $0$ to $2\pi$.  In conjunction with the Madelung transformation
(\ref{e:med}), a pseudo-phase $\psi$ is introduced for the mean flow $\bar{v}$ with $\gamma = -\psi_{z}$ and 
$\bar{v} = \psi_{x}$, as is standard for Whitham modulation theory, in particular for the modulation equations
for the KdV equation \cite{whitham,modproc}.  This results in the consistency equation
\begin{equation}
 \frac{\partial \bar{v}}{\partial z} + \frac{\partial \gamma}{\partial x} = 0,
 \label{e:meancon}
\end{equation}
which is analogous to the wave conservation equation (\ref{e:wavecon}).  We then have that the averaged Lagrangian is
\begin{equation}
 {\cal L} = 2\gamma \bar{\rho} - \bar{\rho} \bar{v}^{2} - 2\frac{\bar{\rho}^{2}}{q} + 
 \left[ \frac{\omega}{2\bar{\rho}} - \frac{1}{4} \frac{k^{2}}{\bar{\rho}} - \frac{1}{2} \frac{k\bar{v}}{\bar{\rho}}
 \right] a^{2} + \ldots 
 \label{e:avlag}
\end{equation}
to $O(a^{2})$.  Note that to obtain the Stokes' expansion dispersion relation (\ref{e:omstokes}) by taking
variations with respect to $a$, the averaged Lagrangian is needed to $O(a^{4})$.  However, this dispersion
relation is known via (\ref{e:omstokes}), so that averaged wave conservation can be directly calculated by
averaging the wave conservation equation (\ref{e:wavecon}) using this dispersion relation.  So these $O(a^{4})$ 
terms are not needed here.

Taking variations of the averaged Lagrangian (\ref{e:avlag}) with respect to the pseudo-phase $\psi$ 
\begin{equation}
 \frac{\partial}{\partial z} \frac{\partial {\cal L}}{\partial \gamma} - \frac{\partial}{\partial x}
 \frac{\partial {\cal L}}{\partial \bar{v}} = 0
 \label{e:vargamma}
\end{equation}
results in the averaged mass conservation equation  
\begin{equation}
 \frac{\partial \bar{\rho}}{\partial z} + \frac{\partial}{\partial x} \left[ \bar{\rho}\bar{v}
 + \frac{k}{4\bar{\rho}} a^{2} \right] = 0,
 \label{e:avmass}
\end{equation}
which can also be obtained by directly averaging the mass conservation equation (\ref{e:masscons}).  Taking
variations with respect to the phase $\varphi$
\begin{equation}
 \frac{\partial}{\partial z} \frac{\partial {\cal L}}{\partial \omega} - \frac{\partial}{\partial x}
 \frac{\partial {\cal L}}{\partial k} = 0
 \label{e:varphase}
\end{equation}
gives the conservation of wave action equation
\begin{equation}
 \frac{\partial}{\partial z} \frac{a^{2}}{\bar{\rho}} + \frac{\partial}{\partial x} 
 \left( \frac{\bar{v}}{\bar{\rho}} + \frac{k}{\bar{\rho}}\right) a^{2} = 0.
 \label{e:waveaction}
\end{equation}
Averaging the conservation of waves equation (\ref{e:wavecon}) gives
\begin{equation}
 \frac{\partial k}{\partial z} + \frac{\partial}{\partial x} \left( k\bar{v} + \frac{1}{2}k^{2} - 
 \frac{k^{2}}{8\bar{\rho}^{2}} a^{2} + \frac{2\bar{\rho}}{q}
 \right) = 0.
 \label{e:waveconsav}
\end{equation}
Note that to obtain the dispersion relation for the original phase $\phi$, rather than for 
$v = \phi_{x}$, the mean term $2\bar{\rho}/q$ needs to be added to the $v$ dispersion relation (\ref{e:disp})
(and (\ref{e:order0})) on integration to account for the solution of the nematic equations with $a=0$, that is 
the constant level solution  \cite{nembore,nemgennady}. 

To obtain the final modulation equation for the modulation variables $\bar{\rho}$, $\bar{v}$, $a$ and
$k$, it is easiest to average the momentum conservation equation (\ref{e:momcons}) directly.  A major 
reason for this is that this conservation equation will be used in Sections \ref{s:tdsw} and \ref{s:vacuum} 
to determine the DSW and its associated resonant wavetrain in the TDSW and VDSW regimes.  This averaged
momentum conservation equation is 
\begin{equation}
 \frac{\partial}{\partial z} \left[ \bar{\rho}\bar{v} + \frac{k}{4\bar{\rho}} a^{2} \right] + 
 \frac{\partial}{\partial x} \left[ \bar{\rho}\bar{v}^{2} + \frac{\bar{\rho}^{2}}{q} + 
 \left( \frac{\bar{v}k}{2\bar{\rho}} + \frac{k^{2}}{4\bar{\rho}} \right) a^{2} \right] = 0 .
 \label{e:avmom}
\end{equation}

Finally, there is the energy conservation equation (\ref{e:energycons}).  The averaged energy conservation
equation will again be needed in Sections \ref{s:tdsw} and \ref{s:vacuum} to determine the DSW solutions in the 
TDSW and VDSW regimes.  This is because shock jump conditions for the Whitham modulation equations will be
used to determine the resonant wavetrain \cite{whitham,patjump} and these are derived from averaged conservation
equations \cite{whitham}.  The averaged energy conservation equation is
\begin{equation}
 \frac{\partial}{\partial z} \left[ \bar{\rho}\bar{v}^{2} + \frac{2}{q} \bar{\rho}^{2} 
 + \left( \frac{k\bar{v}}{2\bar{\rho}} + \frac{k^{2}}{4\bar{\rho}} \right) a^{2} \right]
 + \frac{\partial}{\partial x} \left[ \bar{\rho}\bar{v}^{3} + \frac{4}{q} \bar{\rho}^{2}\bar{v}
 + \left( \frac{3\bar{v}^{2}k}{4\bar{\rho}} + \frac{3\bar{v}k^{2}}{4\bar{\rho}} 
  + \frac{k^{3}}{4\bar{\rho}} + \frac{k}{q} \right) a^{2} \right] = 0.
  \label{e:avenergy}
\end{equation}

The Whitham modulation equations (\ref{e:avmass}), (\ref{e:waveaction}), (\ref{e:waveconsav}) and (\ref{e:avmom}) will 
be used to derive solutions for the resonant wavetrain and to determine the modulational stability of this wavetrain.  
For this purpose, they need to ideally be set in Riemann invariant form if they form a hyperbolic system.  After 
extensive algebra, they can be set in the Riemann variable form \cite{whitham}
\begin{eqnarray}
& & -\frac{32\lambda_{1}\bar{\rho}^{2}a}{k\left( qk^{2} - 10 \bar{\rho}\right)} \frac{d\bar{\rho}}{dz} - 
\frac{8\lambda_{1}\bar{\rho}^{2}\left( qk^{2} - 2\bar{\rho}\right)a}{k^{2}\left( qk^{2} - 
10 \bar{\rho}\right)} \frac{dk}{dz} + \frac{da^{2}}{dz} - \frac{8\lambda_{1} \bar{\rho}^{2}\left(qk^{2} 
+ 2\bar{\rho}\right)a}{k^{2}\left( qk^{2} - 10\bar{\rho}\right)} \frac{d\bar{v}}{dz} = 0 \nonumber \\
& \mbox{on} & \frac{dx}{dz} = \lambda_{1}=\bar{v}+ k + 
{\frac {ka}{4\bar{\rho}}\sqrt {{\frac {20\,\bar{\rho}-2{k}^{2}q}{q{k}^{2} -
2\,\bar{\rho}}}}}, \label{e:lambda1} \\
& & \frac{32\lambda_{1}\bar{\rho}^{2}a}{k\left( qk^{2} - 10 \bar{\rho}\right)} \frac{d\bar{\rho}}{dz} 
+ \frac{8\lambda_{1}\bar{\rho}^{2}\left( qk^{2} - 2\bar{\rho}\right)a}{k^{2}\left( qk^{2} - 10 \bar{\rho}\right)} \frac{dk}{dz} 
+ \frac{da^{2}}{dz} + \frac{8\lambda_{1} \bar{\rho}^{2}\left(qk^{2} + 2\bar{\rho}\right)a}{k^{2}\left( qk^{2} 
- 10\bar{\rho}\right)} \frac{d\bar{v}}{dz} = 0 \nonumber \\
& \mbox{on} & \frac{dx}{dz} = \lambda_{2}= \bar{v}+k - {\frac {ka}{4\bar{\rho}}\sqrt {{\frac {20\,\bar{\rho}-2{k}^{2}q}{q{k}^{2} 
- 2\,\bar{\rho}}}}}, \label{e:lambda2}\\
& & \sqrt{\frac{2}{q}} \frac{1}{\sqrt{\bar{\rho}}} \frac{d\bar{\rho}}{dz} + \frac{a^{2}}{4\bar{\rho}^{2} + 2qk^{2}\bar{\rho} 
- 4\sqrt{2q} k\bar{\rho}^{3/2}} \frac{dk}{dz}
+ \frac{\sqrt{2q} k}{4\bar{\rho}^{3/2}\left( \sqrt{2q} \sqrt{\bar{\rho}} - qk\right)} \frac{da^{2}}{dz} + \frac{d\bar{v}}{dz} 
= 0 \nonumber \\
& \mbox{on} & \frac{dx}{dz} = \lambda_{3}=\bar{v} + \sqrt{ \frac{2}{q}}\sqrt{\bar{\rho}} 
+ \frac{ka^{2}}{2\bar{\rho}}{\frac {2\,\sqrt {2\bar{\rho}}- \sqrt {q} k}{\sqrt {2\bar{\rho}} \left( q{k}^{2}+2\,\bar{\rho}
 \right) -4\sqrt{q}\,k\bar{\rho}}},
 \label{e:lambda3} \\
 & & -\sqrt{\frac{2}{q}} \frac{1}{\sqrt{\bar{\rho}}} \frac{d\bar{\rho}}{dz} + \frac{a^{2}}{4\bar{\rho}^{2} 
 + 2qk^{2}\bar{\rho} + 4\sqrt{2q} k\bar{\rho}^{3/2}} \frac{dk}{dz}
+ \frac{\sqrt{2q} k}{4\bar{\rho}^{3/2}\left( \sqrt{2q} \sqrt{\bar{\rho}} + qk\right)} \frac{da^{2}}{dz} 
+ \frac{d\bar{v}}{dz} = 0 \nonumber \\
 & \mbox{on} & \frac{dx}{dz} = \lambda_{4}= \bar{v} - \sqrt{ \frac{2}{q} } \sqrt{\bar{\rho}}
 + \frac{ka^{2}}{2\bar{\rho}}{\frac {2\,\sqrt {2\bar{\rho}}+\sqrt {q}k }{\sqrt {2\bar{\rho}} \left( q{k}^{2}+2\,\bar{\rho}
 \right) +4\sqrt{q}\,k\bar{\rho}}}
 \label{e:lambda4}
\end{eqnarray}
to $O(a^{2})$.  Note that at the leading order with $a=0$, the two characteristic forms (\ref{e:lambda3}) and (\ref{e:lambda4}) 
are the shallow water equations in Riemann variable form \cite{whitham}, which must be the case.  The system 
(\ref{e:lambda1})--(\ref{e:lambda4}) is hyperbolic if the wavenumber $k$ falls within the restricted range
\begin{equation}
 \bar{u} < \sqrt{\frac{q}{2}} \: k < \sqrt{5} \: \bar{u},
 \label{e:stablemodregion}
\end{equation}
as $\bar{u} = \sqrt{\bar{\rho}}$.  The Stokes wavetrain is then modulationally stable for $k$ in the range 
(\ref{e:stablemodregion}) and unstable otherwise.  The restricted range of the wavenumber for stability leads to the 
resonant wavetrain generated by the DSW to be unstable over a great portion of the range of its existence.  Unfortunately, it 
was not found possible to set the Riemann variable form (\ref{e:lambda1})--(\ref{e:lambda4}) into Riemann invariant form, 
a manifestation of Pfaff's problem.  Outside the range (\ref{e:stablemodregion}) the system (\ref{e:lambda1})--(\ref{e:lambda4}) 
is mixed elliptic-hyperbolic since the characteristics (\ref{e:lambda3}) and (\ref{e:lambda4}), which are those for the 
shallow water equations at leading order, are always real.  This is important as in Sections \ref{s:tdsw} and \ref{s:vacuum} 
the Riemann variables on these characteristics are be used to construct the solutions in the TDSW and VDSW regimes.

In the next sections the solutions for the five non-trivial nematic DSW regimes, the PDSW, RDSW, CDSW, TDSW and VDSW will 
be found and compared with numerical solutions of the nematic equations (\ref{e:eeqn}) and (\ref{e:direqn}).  Different 
analytical methods will be used to derive the DSW solutions in these five regimes.  Comparisons of DSW parameters
with numerical solutions will be made separately as the solution in each regime is found, even though the DSW parameters, 
such as resonant wave height and resonant wavenumber, are common across some, or all, of the regimes and there is no
solution distinction in the numerical parameters.

\section{Perturbed Nematic DSW and RDSW}
\label{s:pdsw}

In this section, the solution for the perturbed DSW (PDSW), illustrated in Figure \ref{f:types}(a), and the radiating DSW,
illustrated in Figure \ref{f:types}(b), will be found based on the Kawahara equation reduction (\ref{e:kdv5nem}) of the 
full nematic equations as these DSW solutions are valid for small jump heights $u_{i} - u_{+}$.  The DSWs in these two 
regimes are similar to the KdV DSW, with solitary waves at the leading edge and linear diffractive waves at the trailing edge.  
While the DSW radiates resonant waves in the RDSW regime, the effect of this loss is small and can be neglected to find 
the DSW solution.  The resonant radiation loss becomes dominate in the CDSW regime.

It was shown that the KdV equation with all the next higher order nonlinear, dispersive and nonlinear-dispersive terms 
of relative order $\alpha$ in the weakly nonlinear, long wave expansion used to derive the KdV equation \cite{whitham} 
\begin{equation}
 \frac{\partial \eta}{\partial T} + 6\eta \frac{\partial \eta}{\partial X} 
 + \alpha C_{1} \eta^{2} \frac{\partial \eta}{\partial X} +
 \frac{\partial^{3} \eta}{\partial X^{3}} + 
 \alpha C_{2} \frac{\partial \eta}{\partial X} 
 \frac{\partial^{2} \eta}{\partial X^{2}} + \alpha C_{3} \eta \frac{\partial^{3} \eta}{\partial X^{3}} + 
 \alpha C_{4} \frac{\partial^{5} \eta}{\partial X^{5}} = 0
 \label{e:kdvhigh}
\end{equation}
can be transformed to the KdV equation ((\ref{e:kdv5}) with $\mu = 1$ and no fifth derivative term)
with error $O(\alpha^{2})$ \cite{perturbkdv}, which is of higher order than the validity of the original equation.  This 
transformation was then used to find the DSW solution of the higher order KdV equation (\ref{e:kdvhigh}) from that for the 
KdV equation \cite{perturbkdv}.  This perturbed DSW solution will be used to find the perturbed DSW solution of the reduction 
(\ref{e:kdv5nem}) of the nematic equations.  The KdV reduction (\ref{e:kdv5nem}) of the nematic equations does not contain 
the higher order terms $u_{1}^{2}u_{1\xi}$, $u_{1\xi}u_{1\xi\xi}$ and $u_{1}u_{1\xi\xi\xi}$ of the general higher order KdV equation 
(\ref{e:kdvhigh}).  The reason for this can most easily be seen on expanding the nematic dispersion relation (\ref{e:disp}) in 
the long wave limit $k \ll 1$, giving 
\begin{equation}
 \omega = k\left(\sqrt{\frac{2}{q}}\bar{u} +  \bar{v}\right) -  \frac{c}{4} \left( \frac{\nu }{q} 
 - \frac{q}{4 \bar \rho} \right)  k^{3} 
  + \frac{c}{32}\left( \frac{3 \nu^2}{q^2} + \frac{\nu}{\bar \rho} - \frac{q^2}{16 \bar \rho^2} \right) k^5 + \ldots,
 \label{e:kdv5disp}
\end{equation}
with $\bar{u} = \sqrt{\bar{\rho}}$.  We note that this dispersion relation does not contain a term $\bar{u}^{2}k$, so that 
there is no higher order nonlinear term $u_{1}^{2}u_{1\xi}$.  In addition, the fifth derivative term in (\ref{e:kdvhigh}) was 
obtained by including the fifth derivative from higher order in the expansion (\ref{e:ukdv})--(\ref{e:thetakdv}) as the 
coefficient $\epsilon^{2}\nu^{2}$ can be $O(1)$ as $\nu$ is large.  In a similar manner, the higher order nonlinear/dispersive 
terms $u_{1\xi}u_{1\xi\xi}$ and $u_{1}u_{1\xi\xi\xi}$ would be at most $O(\epsilon^{2} \nu)$ and so can be ignored.

The transformation of \cite{perturbkdv} then gives that the PDSW is a modulated, perturbed cnoidal wave of amplitude $a$ 
(measured peak to trough), wavenumber $k$ and mean height $\bar{u}$.  On identifying $u_{0} = u_{+}$ and 
$\epsilon^{2} = u_{i} - u_{+}$ in the perturbation expansions (\ref{e:ukdv})--(\ref{e:thetakdv}), we have from 
\cite{perturbkdv} that the PDSW solution is given by
\begin{eqnarray}
 a & = & 2(u_{i} - u_{+}) m + 2C_{4} (u_{i} - u_{+})^{2} m \left( \frac{8}{3} - m \right), \label{e:amppdsw} \\
 k & = & \frac{\pi \sqrt{2(u_{i} - u_{+})}}{K(m)\sqrt{u_{+}}\sqrt{\frac{\nu}{q} - \frac{q}{4u_{+}^{2}}}}
 \left[ 1 - \frac{1}{3}C_{4} \left( 8m^{2} - 14m + 11\right) \left( u_{i} - u_{+} \right)\right], \label{e:kpdsw} \\
 \bar{u} & = & 2u_{+} -u_{i} + (u_{i} - u_{+})\left( 2\frac{E(m)}{K(m)} + m \right) 
 \nonumber \\
 & & \mbox{} + \frac{2}{9} C_{4} (u_{i} - u_{+})^{2} \left[ 2 - 5m + 3m^{2} + 2(2m-1)\frac{E(m)}{K(m)} \right] \nonumber \\
 & & \mbox{} + \frac{32}{9} C_{4}(u_{i} - u_{+})^{2} \left[ 3\left( 1 - \frac{E(m)}{K(m)}\right)^{2} - 
 2\left( 1 - \frac{E(m)}{K(m)} \right) (1+m) + m \right] \label{e:ubarpdsw} 
\end{eqnarray}
on
\begin{eqnarray}
 \frac{x}{z} & = &  \sqrt{\frac{2}{q}} \left[ u_{+} + \frac{1}{2}\lambda - \frac{1}{3} C_{4} \lambda_{1} 
 + \frac{1}{6} C_{4} \lambda^{2}
 - \frac{4}{3} C_{4} (u_{i} - u_{+})^{2} [m(2-m)-1] \right. \nonumber \\
 & & \mbox{} - \frac{4}{3} C_{4} (u_{i} - u_{+}) \left( m-1
\left.  + \frac{2E(m)}{K(m)}\right) (\lambda - V) \right], \label{e:xzpdsw} 
\end{eqnarray}
with
\begin{equation}
 V = 2(u_{i} - u_{+})(1+m) - \frac{4}{3}C_{4}(u_{i} - u_{+})^{2} (3m-3m^{2}-2), \label{e:updsw} 
\end{equation}
and
\begin{equation}
 \lambda = V - \frac{4(u_{i} - u_{+})m(1-m)K(m)}{E(m) - (1-m)K(m)}, \quad 
 \lambda_{1} = 2(u_{i} - u_{+})^{2} \left[ 1+m - \frac{2m(1-m)K(m)}{E(m) - (1-m)K(m)} \right].
 \label{e:lambda1simple}
\end{equation}
Note that various typographical errors in the asymptotic expressions of \cite{perturbkdv} have been corrected to obtain the 
expressions (\ref{e:amppdsw})--(\ref{e:lambda1simple}).   The coefficient $C_{4}$, which comes from rescaling the fifth order 
KdV equation (\ref{e:kdv5}) to the higher order KdV equation (\ref{e:kdvhigh}), is 
\begin{equation}
 C_{4} = \frac{3\nu^{2}}{2q^{2}}\left[ \frac{\nu}{q} - \frac{q}{4u_{+}^{2}} \right]^{-2}.
 \label{e:c4}
\end{equation}
Here, $K(m)$ and $E(m)$ are complete elliptic integrals of the first and second kinds of modulus $m$, respectively.
The limit $m=0$ corresponds to the trailing linear, diffractive wave edge of the PDSW and $m=1$ corresponds to the 
leading, solitary wave edge.  The DSW then lies within the region
\begin{equation}
 s_{i} = \sqrt{\frac{2}{q}} \left[ 4u_{+} - 3u_{i} + \frac{64}{3} C_{4}(u_{i} - u_{+})^{2} \right] \le
 \frac{x}{z} \le s_{+} = \sqrt{\frac{2}{q}} \left[ 2u_{i} - u_{+} + \frac{4}{3} C_{4} (u_{i} - u_{+})^{2} \right]. 
 \label{e:pdswbound}
\end{equation}
The amplitude $a_{s}$ of the solitary wave at the leading edge of the DSW can be found from the amplitude expression 
(\ref{e:amppdsw}) on setting $m=1$.  This gives 
\begin{equation}
 a_{s} = 2\left( u_{i} - u_{+} \right) + \frac{10}{3} C_{4} \left( u_{i} - u_{+} \right)^{2},
 \label{e:as}
\end{equation}
so that the height $H_{s}$ of the lead solitary wave of the DSW from $u=0$ is
\begin{equation}
 H_{s} = a_{s} + u_{+} = 2u_{i} - u_{+} + 
 \frac{10}{3} C_{4} \left( u_{i} - u_{+} \right)^{2}.
 \label{e:hs}
\end{equation}

\begin{table}
\centering
 \begin{tabular}{|c|c|c|} \hline 
  DSW type         &         numerical existence interval        &    theoretical existence interval  \\ \hline
  PDSW       &         $0.76< u_{+} < 1.0$          &      $0.73 < u_{+} < 1.0$        \\ \hline
  RDSW       &         $0.70 < u_{+} < 0.76$  &        $0.70 < u_{+} < 0.73$   \\ \hline
  CDSW  &  $0.44 < u_{+} < 0.70$     &        $0.44 < u_{+} < 0.70$                 \\ \hline
  TDSW  &  $0.22 < u_{+} < 0.44$  & $0.24 < u_{+} < 0.44$  \\ \hline
  VDSW  &    $0 < u_{+} < 0.22$               &   $0 < u_{+} < 0.24$   \\  \hline
  Dam break &      $u_{+} = 0$                  &  $u_{+} = 0$                            \\ \hline
 \end{tabular}
\caption{Comparisons between numerical and theoretical existence regions for the six nematic bore types
described in Section \ref{s:types}.  Here $u_{-} = 1.0$, $\nu = 200$ and $q = 2$.}
\label{t:regions}
\end{table}

The Riemann invariant $R_{-}$ on the characteristic $C_{-}$ given by (\ref{e:cm}) of the non-dispersive equations is 
conserved through the DSW \cite{elreview,chaos}.  As $\rho = u_{+}^{2}$ and $v=0$ at the leading edge of the DSW, 
$R_{-} = -2\sqrt{2}u_{+}/\sqrt{q} = v_{i} - 2\sqrt{2}u_{i}/\sqrt{q}$, so that on using expression (\ref{e:v2expr}) 
from the expansion wave solution we have \cite{nembore,nemgennady}.  
\begin{equation}
  u_{i} = \frac{1}{2}\left( u_{-} + u_{+} \right), \quad v_{i} = \frac{\sqrt{2}}{\sqrt{q}} \left( u_{-} - u_{+} \right).
 \label{e:ui}
\end{equation}
This completes the solution for the PDSW.  

\begin{figure}[t]
\centering
   \includegraphics[width=0.33\textwidth,angle=270]{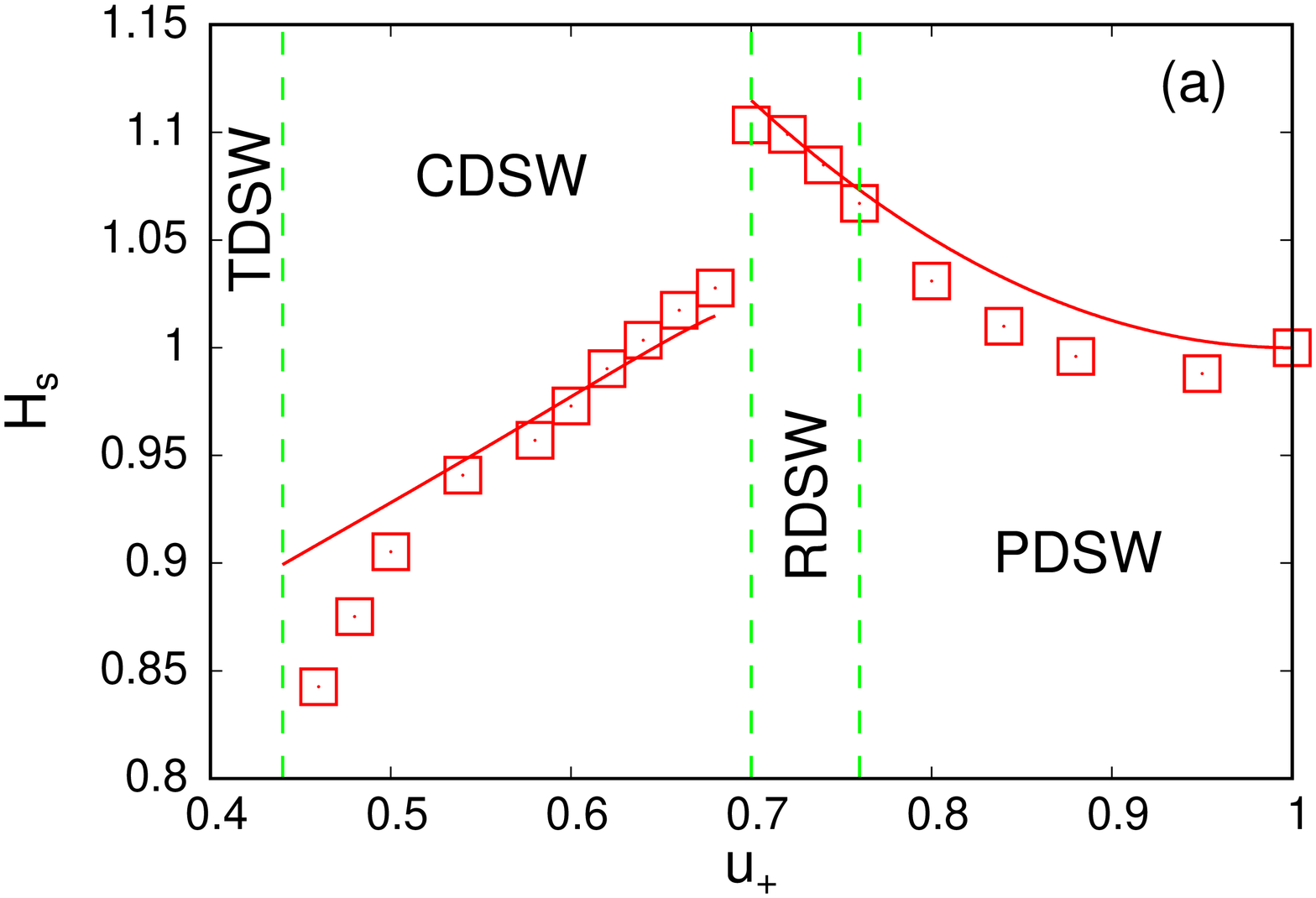}
   \includegraphics[width=0.33\textwidth,angle=270]{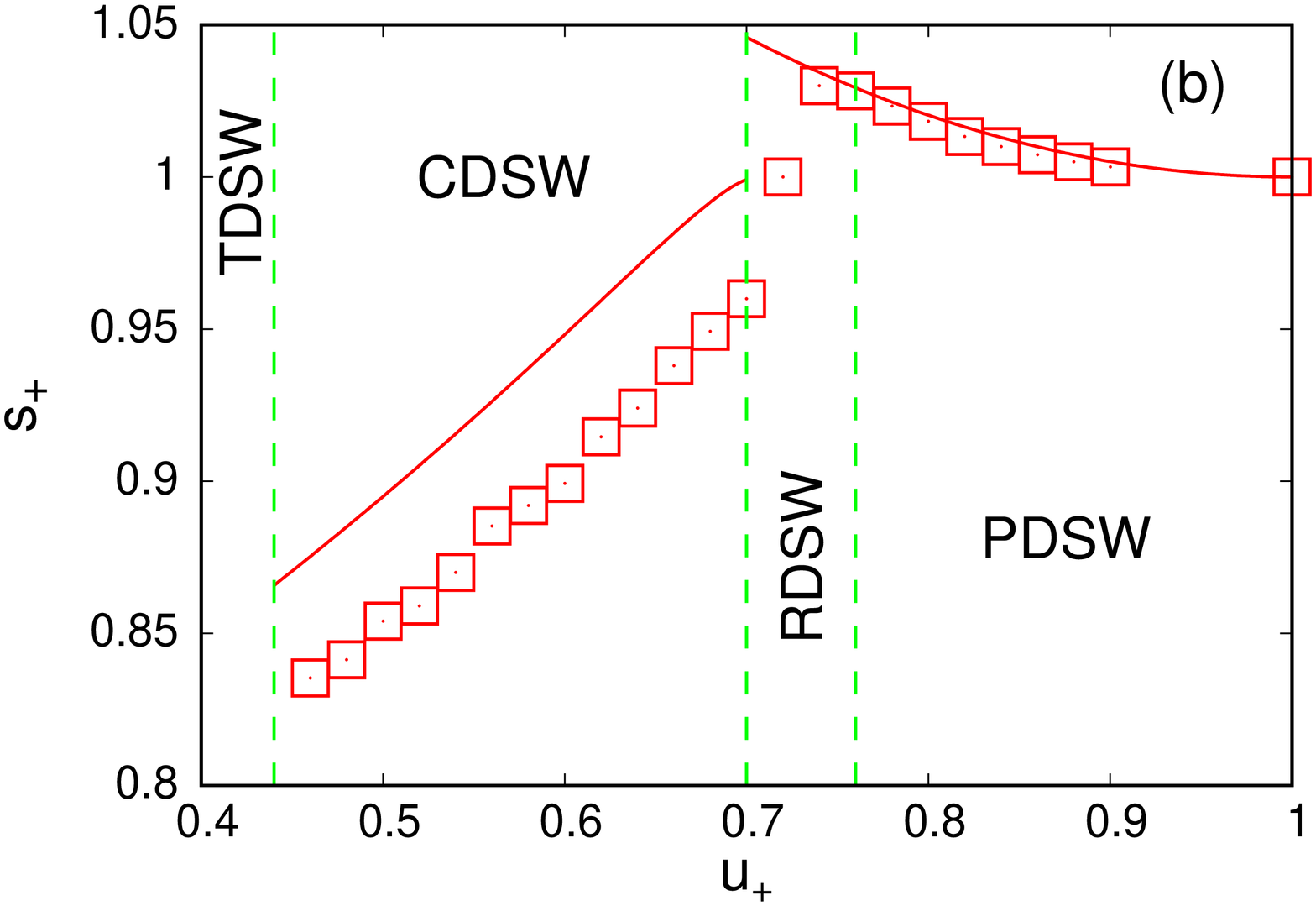}
      \includegraphics[width=0.33\textwidth,angle=270]{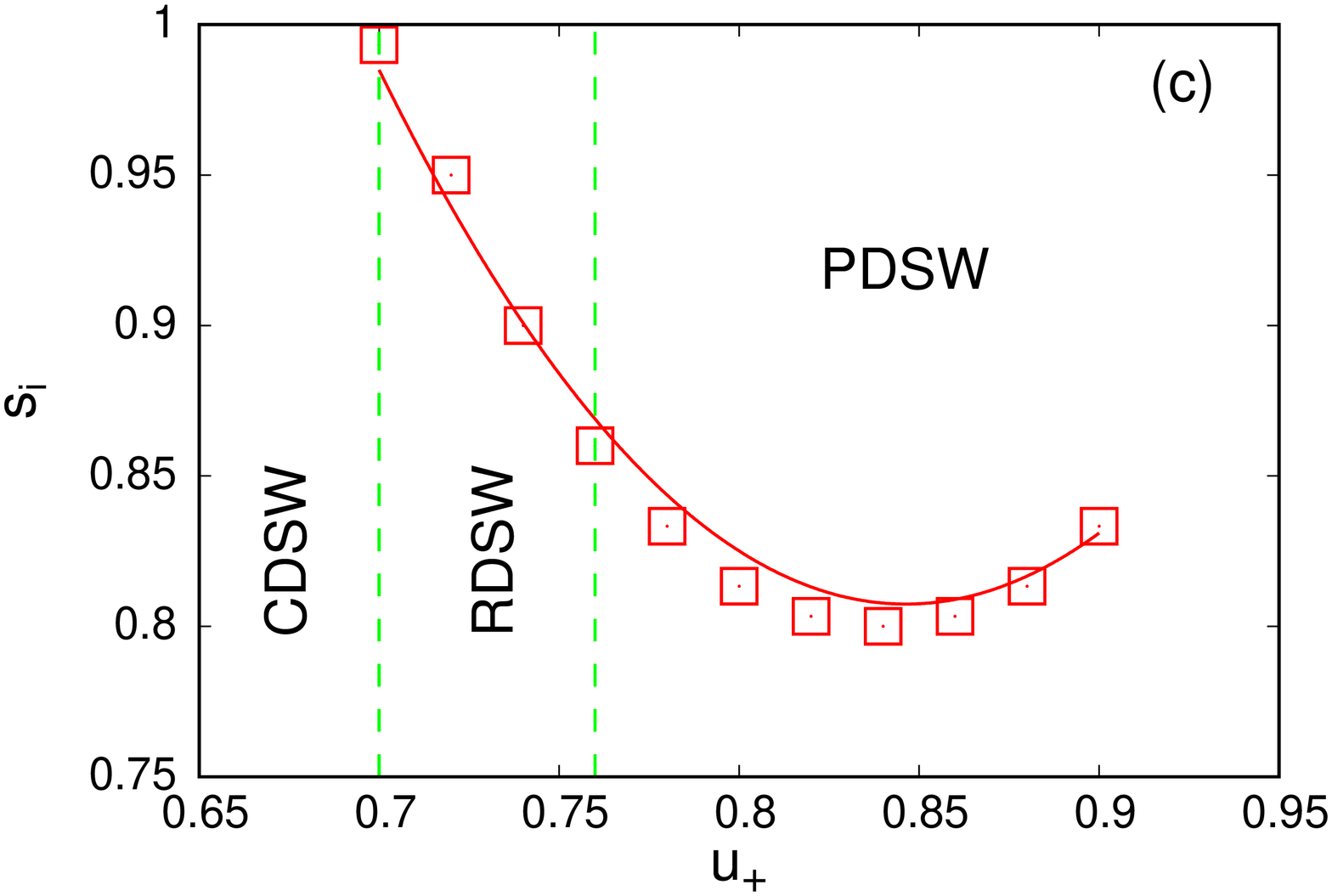}
\caption{Comparisons between numerical solutions of the nematic equations (\ref{e:eeqn}) and (\ref{e:direqn}) and the 
DSW solutions for the PDSW, RDSW and CDSW regimes.  Numerical solution:  red boxes; analytical solution:  red (solid) line.  
(a) lead solitary wave height $H_{s}$ (\ref{e:hs}), (b) velocity $s_{+}$ of leading edge of DSW, (c) velocity $s_{i}$ of 
trailing edge of DSW.  Here $u_{-} = 1.0$, $\nu =200$ and $q = 2$.} 
\label{f:pdswrdsw}
\end{figure}

As previously found \cite{nembore,nemgennady}, a resonance between diffractive radiation and the DSW can occur 
when the 
phase velocity of the diffractive radiation equals the velocity $s_{+}$ of the lead solitary wave of the DSW.  As can be 
seen from Figure \ref{f:types}(b) this resonant radiation is short wave relative to the DSW, so that the appropriate 
linear dispersion relation is (\ref{e:displargenu}).  Resonance then occurs for the wavenumber $k_{r}$, with
\begin{equation}
 k_{r} = s_{+} + \left( s_{+}^{2} - \frac{4}{q}u_{+}^{2} \right)^{1/2} ,
 \label{e:krdsw}
\end{equation}
on taking $v_{+}=0$, which is assuming that the resonant wavetrain sits on the initial level $u_{+}$ ahead.  
If $s_{+} \le 2u_{+}/\sqrt{q}$, then there is no resonant wavetrain ahead of the DSW, which is the PDSW regime 
illustrated in Figure \ref{f:types}(a).  If $s_{+} > 2u_{+}/\sqrt{q}$, then the DSW sheds a resonant wavetrain ahead
of it as the group velocity $c_{g} = k_{r}$ of the resonant wavetrain is greater than its phase velocity, 
$c < c_{g}$ \cite{nembore,nemgennady}, on using (\ref{e:pdswbound}) and (\ref{e:displargenu}).  
Resonance can then 
occur if $s_{+} \ge 2u_{+}/\sqrt{q}$.  If this condition is not satisfied, then there is just a DSW and we have the PDSW regime.    

When the resonance condition (\ref{e:krdsw}) has a solution, we have the RDSW regime for the nematic bore, as illustrated 
in Figure \ref{f:types}(b).  It was found that the RDSW regime is also well described by the perturbed KdV DSW solution 
(\ref{e:amppdsw})--(\ref{e:xzpdsw}).  This is because the resonant radiation shed by the DSW is of small amplitude, as 
can be seen from Figure \ref{f:types}(b), and the existence regime for this type is small, as seen from Table 
\ref{t:regions}.  The resonant wavetrain has a major effect in the CDSW regime, which is dealt with in the next section, 
as the resonant wavetrain acts as a damping on the DSW and the DSW is unstable in this regime. 

Figure \ref{f:pdswrdsw} shows comparisons between numerical solutions and the perturbed KdV DSW solution 
(\ref{e:amppdsw})--(\ref{e:xzpdsw}) for the height $H_{s} = u_{+} + a_{s}$ and velocity $s_{+}$ of the lead solitary wave 
and the velocity $s_{i}$ of the trailing edge of the DSW.  It can be seen that there is excellent agreement for the 
lead solitary wave height over both the PDSW and RDSW regimes.  When a resonant wavetrain is present, the amplitude of 
the lead wave of the DSW oscillates as radiation propagates out of the DSW \cite{pat}.  In this case, the average lead 
wave amplitude is taken for the numerical comparisons.  Figure \ref{f:pdswrdsw}(b) shows that there is also excellent 
agreement for the velocity $s_{+}$ of the leading edge.  There is some disagreement in the lead wave height and velocity 
at the transition from the RDSW to the CDSW regime, but this is to be expected as the DSW drastically changes form.  The 
leading edge velocity does not show much variation from $u_{-}$ in these PDSW and RDSW regimes.  Setting $C_{4}=0$ in the 
perturbed KdV DSW solution (\ref{e:amppdsw})--(\ref{e:xzpdsw}) gives the standard KdV DSW solution, for which 
$H_{s}=u_{-}$ and $s_{+}=u_{-}$ on using (\ref{e:ui}) for $u_{i}$, as found in a previous study \cite{nembore}.  It is 
then seen that the higher order dispersion in the Kawahara equation (\ref{e:kdv5nem}) is necessary to fully account for 
the lead solitary wave in the PDSW and RDSW regimes, although the additional effect is not large.  Figure \ref{f:pdswrdsw}(c) 
shows a comparison between numerical solutions and the perturbed KdV DSW solution for the trailing edge velocity $s_{i}$.  The 
comparison is again very good, in particular for the turning point in the trailing edge velocity.  The implications of
this turning point will be discussed below.

The wavenumber of the resonant wavetrain in the RDSW regime is given by the resonance condition (\ref{e:krdsw}), which 
has a solution if $s_{+} \ge 2u_{+}/\sqrt{q}$.  Table \ref{t:regions} gives comparisons between the existence intervals
for the six DSW regimes as given by numerical solutions and by the analytical solutions for the parameter
choices $u_{-}=1$, $\nu = 200$ and $q = 2$.  In particular, it can be seen that there is good agreement for the intervals
of existence for the PDSW solution, which does not have an associated resonant wavetrain as
$s_{+} < 2u_{+}/\sqrt{q}$.  There is excellent agreement for the existence region of the RDSW, which has an associated 
resonant wavetrain.  Note that the lower bound for the RDSW regime is connected with the existence of the CDSW regime, which is 
dealt with in Section \ref{s:cross}.

\begin{figure}
\centering
   \includegraphics[width=0.33\textwidth,angle=270]{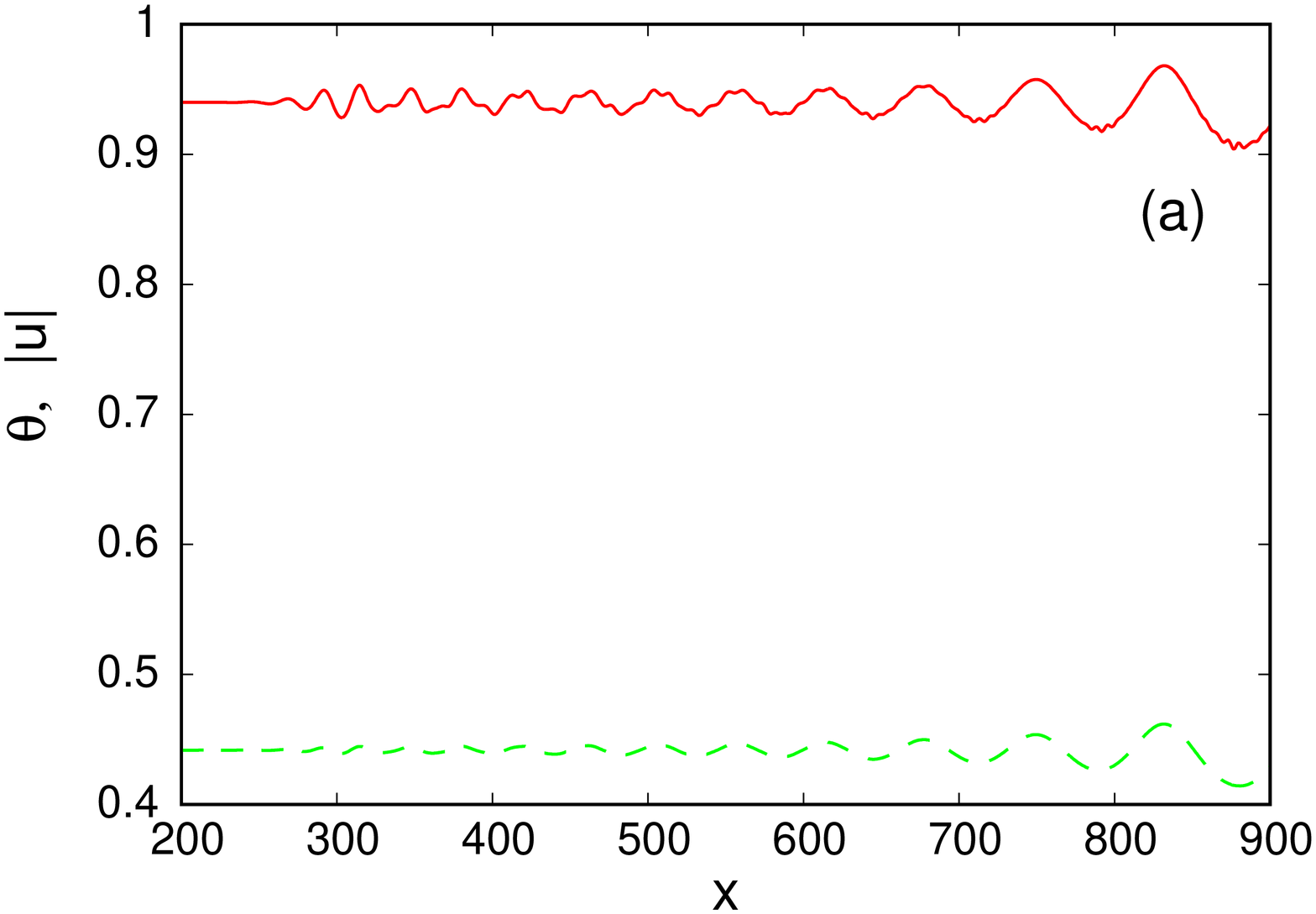}
      \includegraphics[width=0.33\textwidth,angle=270]{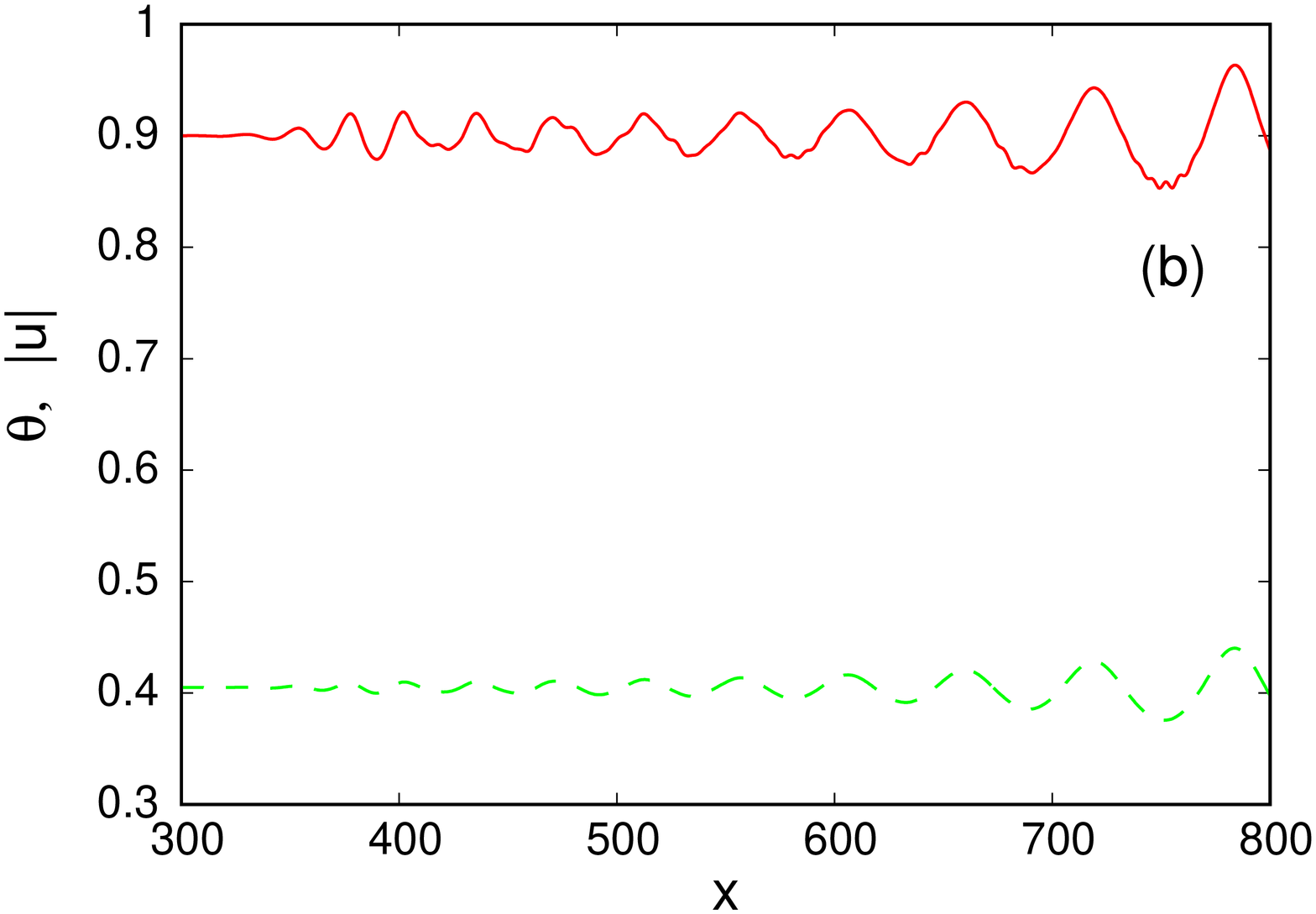}
\caption{Numerical solutions of the nematic equations (\ref{e:eeqn}) and (\ref{e:direqn}) for the 
initial condition (\ref{e:ic}) with $u_{-}=1$.  Red (solid) lines:  $|u|$ at $z=1000$; green (dashed)
lines $\theta$ at $z=1000$.  (a) PDSW with $u_{+}= 0.88$, (b) RDSW with $u_{+} = 0.8$.  Here $u_{-} = 1.0$, 
$\nu =200$ and $q = 2$.} 
\label{f:pdswrdswdetail}
\end{figure}

At this point the admissibility conditions for the existence of a DSW need to be checked 
\cite{elreview,lowman}.  These relate to the genuine nonlinearity and hyperbolicity of the associated Whitham
modulation equations upon which DSW solutions are ultimately based.  The breakdown of these lead to linear degeneracy in the 
first case and modulational instability in the second case, so that the standard DSW solution, as derived above, breaks down.  
The admissibility conditions require that the derivatives of the velocities of the trailing edge $s_{i}$ and of the 
leading edge $s_{+}$ do not have turning points as functions of $(u_{-}$,$u_{+})$.  It is found from the leading edge velocity 
expression (\ref{e:pdswbound}) for $s_{+}$ that these derivatives do not vanish at the leading edge, so there is no breakdown 
of the DSW structure at the leading edge.  At the trailing edge, the derivative of $s_{i}$ with respect to $u_{-}$ vanishes when 
$u_{+} = u_{-} -3/32$, which is $u_{+} = 0.90625$ for $u_{-}=1$, and the derivative with respect to $u_{+}$ vanishes when 
$u_{+} = u_{-} - 5/32$, which is $u_{+} = 0.84375$ for $u_{-}=1$.  In these calculations, $C_{4}$ given by (\ref{e:c4}) has been 
approximated by $C_{4} = 3/2$ as $\nu$ is large.  The turning point with respect to $u_{-}$ means that after this point a simple 
wave solution of the Whitham modulation equations governing the DSW is not possible, while the vanishing of the derivative with 
respect to $u_{+}$ means that the Whitham modulation equations lose hyperbolicity, so that the modulated wavetrain is not stable.  
The breakdown of these admissibility conditions is usually clearly mirrored in numerical solutions by non-standard DSW behaviour, 
such as the generation of a multi-phase wavetrain or wavetrain instability \cite{lowman}.  However, for the nematic DSW there is 
no clear evidence of such behaviour for values of $u_{+}$ below the turning points, as seen in Figures \ref{f:types}(a) and (b).
These figures show some evidence of instability at the trailing edge of the DSW, but it is minor.  

Figure \ref{f:pdswrdswdetail} shows details of the trailing edges of DSWs on either side of the predicted onset of modulational
instability at $u_{+} = 0.84375$.  It can be seen that there is some non-uniform modulation of the trailing edges, with
a modulated wavepacket, but there is no distinct change in the behaviour of the DSW, as found for other DSWs for 
which the admissibility conditions are not satisfied \cite{elreview,lowman}.  The reason for this is the highly nonlocal 
response of the nematic.  It has been shown theoretically and verified experimentally that high nonlocality acts to suppress 
modulational instability \cite{neminstab3,neminstab1}.  This greatly delays the onset of instability, so much so that 
theoretically unstable nematic wavetrains show no instability over experimental propagation distances ($\sim 1mm$).  

\begin{figure}
\centering
   \includegraphics[width=0.5\textwidth,angle=270]{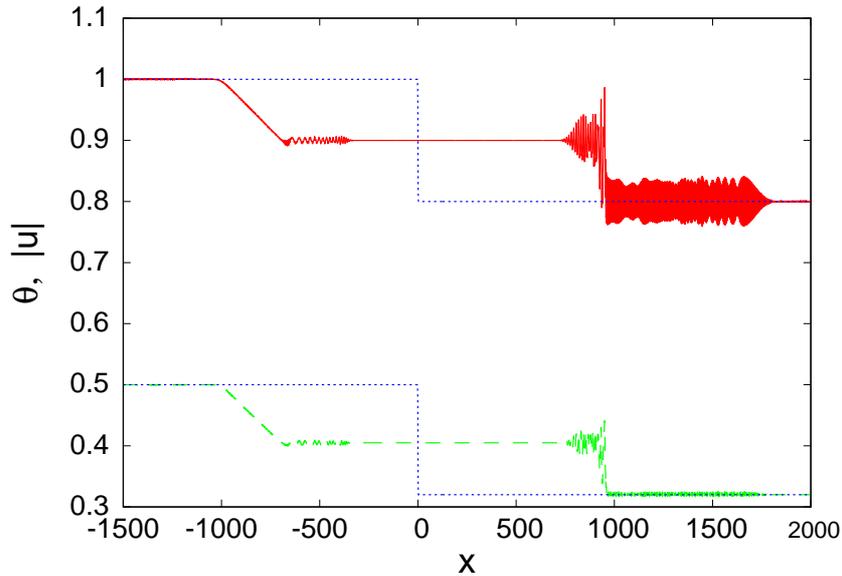}
\caption{Numerical solution of the nematic equations (\ref{e:eeqn}) and (\ref{e:direqn}) for the 
initial condition (\ref{e:ic}).  Red (solid) line:  $|u|$ at $z=1000$; green (dashed)
line $\theta$ at $z=1000$;  blue (dotted) lines:  $|u|$ at $z=0$ (upper) and $\theta$ at $z=0$ (lower).
Here $u_{-} = 1.0$, $u_{+}=0.8$, $\nu =10$ and $q = 2$.} 
\label{f:nu10}
\end{figure}

The stability of the Stokes' wave solution of the Kawahara equation (\ref{e:kdv5}) has been studied \cite{pat}.  This work 
can be used to determine the stability of the trailing edge of the DSW, where the wavenumber is $k_{i}$ given by 
(\ref{e:kpdsw}) with $m=0$.  Rescaling the nematic KdV equation (\ref{e:kdv5nem}) to the standard form (\ref{e:kdv5}) of 
the Kawahara equation gives the scaled wavenumber $\tilde{k}_{i}$ at the trailing edge of the DSW as
\begin{equation}
 \tilde{k}_{i} = 2 \left[ 1 - \frac{11}{6}C_{4} \left( u_{-} - u_{+} \right) \right] .
 \label{e:kmrho}
\end{equation}
The work of \cite{pat} then gives that the trailing edge of the DSW is stable if 
\begin{equation}
 15\tilde{k}_{i}^{2} \left[ \frac{6}{5\tilde{k}_{i}^{2} - 3/C_{4}} + \frac{1}{5\tilde{k}_{i}^{2} - 1/C_{4}} \right] + 9 > 0.
 \label{e:modistab}
\end{equation}
For the parameter values $\nu=200$ and $q=2$, this stability condition gives that the trailing edge of the DSW is unstable 
for $0.7223 \le u_{+} \le 0.7547$ and $0.6938 \le u_{+} < 0.7080$.  The onset of instability corresponds to the transition from 
the PDSW to the RDSW regime, as seen from Table \ref{t:regions}, and the onset of the generation of resonant radiation.  Again, 
numerical solutions show no evidence of any major instability in this region.  The suppression of modulational instability due 
to the high nonlocality \cite{neminstab3,neminstab1} is again presumed to be the reason.  This conjecture about the stabilising 
role of nonlocality is verified by the solution shown in Figure \ref{f:nu10}, which is for the same initial condition and 
parameter values as Figure \ref{f:types}(a), except that the nonlocality has been lowered to $\nu=10$ from $\nu=200$.  The PDSW 
of Figure \ref{f:types}(a) is stable, but Figure \ref{f:nu10} shows that lowering the nonlocality results in a dramatic change 
in the DSW.  It is now unstable, in accord with the admissibility conditions and the modulational instability condition of 
\cite{pat}.  In addition, there is now a resonant wavetrain.  The resonant condition (\ref{e:krdsw}) is valid in the limit 
$\nu \gg 1$, so is not strictly applicable for $\nu=10$.  Further decrease of the nonlocality $\nu$ is not possible as then 
the coefficient of the third derivative in the nematic Kawahara equation (\ref{e:kdv5nem}) becomes negative and the DSW changes 
from a positive polarity KdV-type to a negative polarity NLS-type DSW consisting of dark, rather than bright, waves. 

The nematic DSW structure instability is driven by the instability of the resonant wavetrain.  The Whitham 
modulation equations (\ref{e:lambda1})--(\ref{e:lambda4}) for the resonant wavetrain become elliptic, so that Benjamin-Feir 
instability arises and the underlying Stokes' wave solution for the resonant wavetrain becomes unstable.  The Whitham 
modulation equations for the nematic Stokes' wave are hyperbolic for wavenumbers in the range (\ref{e:stablemodregion}), so 
that the weakly nonlinear Stokes' wavetrain is stable in this region.  Using the resonance condition (\ref{e:krdsw}) with
$k=k_{r}$, $s_{+}$ given by (\ref{e:pdswbound}) and $\bar{u} = u_{+}$ with $q=2$ and $\nu=200$, this gives that the 
resonant wavetrain is unstable if $u_{+} < 0.674$, which from Table \ref{t:regions} is the CDSW regime.  However, numerical
solutions show that the resonant wavetrain in the RDSW regime is unstable, see Figure \ref{f:types}(b), for instance.  The 
reason for this disagreement between theory and numerical solutions will be discussed in detail in Sections 
\ref{s:cross}--\ref{s:vacuum}.  It is sufficient to note at this stage that when the resonant wavetrain is unstable, 
the resonance condition (\ref{e:krdsw}) gives a poor prediction of the resonant wavenumber as the wavetrain ceases to
be a uniform wavetrain with a single wavenumber, but gives a good prediction when the resonant wavetrain is stable, see 
Figure \ref{f:resonantwave}(c).


\begin{figure}
\centering
 \includegraphics[width=0.3\textwidth,angle=270]{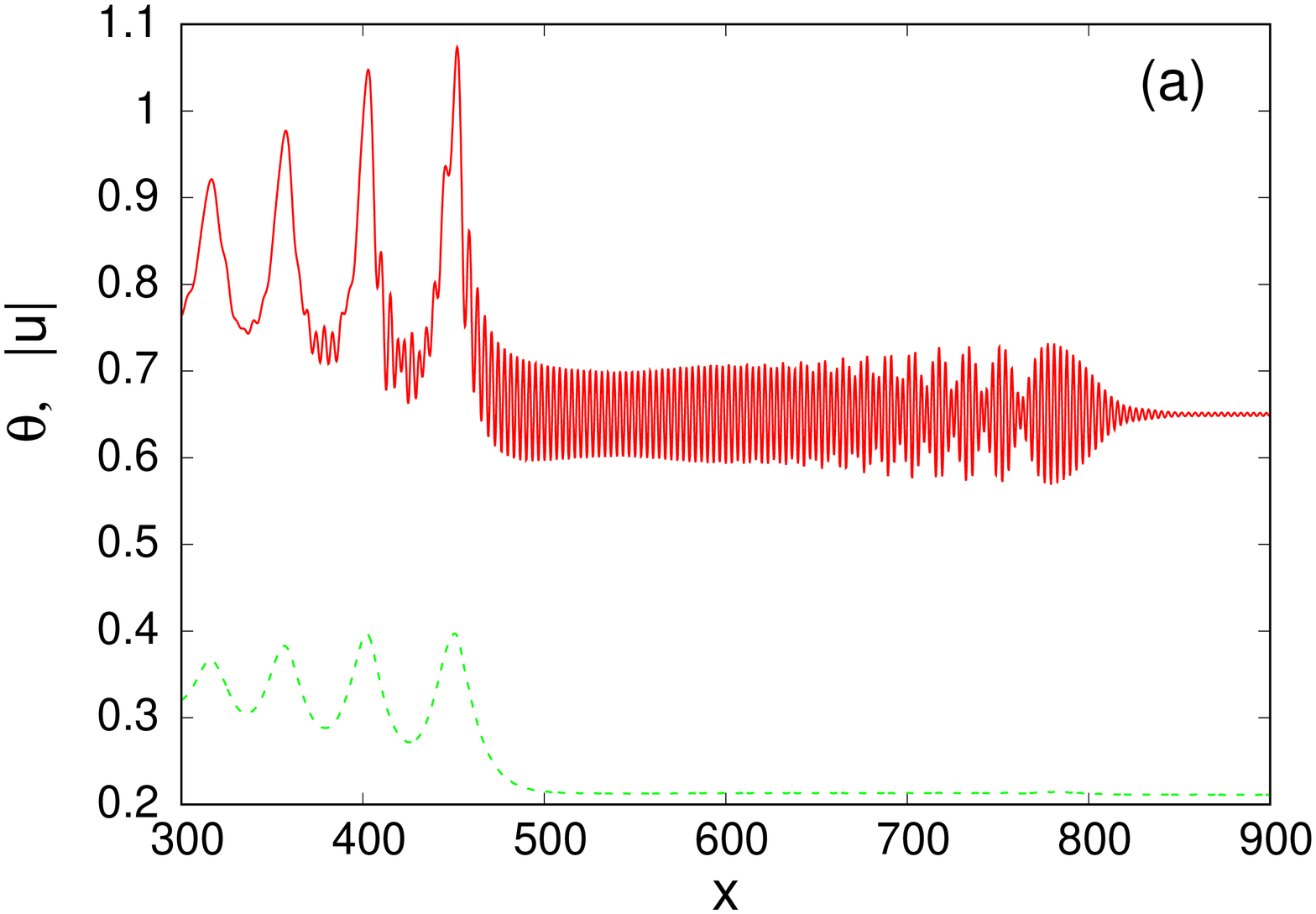}
 \includegraphics[width=0.3\textwidth,angle=270]{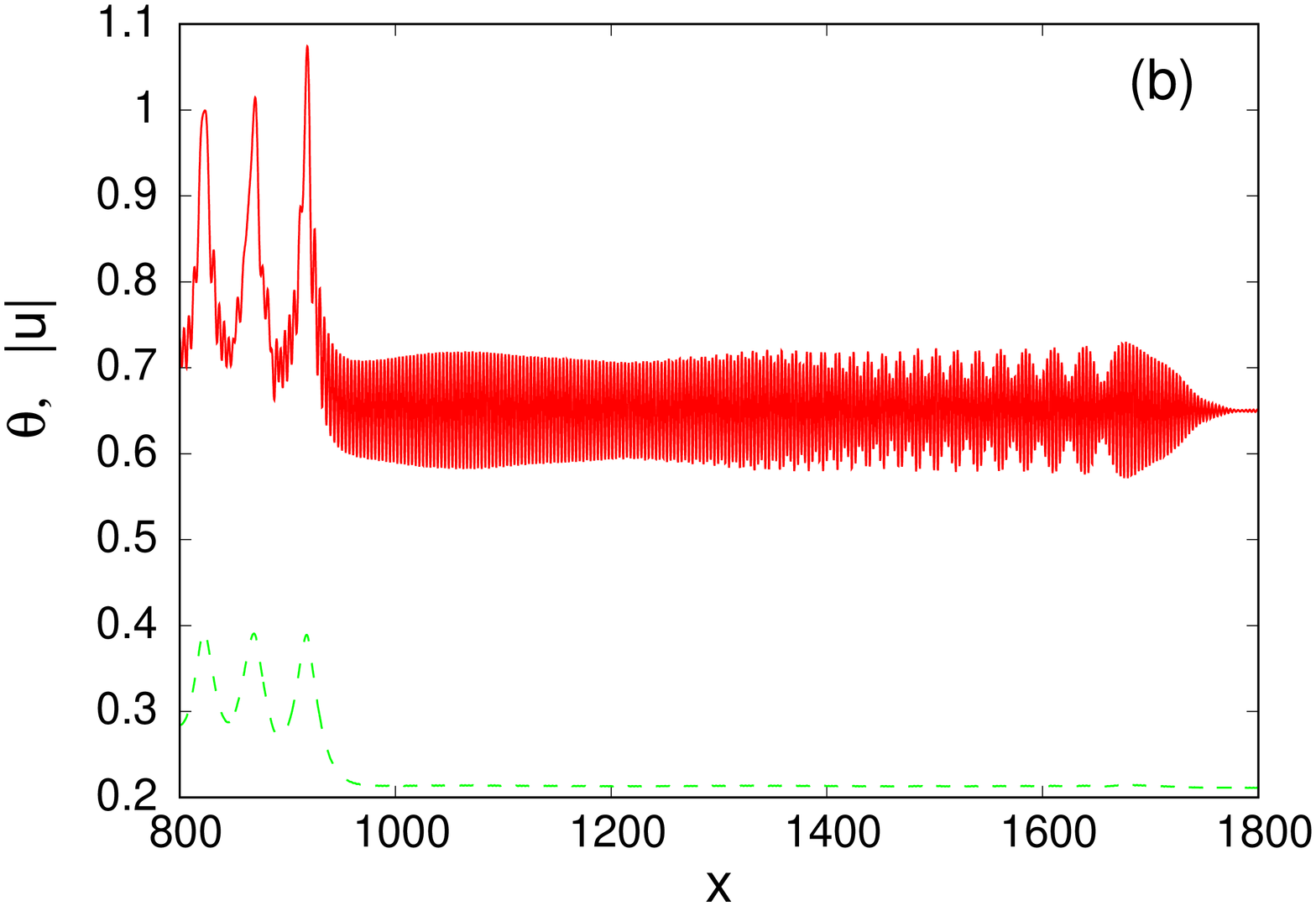}
 
 \includegraphics[width=0.3\textwidth,angle=270]{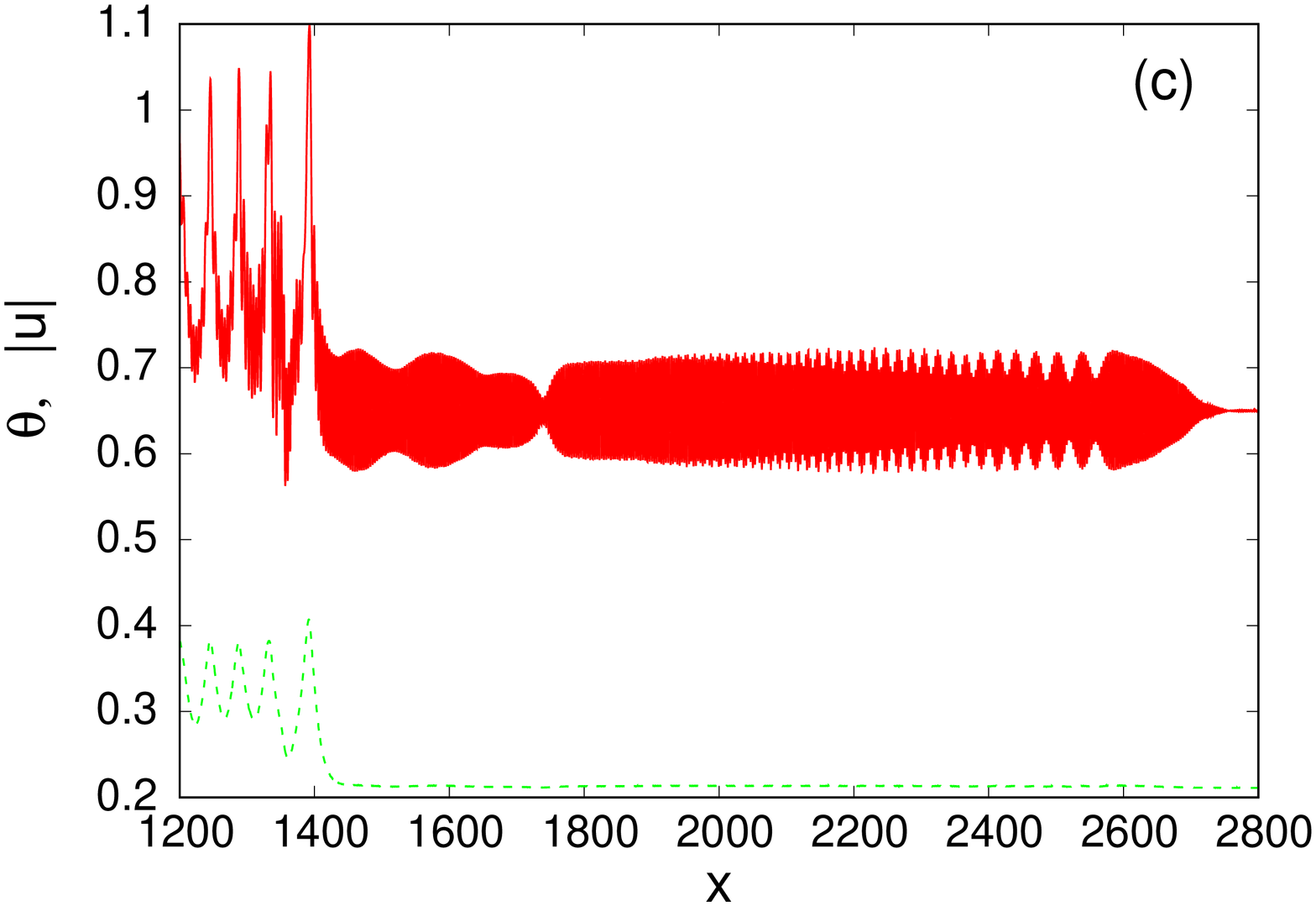}
\caption{Numerical solutions of the nematic equations (\ref{e:eeqn}) and (\ref{e:direqn}).  Red (solid) lines:  $|u|$; 
green (dashed) lines: $\theta$.  Solutions at (a) $z=500$, (b) $z=1000$, (c) $z=1500$.  Here $u_{-} = 1.0$, $u_{+} = 0.65$, 
$q=2$ and $\nu =200$.} 
\label{f:unstable}
\end{figure}

\begin{figure}
\centering
   \includegraphics[width=0.95\textwidth]{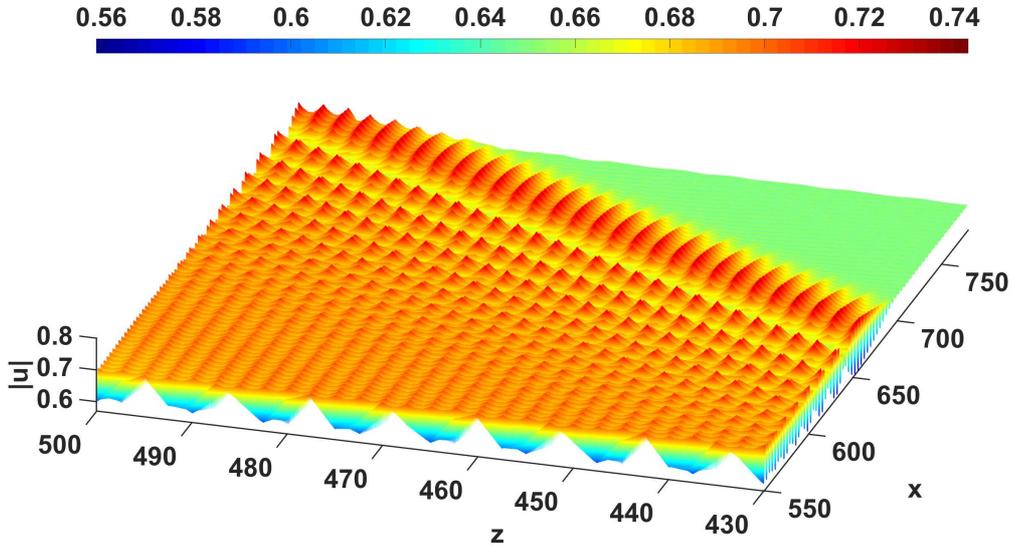}
\caption{Numerical solution of the nematic equations (\ref{e:eeqn}) and (\ref{e:direqn}) for $|u|$ showing evolution of
resonant wavetrain instability.  Here $u_{-} = 1.0$, $u_{+} = 0.65$, $\nu =200$ and $q = 2$.} 
\label{f:unstableevolve}
\end{figure}

Figure \ref{f:unstable} shows this modulational instability of the resonant wavetrain for $u_{+}=0.65$.  The instability of 
the resonant wavetrain is now clear, in contrast to the example shown in Figure \ref{f:types}(b), with the breakup of the resonant 
wavetrain as it propagates in $z$ now prominent.  Each wave of the DSW generates a resonant wave as a DSW is a modulated 
wavetrain \cite{pat}.  Figure \ref{f:unstable} shows that these resonant waves within the DSW are also unstable.  Figure 
\ref{f:unstableevolve} further details the evolution of the instability of the resonant wavetrain and shows a transition from a 
uniform wavetrain to a series of wavepackets.  This detailed modulational instability evolution closely resembles 
experimental photographs of Benjamin-Feir instability for water waves \cite{vandyke}.

\begin{figure}
\centering
 \includegraphics[width=0.6\textwidth,angle=270]{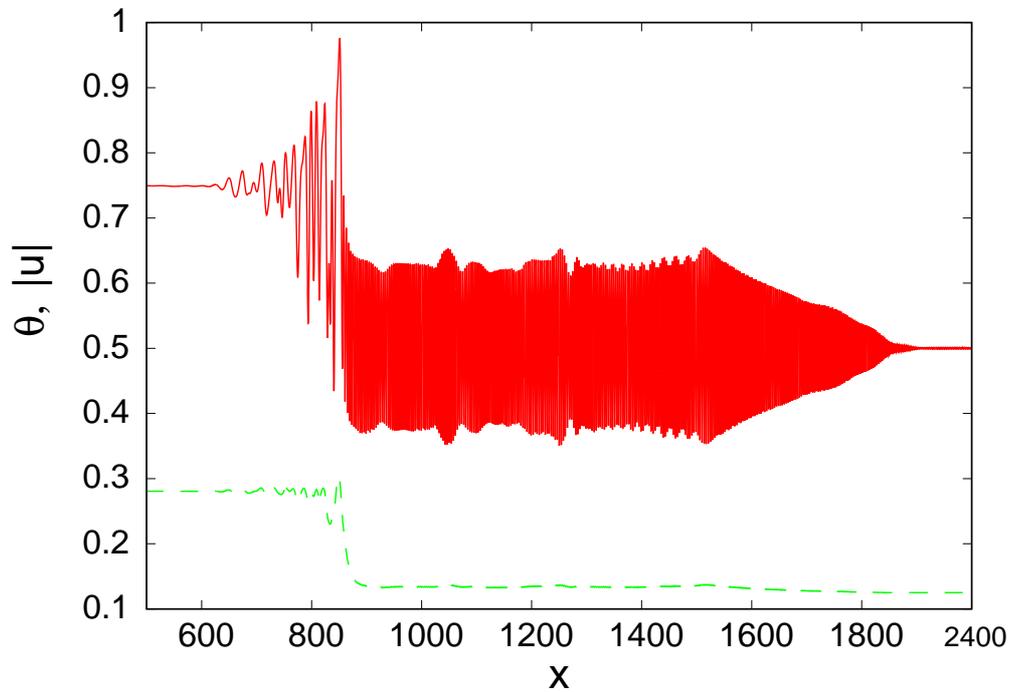}
 \caption{Numerical solutions of nematic equations (\ref{e:eeqn}) and (\ref{e:direqn}) at $z=1000$.  Red (solid) lines:  $|u|$; 
green (dashed) lines $\theta$.  Here $u_{-} = 1.0$, $u_{+} = 0.5$, 
$q=2$ and $\nu =200$.} 
\label{f:cdswdetail}
\end{figure}

The Stokes' wave expansion (\ref{e:rhostokes})--(\ref{e:omstokes}) and the associated Whitham modulation equations 
(\ref{e:lambda1})--(\ref{e:lambda4}) are weakly nonlinear and hold for relatively small amplitudes.  As the amplitude of 
the resonant wavetrain rises, it is found numerically that it restabilises, as illustrated in Figure \ref{f:cdswdetail}, which 
is an expanded view of the DSW of Figure \ref{f:types}(c).  It can be seen that while the resonant wavetrain is not uniform, 
as for the fifth order KdV resonant wavetrain \cite{pat}, the degree of modulation is much reduced over that of Figure 
\ref{f:unstable}.  This resonant wavetrain modulational instability is significant in the CDSW regime, which is analysed in 
the next section.

\section{Nematic Crossover DSW}
\label{s:cross}

The solution for the crossover DSW (CDSW) illustrated in Figure \ref{f:types}(c) will now be derived.  As can be seen from 
Figure \ref{f:types}(c), moving to the CDSW regime results in the KdV-type DSW in the PDSW and RDSW regimes becoming unstable, 
with the waves in the DSW ceasing to have a near linear amplitude decrease, as seen in Figure \ref{f:cdswdetail}, which gives 
a detail view of the DSW of Figure \ref{f:types}(c). 
The numerical leading edge velocity $s_{+}$ of Figure \ref{f:pdswrdsw}(b) shows that it has a maximum in $u_{+}$ around the boundary between the RDSW and CDSW regimes, so that the admissibility conditions discussed above give that the DSW becomes unstable, matching the numerical behaviour.  Figure \ref{f:types}(c) shows that for a large part of the CDSW, the waves have nearly equal amplitudes, with 
a rapid drop near its trailing edge.  As discussed, this detailed figure also shows that the resonant wavetrain is non-uniform, 
in stark contrast with the resonant DSW solution of the 5th order KdV equation, equation (\ref{e:kdv5}) with $\mu=0$, for 
which the resonant wavetrain has a constant amplitude with a front consisting of a partial DSW which takes the wavetrain 
down to the initial level $u_{+}$ ahead \cite{patkdv}.  The same DSW structure as the present CDSW was found for the 
(unstable) DSW for the focusing NLS equation \cite{tovbis}.  The DSW for this equation consists of a train of nearly 
equal amplitude waves, with a rapid decrease to the level $u_{-}$ behind at the trailing edge.  The work of \cite{tovbis} found that the major portion of unstable DSWs consists of a train of nearly equal amplitude waves, 
justifying the approximate theory of \cite{boreapprox} which approximates DSWs as a train of equal amplitude solitary waves, 
with the amplitude, width and spacing of the solitary waves found from conservation laws, motivated by the work of \cite{resflow} 
on the transcritical flow of a fluid over topography.  The approximate method of \cite{boreapprox} will then be used here 
to derive a solution for the nematic CDSW.

\begin{figure}[t]
\centering
   \includegraphics[width=0.33\textwidth,angle=270]{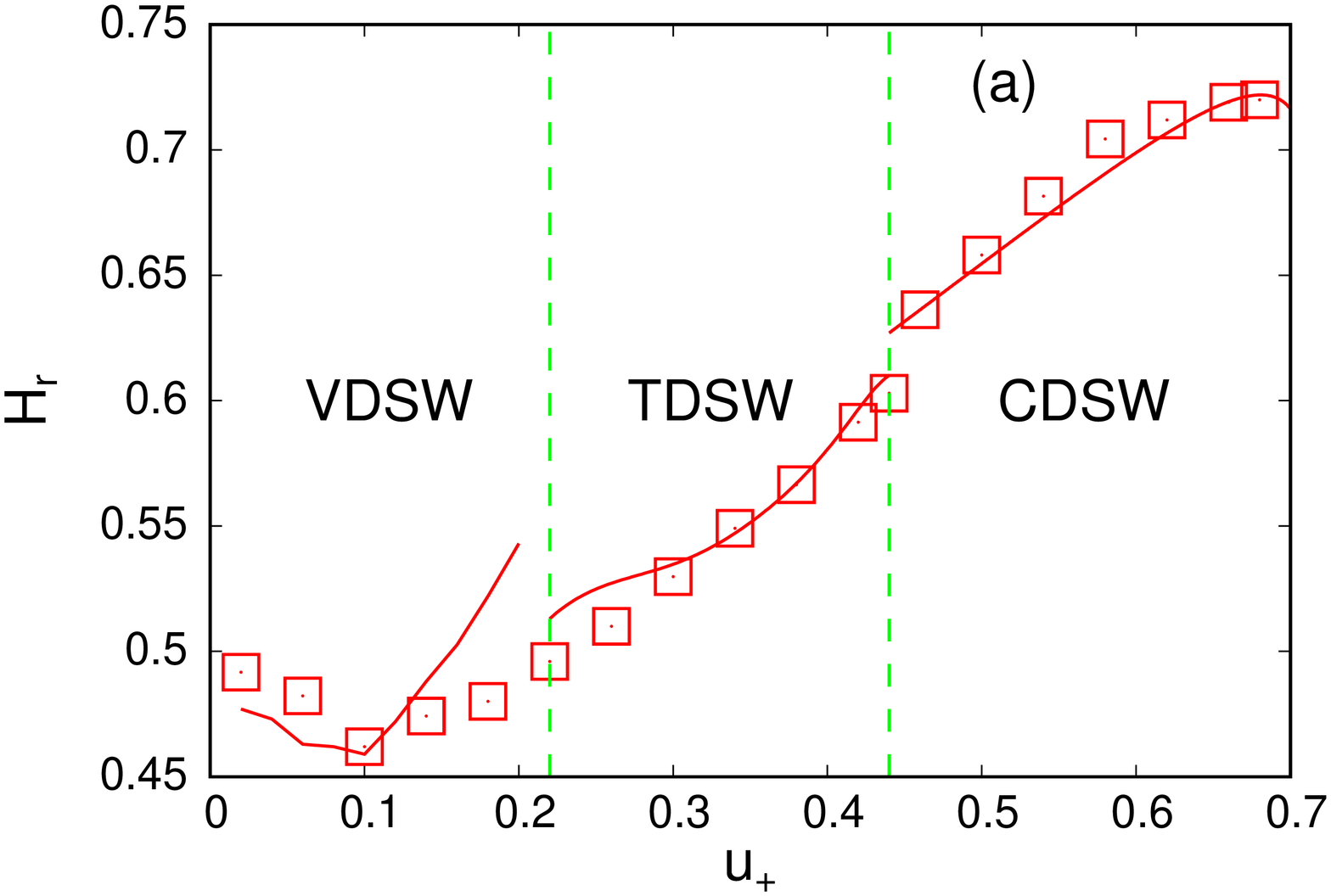}
      \includegraphics[width=0.33\textwidth,angle=270]{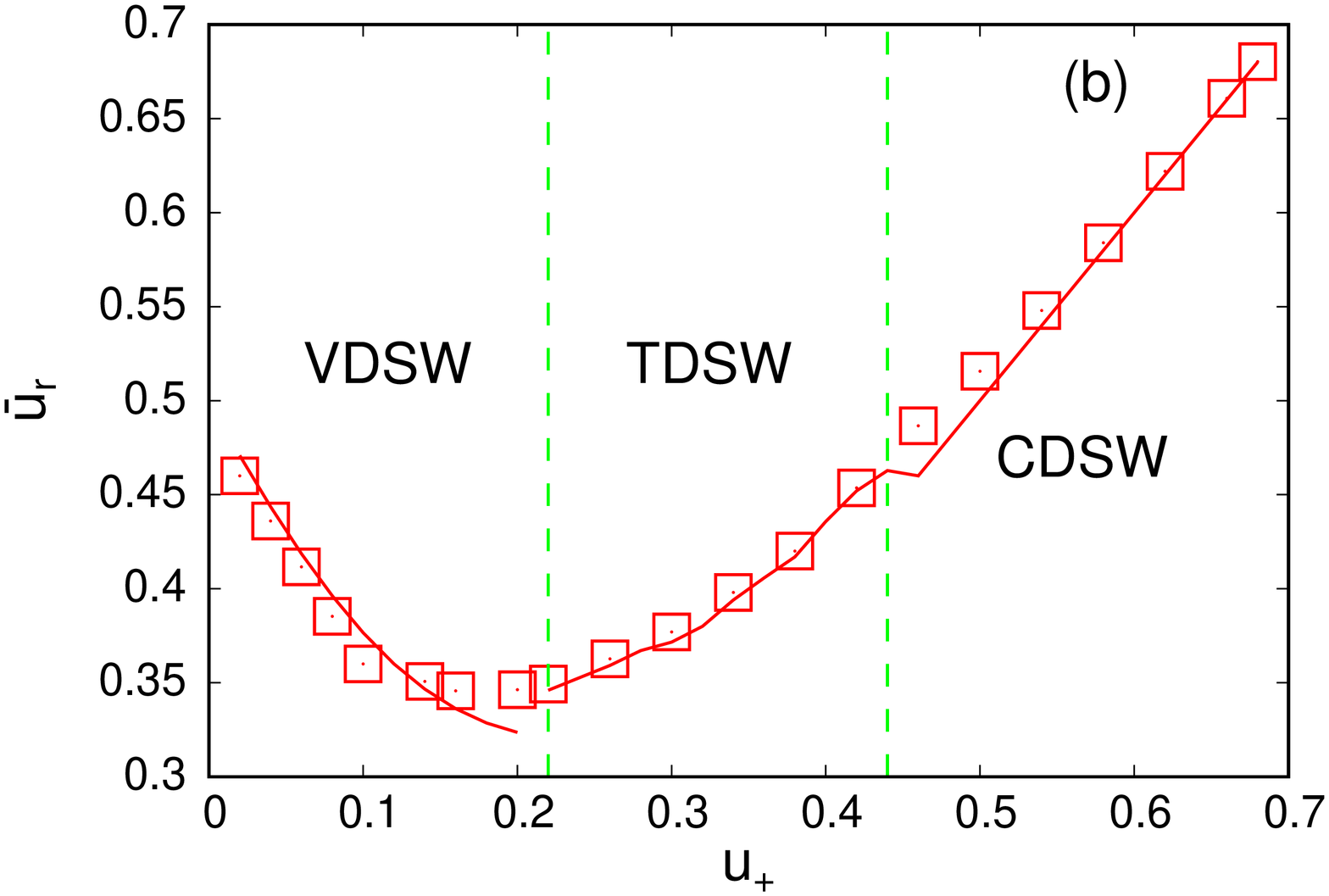}
      \includegraphics[width=0.33\textwidth,angle=270]{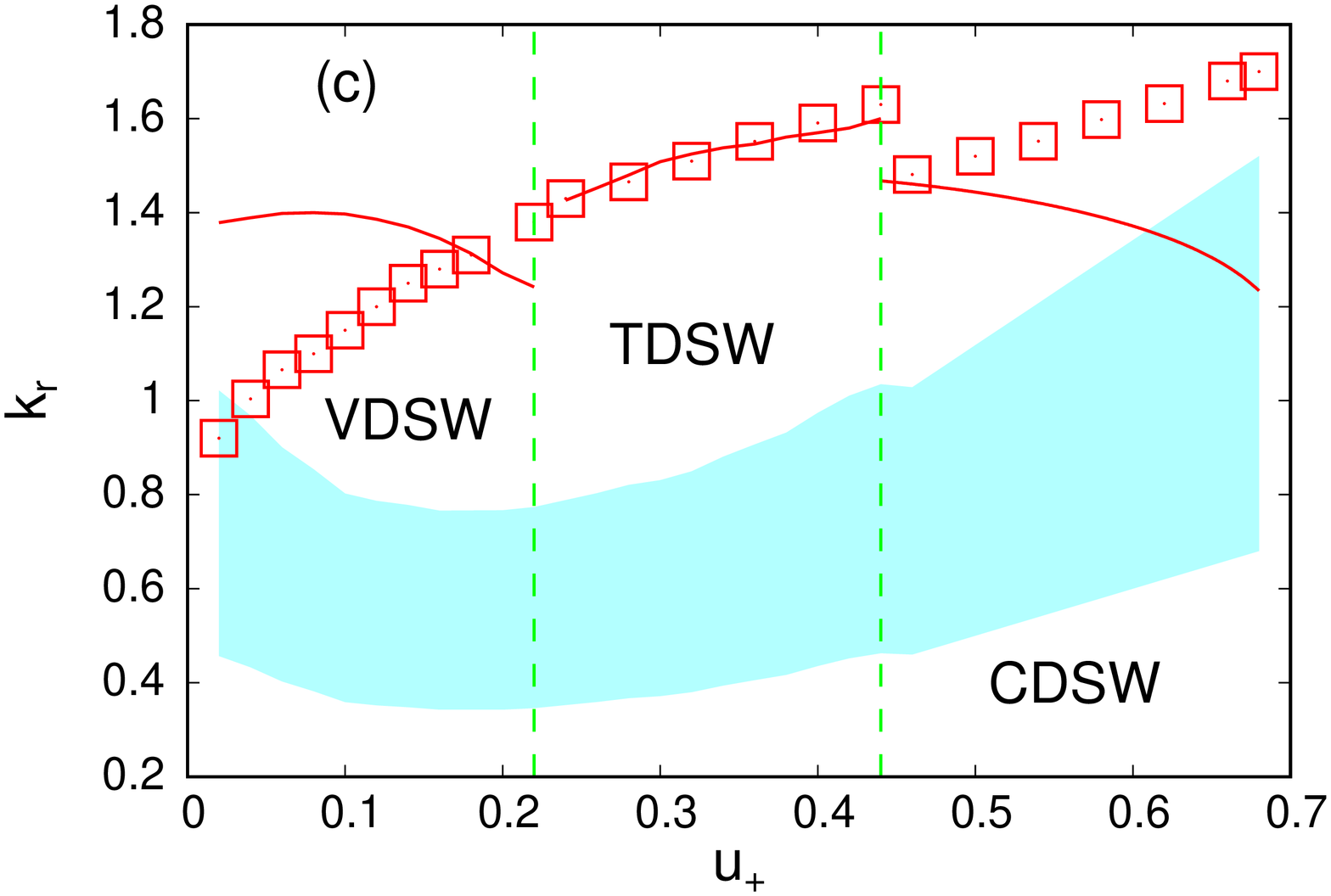}
\caption{Resonant wavetrain parameters as given by numerical solutions of the nematic equations (\ref{e:eeqn}) and (\ref{e:direqn}) and the DSW solutions for the CDSW, TDSW and VDSW regimes.  Numerical solution:  red boxes; analytical solution:  red (solid)
line.  (a) height $H_{r}$ of resonant wavetrain, (b) mean $\bar{u}_{r}$ of resonant wavetrain,
(c) wavenumber $k_{r}$ of resonant wavetrain.  The blue (shaded) region is the region (\ref{e:stablemodregion}) of 
modulational stability for the resonant wave.  Here $u_{-} = 1.0$, $\nu =200$ and $q = 2$.} 
\label{f:resonantwave}
\end{figure}

For simplicity, let us express the nematic KdV equation (\ref{e:kdv5nem}) as
\begin{equation}
\frac{\partial u_{1}}{\partial \eta} + B_{2}u_{1}\frac{\partial u_{1}}{\partial \xi}
+ B_{3} \frac{\partial^{3}u_{1}}{\partial \xi^{3}} 
+ \epsilon^{2}B_{4} \frac{\partial^{5} u_{1}}{\partial \xi^{5}} = 0,
\label{e:kdv5nemb} 
\end{equation}
where 
\begin{equation}
 B_{2} = 3\sqrt{\frac{2}{q}}, \quad
 B_{3} = \sqrt{\frac{2}{q}}\frac{u_{+}}{4}\left( \frac{\nu }{q} - \frac{q}{4u_{+}^{2}} \right), \quad 
 B_{4} = \sqrt{\frac{2}{q}}\frac{3\nu^{2}u_{+}^{2}}{16q^{2}}.
 \label{e:cs}
\end{equation}
This nematic KdV equation has the ``mass'' (optical power) conservation equation
\begin{equation}
 \frac{\partial}{\partial \eta} u_{1} +
 \frac{\partial}{\partial \xi} \left[ \frac{1}{2}B_{2} u_{1}^{2} + B_{3} u_{1\xi\xi} 
 + \epsilon^{2}B_{4} u_{1\xi\xi\xi\xi} \right] = 0
 \label{e:kdvmass}
\end{equation}
and the energy conservation equation
\begin{equation}
 \frac{\partial}{\partial \eta} \frac{1}{2}u_{1}^{2} + \frac{\partial}{\partial \xi} \left[ \frac{1}{3}B_{2}u_{1}^{3} 
 + B_{3}u_{1}u_{1\xi\xi}
 - \frac{1}{2}B_{3} u_{1\xi}^{2} + \epsilon^{2} B_{4} u_{1}u_{1\xi\xi\xi\xi} - \epsilon^{2}B_{4} u_{1\xi}u_{1\xi\xi\xi} 
 + \frac{1}{2}\epsilon^{2}B_{4}u_{1\xi\xi}^{2} \right] = 0.
 \label{e:kdvenergy}
\end{equation}
These will now be used to find an approximate solution for the nematic CDSW by assuming
that it consists of a uniform series of solitary waves \cite{boreapprox}, as discussed above.

Let us assume that the CDSW at position $\eta$ consists of $N$ equal solitary waves of 
amplitude $\tilde{a}_{s}$ and width $\tilde{w}_{s}$, where we shall use tildes to denote scaled variables in the moving 
and stretched coordinates $(\xi,\eta)$.  It is also assumed that the CDSW sheds a uniform resonant wavetrain of 
amplitude $\tilde{a}_{r}$ which propagates ahead of it.  Then as $\xi \to -\infty$, $u_{1} \to 1$ and as 
$\xi \to \infty$, $u_{1} \to \tilde{a}_{r} \cos \left( \tilde{k}_{r} \xi - \tilde{\omega}_{r}\eta \right)$, since 
$|u| = u_{+} + \epsilon^{2} u_{1}$ with $\epsilon^{2} = u_{i} - u_{+}$. 
Integrating the mass and energy equations (\ref{e:kdvmass}) and (\ref{e:kdvenergy}) over the CDSW, we have 
\begin{equation}
 N \int_{-\infty}^{\infty} u_{1} \: d\xi = \left[ \frac{1}{2} B_{2} - \frac{1}{4} B_{2} \tilde{a}_{r}^{2} \right] \eta , 
 \quad N \int_{-\infty}^{\infty} \frac{1}{2} u_{1}^{2} \: d\xi = \left[ \frac{1}{3} B_{2} 
 - \frac{1}{4} \tilde{c}_{g} \tilde{a}_{r}^{2} \right] \eta .
 \label{e:cdswmassen}
\end{equation}
In calculating the flux terms due to the resonant wavetrain on the right hand sides of (\ref{e:cdswmassen}), the averages of
the resonant wavetrain and its square over a period have been used as the resonant wave is high frequency relative to the 
CDSW, see Figure \ref{f:types}(c).  Also, $\tilde{c}_{g}$ is the group velocity of the shed radiation in the moving and 
stretched $(\xi,\eta)$ coordinates.  In calculating the energy conservation expression of (\ref{e:cdswmassen}) (the second) 
this group velocity is not that for the Kawahara equation (\ref{e:kdv5nem}), which is the group velocity for long waves in the 
nonlocal limit $\nu$ large, but that for the shed radiation, which is short wave radiation with the dispersion relation 
(\ref{e:displargenu}).  It can be seen from these conservation relations that the shed radiation leaks mass and energy 
from the CDSW, which is the reason for the rapid decrease in its amplitude as $u_{+}$ decreases, as seen in Figure 
\ref{f:pdswrdsw}(a).  

The integrals in the conservation expressions (\ref{e:cdswmassen}) are $N$ times the integrals for a single solitary wave.  
Due to the high number of derivatives in the Kawahara equation (\ref{e:kdv5nemb}), no exact solitary wave solutions have been 
derived for it.  However, as for the PDSW solution of Section \ref{s:pdsw}, the perturbation solution of 
\cite{perturbkdv,solapprox} which transforms the Kawahara equation to the standard KdV equation can be used to find a perturbed 
solitary wave solution of the Kawahara equation.  This transformation gives that the solitary wave solution of (\ref{e:kdv5nemb}) 
is
\begin{equation}
 u_{1} = \tilde{a}_{s} \sech^{2} \Theta - 5\epsilon^{2}C_{4} \tilde{a}_{s}^{2} \sech^{2} \Theta 
 + \frac{15}{2} \epsilon^{2} C_{4}\tilde{a}_{s}^{2} \sech^{4} \Theta,
 \label{e:highsol}
\end{equation}
with the width $\tilde{w}_{s}$ of the solitary wave and its phase $\Theta$ given by
\begin{equation}
\tilde{w}_{s} = \frac{\sqrt{ 6B_{3}}}{\sqrt{B_{2}}} \frac{\sqrt{2}}{\sqrt{\tilde{a}_{s}}}, \quad 
\Theta = \frac{\xi - \tilde{V}_{s}\eta}{\tilde{w}_{s}}, \quad 
\tilde{V}_{s} = \frac{1}{3}B_{2} \tilde{a}_{s}\left(1 + 2\epsilon^{2}C_{4}\tilde{a}_{s} \right).
\label{e:Theta}
\end{equation}
With this perturbation solution for the solitary wave, the mass and energy of a single solitary wave can be calculated as
\begin{equation}
 \int_{-\infty}^{\infty} u_{1} \: d\xi = 
 2\tilde{a}_{s}\tilde{w}_{s}
 \label{e:masssol}
\end{equation}
and
\begin{equation}
 \int_{-\infty}^{\infty} \frac{1}{2}u_{1}^{2} \: d\xi = \frac{2}{3}\tilde{a}_{s}^{2}\tilde{w}_{s} 
 + \frac{4}{3}\epsilon^{2}C_{4} \tilde{a}_{s}^{3}\tilde{w}_{s},
 \label{e:energysol}
\end{equation}
respectively.  Note that in the calculation of the energy, only terms up to $O(\epsilon^{2})$ have been retained, consistent with the order of the transformation \cite{perturbkdv,solapprox}.  
Dividing the mass and energy relations (\ref{e:cdswmassen}) hence gives
\begin{equation}
  \frac{1}{3}\tilde{a}_{s} + \frac{2}{3} \epsilon^{2}C_{4}\tilde{a}_{s}^{2}
 =
 \frac{2}{3}\frac{1 - \frac{3}{4}\frac{\tilde{c}_{g} \tilde{a}_{r}^{2}}{B_{2}}}{1 - \frac{1}{2}\tilde{a}_{r}^{2}}.
 \label{e:cdswsolamp}
\end{equation}

Equation (\ref{e:cdswsolamp}) determines the scaled amplitude $\tilde{a}_{s}$ of the solitary waves of the CDSW once 
the scaled amplitude $\tilde{a}_{r}$ of the shed resonant radiation is known.  In a previous analysis of the nematic DSW 
a WKB solution for the shed resonant radiation was found which was found to be in excellent agreement with numerical 
solutions \cite{nemgennady}.  A WKB solution was used as the resonant wavetrain is short wave relative to the DSW.  This 
solution for the resonant wavetrain was given in terms of the original variables $u$ and $(x,z)$.  Returning the amplitude relation (\ref{e:cdswsolamp}) 
to the original variables using $a_{s} = \epsilon ^{2} \tilde{a}_{s} = (u_{i} - u_{+}) \tilde{a}_{s}$ and 
$a_{r} = \epsilon^{2}\tilde{a}_{r} = (u_{i} - u_{+}) \tilde{a}_{r}$ and using the resonant wavetrain solution of \cite{nemgennady} gives that the amplitude of the CDSW solitary waves is
determined by
\begin{equation}
a_{s} \left[ 1 + 2 C_{4}a_{s} \right]
= 2\left(u_{i} - u_{+}\right) 
 \frac{1- \frac{3}{4} \frac{\left( c_{g}-\sqrt{\frac{2}{q}} \: u_{+} \right) a_{r}^{2}}{B_{2}(u_{i} - u_{+})^{3}}}{1 - \frac{1}{2} \frac{a_{r}^{2}}{(u_{i} - u_{+})^{2}}},
 \label{e:cdswsolamps}
\end{equation}
with the resonant radiation's amplitude $a_{r}$ given by \cite{nemgennady}
\begin{equation}
 a_{r} = \frac{u_{i} - u_{+}}{1 + 
 \frac{2u_{+}k_{r}a_{s}}{qs_{+}(k_{r} - s_{+})^{2}}}.
 \label{e:ar}
\end{equation}
Here the group velocity $c_{g} = k_{r}$, see (\ref{e:displargenu}) as $v_{+}=0$, has been used and the velocity of the 
solitary waves of the CDSW in the original variables is
\begin{equation}
 s_{+} = \sqrt{\frac{2}{q}} \: u_{+} + \frac{1}{3}B_{2}a_{s} \left( 1 + 2C_{4}a_{s}\right)
 \label{e:spcdsw}
\end{equation}
as the $\xi$ frame moves with velocity $u_{+}\sqrt{2/q}$.  The resonant wavenumber $k_{r}$ is determined by the 
resonance condition (\ref{e:krdsw}).  Equations (\ref{e:cdswsolamp}) and (\ref{e:krdsw}), with (\ref{e:ar}), were solved 
numerically using Newton's method to determine the amplitude $a_{s}$ and velocity $s_{+}$ of the solitary waves of the CDSW 
and the amplitude $a_{r}$ and wavenumber $k_{r}$ of the resonant wavetrain.  It was found that the resonant wave amplitude 
$a_{r}$ is negligible for $u_{+} > 0.70$.  This is then the upper limit of the CDSW regime, in excellent agreement with the 
upper CDSW regime limit of Table \ref{t:regions} found from numerical solutions.  The lower limit of the CDSW regime will 
be discussed in the next section dealing with the TDSW regime.

Figure \ref{f:pdswrdsw}(a) shows the height $H_{s} = u_{+} + a_{s} + (5/2)C_{4}a_{s}^{2}$ of the solitary waves of the CDSW
(see (\ref{e:highsol})) as given by numerical solutions and the equal amplitude approximation of this section.  It can be seen 
that there is excellent agreement, except near the boundary with the TDSW regime.  This is expected since as the height of the 
jump $u_{i} - u_{+}$ grows, the KdV approximation becomes less valid.  In addition, as the TDSW regime is approached, the form 
of the DSW changes fundamentally, with the waves of the DSW disappearing, except for one wave at the leading edge, see Figure 
\ref{f:types}(d).   Figure \ref{f:types}(c) shows that the amplitudes of the individual waves of the CDSW have a random variation 
as the CDSW is unstable.   To calculate $H_{s}$ from the numerical solutions, an average was then taken over the CDSW, which is 
consistent with the equal amplitude approximation.  Figure \ref{f:pdswrdsw}(a) shows a rapid decrease in the height of the CDSW 
as $u_{+}$ decreases.  However, the amplitude $a_{r}$ of the shed radiation grows, as can be seen from Figure 
\ref{f:resonantwave}(a), since $a_{r} = H_{r} - u_{+}$.  The shed resonant radiation acts as a damping on the DSW and leads to 
its decay over the non-resonant PDSW case.  Figure \ref{f:pdswrdsw}(b) shows comparisons for the velocity $s_{+}$ of the leading 
edge of the CDSW.  The agreement is not as good as for the solitary wave height, but the difference between the numerical and 
analytical solutions is about $4\%$.   There are no results presented for the trailing edge velocity, as for the PDSW and RDSW 
cases, as the equal amplitude approximation cannot give results for the linear trailing edge.

\begin{figure}[t]
\centering
   \includegraphics[width=1.0\textwidth]{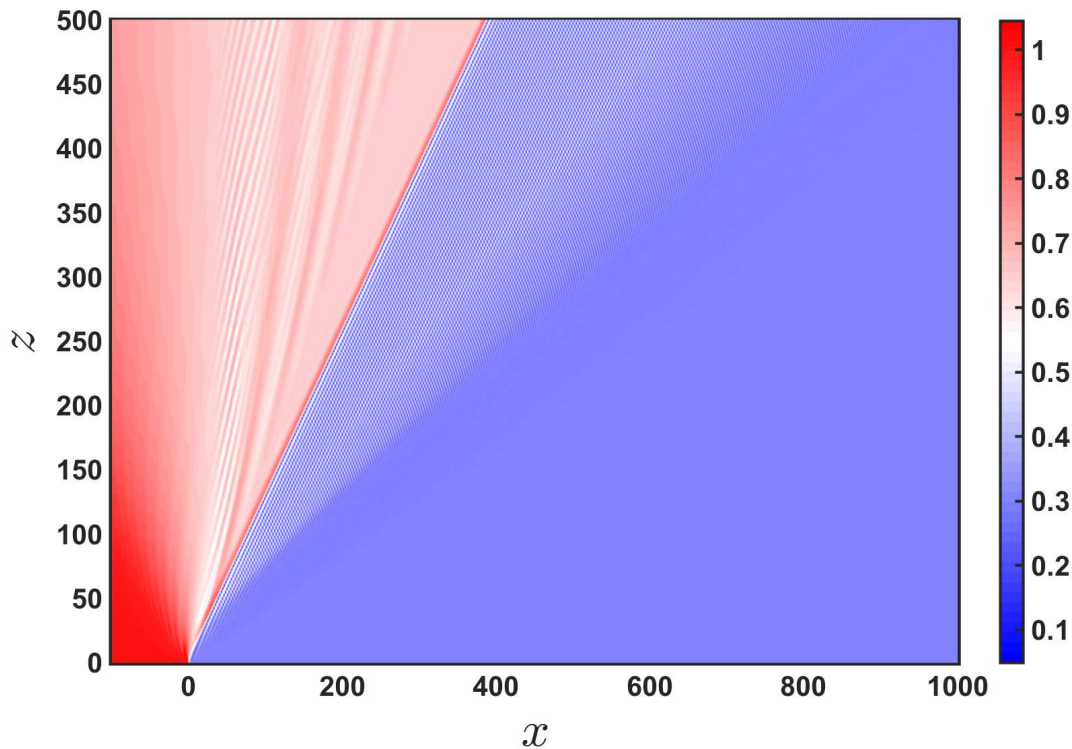}
\caption{Numerical solution of the nematic equations (\ref{e:eeqn}) and (\ref{e:direqn}) for $|u|$ in the TDSW regime showing 
the Whitham shock linking the resonant wavetrain with the intermediate shelf.
Here $u_{-} = 1.0$, $u_{+} = 0.3$, $\nu =200$ and $q = 2$.} 
\label{f:shockevolve}
\end{figure}

Figure \ref{f:resonantwave}(a) shows comparisons between numerical and analytical solutions for the height 
$H_{r} = u_{+} + a_{r}$ of the resonant wavetrain.  There is again excellent agreement, as expected as the WKB solution was 
previously found to be in excellent agreement with numerical solutions \cite{nemgennady}.  It can be seen that as the 
RDSW regime is approached the amplitude $a_{r} = H_{r} - u_{+}$ of the resonant wavetrain goes to zero, as required.  As 
discussed in Section \ref{s:pdsw} the resonant wavetrain in the CDSW regime is unstable when its amplitude is low and 
stabilises at high enough amplitude.  For the comparisons of Figure \ref{f:resonantwave}(a) the amplitudes and heights of 
low amplitude numerical resonant waves were determined before instability broke the wavetrain up.  In this regard, a unit 
distance in the non-dimensional variable $z$ corresponds to about $2\mu m$ for optical beams in the near infrared 
\cite{curvenature,curvejosab,curvepra}.  Typical experimental nematic cell lengths are $\sim 1mm$, which is $z=500$.  As 
shown in Figure \ref{f:unstable} the breakup of the resonant wavetrain in the vicinity of the CDSW occurs after $z=1000$.  
This is well beyond the experimental distance at which optical solitary waves can be observed, due to their decay due to 
scattering losses \cite{PR,Wiley}.  Numerical solutions show that the mean height of the resonant wavetrain is given by $\bar{u}_{r} = \sqrt{q\theta_{r}}$.  The resonant wavetrain is rapidly varying.  Averaging the director equation (\ref{e:direqn}) then gives this mean height expression.  Figure \ref{f:resonantwave}(b) shows a comparison between the mean height $\bar{u}_{r}$ of 
the resonant wavetrain as given by numerical solutions and by the theoretical CDSW value $\bar{u}_{r}=u_{+}$, with 
excellent agreement seen.  Finally, Figure \ref{f:resonantwave}(c) shows comparisons for the wavenumber $k_{r}$ of the resonant 
wavetrain.  The numerical values of $k_{r}$ were determined by averaging over the resonant wavetrain.  Unlike the resonant 
wave height and mean level, the comparison is poor, except near the TDSW regime where the resonant wavetrain restabilises.  The 
reason is that the resonant wavetrain is unstable over nearly all of the CDSW regime and does not 
consist of a single dominant wavenumber, as seen in Figures \ref{f:unstable} and \ref{f:unstableevolve}.  


\section{Nematic TDSW}
\label{s:tdsw}

As the initial level ahead $u_{+}$ decreases in the CDSW regime, the effect of the shed resonant wavetrain grows until the 
DSW itself ceases to exist and there is just a large amplitude resonant wavetrain with a small amplitude wave at its trailing 
edge linking it to the intermediate level, as shown in Figure \ref{f:types}(d).   
This is similar to the behaviour of the DSW for the Kawahara equation (\ref{e:kdv5}) for which the DSW ceases to exist for 
large enough initial steps (or small enough $\mu$).  The DSW is replaced by what is termed a TDSW, a travelling dispersive shock 
wave \cite{patkdv}.  This work on the Kahawara equation shows that in the TDSW regime the DSW largely disappears, with a 
single remnant negative polarity solitary wave connecting the resonant wavetrain to the intermediate level $u_{i}$.  The 
same structure is seen in Figure \ref{f:types}(d) for the nematic TDSW regime.  

It was shown for the Kawahara equation that in the TDSW regime the resonant wavetrain is linked to the level $u_{+}$ ahead 
by a partial DSW \cite{pat}.  A partial DSW is a modulated wavetrain which takes a uniform wavetrain down to a constant level, 
in contrast to a full DSW which is a modulated wavetrain which connects two uniform levels \cite{kdvbound1,kdvbound2}.
As stated, a negative polarity solitary wave connects the resonant wavetrain to the level $u_{i}$ behind \cite{patkdv}.  As 
there is no known solitary wave solution of the Kawahara equation of either polarity, this connection was done numerically 
and by using approximate theory in previous work \cite{patkdv,pat}.  However, it has been recently realised that the determination 
of this connecting negative polarity solitary wave is not necessary \cite{patjump}.  The connection between the resonant wavetrain 
and the level behind can be treated as a Whitham shock, the wavetrain equivalent of a gas dynamic shock, for the Whitham modulation 
equations governing the resonant wavetrain, confirming speculation by Whitham when he originally developed modulation theory 
\cite{whitham,modproc}.  This shock-like nature of the connection can be seen for the nematic TDSW illustrated in Figure 
\ref{f:shockevolve}, where the discontinuous connection between the resonant wave and the uniform level behind is clearly visible.  

For the modulation equations to have a Whitham shock, they need to form a hyperbolic system.  The Whitham modulation equations 
(\ref{e:lambda1})--(\ref{e:lambda4}) for the Stokes' wave are hyperbolic in the restricted interval (\ref{e:stablemodregion}) 
and Figure \ref{f:resonantwave}(c) shows that the resonant wavetrain falls outside this region over nearly all the CDSW and 
all the TDSW existence intervals.  However, as shown in Figure \ref{f:types}(d), the resonant wavetrain is stable in the TDSW 
regime.  As discussed above, the Stokes' wave modulation equations are valid in the weakly nonlinear limit and the numerical 
results show that they do not correctly predict modulational stability for the resonant TDSW wave as it has relatively large 
amplitude.  As the full Whitham modulation equations are not known, the Stokes' wave modulation equations will be used to find 
jump conditions for the connection between the intermediate level $u_{i}$ and the resonant wavetrain in the TDSW regime.

As for the compressible gas equations, the appropriate modulation equations to determine the wave shock jump conditions are 
the mass (\ref{e:avmass}), momentum (\ref{e:avmom}) and energy equations (\ref{e:avenergy}) \cite{whitham}.  Consistent with 
the TDSW structure illustrated in Figure \ref{f:types}(d), ahead of the Whitham shock there is the resonant wavetrain with 
wavenumber $k=k_{r}$, mean height $\bar{u} = \bar{u}_{r}$ ($\bar{\rho} = \bar{\rho}_{r} = \bar{u}_{r}^{2}$), flow
$\bar{v}_{r}$ and amplitude $a = a_{r}$.  Behind the Whitham shock there is no wavetrain, so the wavenumber $k=0$, the amplitude 
$a=0$ and the mean height is $\bar{u} = u_{i}$.  Let us denote the Whitham shock velocity by $U_{\mbox{shock}}$.  Then the mass 
(\ref{e:avmass}), momentum (\ref{e:avmom}) and energy (\ref{e:avenergy}) modulation equations give the jump conditions
\begin{equation}
-\left[ \bar{\rho}_{r} - \rho_{i} \right] U_{\mbox{shock}} + \left[ \bar{\rho}_{r}\bar{v}_{r} 
+ \frac{k_r a_{r}^{2}}{4\bar{\rho}_r} - \rho_{i}v_{i} \right] = 0,
\label{e:massjump}
\end{equation}
\begin{equation}
-\left[ \bar{\rho}_r \bar{v}_r + \frac{a^{2}_{r} k_r}{4\bar{\rho}_r} - \rho_i v_i \right] U_{\mbox{shock}}
+ \left[ \frac{\bar{\rho}^2_r}{q} + \bar{v}^2_r\bar{\rho}_r + \frac{a^{2}_{r}k_r^2}{4\bar{\rho}_r} + 
\frac{a^{2}_{r} \bar{v}_r k_r}{2\bar{\rho}_r} - \frac{\rho^2_i}{q} - v^2_i\rho_i \right] = 0,
\label{e:momjump}
\end{equation}
and
\begin{eqnarray}
& & \mbox{} -\left[ \frac{2\bar{\rho}^2_r}{q} + \bar{v}^2_r \bar{\rho}_r + \frac{a^{2}_{r} k^2_r}{4\bar{\rho}_r}
+ \frac{a^{2}_{r} \bar{v}_r k_r}{2\bar{\rho}_r} - \frac{2\rho^2_i}{q} - v^2_i  \rho_i \right] U_{\mbox{shock}} \nonumber \\
& & \mbox{} + \left[ \frac{4\bar{\rho}^2_r \bar{v}_r}{q} + \bar{\rho}_r \bar{v}^3_r + \frac{a^{2}_{r} k_r}{q} 
+  \frac{3a^{2}_{r} k^2_r \bar{v}_r}{4\bar{\rho}_r} + \frac{3a^{2}_{r} \bar{v}^2_r k_r}{4\bar{\rho}_r} 
+ \frac{a^{2}_{r} k^3_r}{4\bar{\rho}_r} - \frac{4\rho^2_i v_i}{q} - \rho_i v^3_i \right] = 0,
\label{e:energyjump}
\end{eqnarray}
respectively \cite{whitham}.  Here, $v_{i}$ on the intermediate level is 
related to $\rho_{i}$ by (\ref{e:v2expr}) for the expansion wave solution from the initial level $u_{-}$ behind.
In addition, there is the resonance condition obtained from the Stokes' wave dispersion relation (\ref{e:omstokes}) 
\begin{equation}
 U_{\mbox{shock}} = \bar{v}_{r} + \frac{1}{2}k_{r} + \frac{2\bar{\rho}}{qk_{r}} - \frac{k_{r}a_{r}^{2}}{8\bar{\rho}_{r}^{2}}.
 \label{resstokes}
\end{equation}
These jump conditions and the resonance condition form four equations for the six unknown parameters $\rho_{i}$, $\bar{\rho}_{r}$, 
$\bar{v}_{r}$, $k_{r}$, $a_{r}$ and $U_{\mbox{shock}}$.  For the fifth order KdV equation, (\ref{e:kdv5}) with $\mu = 0$, the 
system was completed by matching the resonant wavetrain to the partial DSW at its leading edge, which brings the solution 
back to the level $u_{+}$.  This partial bore was found as a simple wave solution of the modulation equations of the fifth 
order KdV equation in Riemann invariant form.  Even when the Stokes' wave modulation equations are hyperbolic, they cannot 
be expressed in Riemann invariant form.  So to complete the solution for the TDSW regime, some assumptions based on 
numerical solutions will be made.

\begin{figure}
\centering
\includegraphics[width=0.5\textwidth,angle=270]{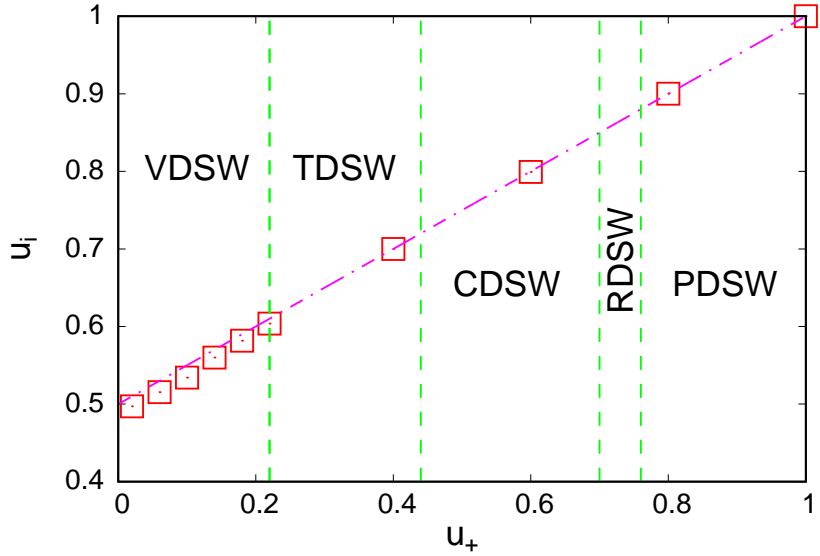}
\caption{Intermediate level $u_{i}$ as given by numerical solutions of the nematic equations (\ref{e:eeqn}) and 
(\ref{e:direqn}), the Riemann invariant value (\ref{e:ui}) for the PDSW, RDSW, CDSW and TDSW regimes and the 
VDSW value (\ref{e:uivdsw}).  Numerical solution:  red boxes; theoretical values: pink (dot-dashed) line.  
Here $u_{-} = 1.0$, $\nu =200$ and $q = 2$.} 
\label{f:intermediate}
\end{figure}

Figure \ref{f:intermediate} shows the level $u_{i}$ of the shelf between the expansion wave from $u_{-}$ and the resonant 
wavetrain.  It can be seen that in the TDSW regime, it is given by the average (\ref{e:ui}), $(1+u_{+})/2$ in this case, based 
on the conservation of the Riemann invariant $R_{-}$ for the non-dispersive shallow water equations, given by (\ref{e:cm}), 
through the Whitham shock.  The reason that this is an excellent approximation is that the Whitham shock is relatively weak, as 
seen from Figure \ref{f:types}(d), so that this Riemann invariant is conserved through the shock to leading order \cite{whitham}.  
It will then be assumed that $u_{i}$ is given by (\ref{e:ui}).  Consistent with this, it is assumed from the shallow water 
Riemann invariant $R_{-}$ given by (\ref{e:cm}) that 
\begin{equation}
 \bar{v}_{r} = 2\sqrt{\frac{2}{q}} \left( \sqrt{\bar{\rho}_{r}} - \sqrt{\rho_{+}} \right) = 
 2\sqrt{\frac{2}{q}} \left( \sqrt{\bar{\rho}_{r}} - u_{+} \right).
 \label{e:vrcm}
\end{equation}
As noted in Section \ref{s:modtheory} the shallow water characteristics (\ref{e:lambda3}) and (\ref{e:lambda4}) are always 
real, so the Riemann invariants on these can be used to propagate the solution.  This assumption is based on ignoring the wave 
amplitude correction in the Riemann variable (\ref{e:lambda4}) as this amplitude is small.  With these assumptions, the jump 
conditions (\ref{e:massjump})--(\ref{e:energyjump}) were solved numerically using Newton's method.

\begin{figure}[t]
\centering
\includegraphics[width=0.6\textwidth,angle=270]{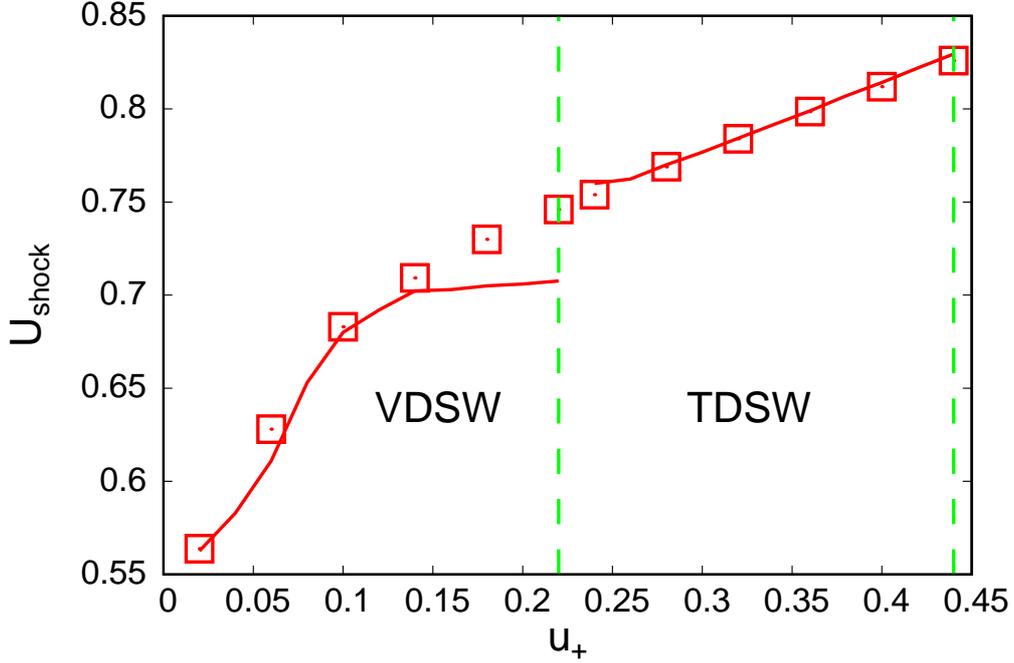}
   \caption{Modulation shock velocity $U_{\mbox{shock}}$ in the TDSW and VDSW regimes as given by numerical solutions of 
   the nematic equations (\ref{e:eeqn}) and (\ref{e:direqn}) and the TDSW and VDSW regime solutions.  Numerical solution:  
   red boxes; analytical solution:  red (solid) line.  Here $u_{-} = 1.0$, $\nu =200$ and $q = 2$.} 
\label{f:shock}
\end{figure}

Figures \ref{f:resonantwave}(a) and (b) give comparisons between numerical solutions and the results from the shock 
jump conditions for the resonant wave height $H_{r} = a_{r} + \bar{u}_{r}$ and the resonant wave mean level $\bar{u}_{r}$ in 
the TDSW regime.  It can be seen that there is excellent agreement for both parameters with numerical solutions, with some 
slight deviation for the resonant wave height when it transitions to the VDSW regime.  As for the equivalent transitions 
for the CDSW regime, this is to be expected as the resonant wave drastically changes form in the VDSW regime, as seen 
from Figure \ref{f:types}(e).  Figure \ref{f:resonantwave}(c) shows a comparison for the resonant wavenumber $k_{r}$ and 
nearly perfect agreement across the TDSW regime is seen, unlike for the CDSW regime.  The higher resonant wave amplitude 
in the TDSW regime has stabilised the wavetrain, so that the assumption of a single dominant resonant wavenumber is now valid.  
For $u_{+} > 0.44$, the Whitham shock velocity $U_{\mbox{shock}}$ is greater than the (linear) group velocity $c_{g}$ of the 
resonant wavetrain, which is unphysical.  This value of $u_{+}$ is then the boundary between the TDSW and CDSW regimes, 
in excellent agreement with numerical results, as tabulated in Table \ref{t:regions}.

Figure \ref{f:shock} shows a comparison between numerical solutions and the Whitham shock jump conditions for the shock 
velocity $U_{\mbox{shock}}$.  Again there is near perfect agreement, except for a slight deviation near the transition
to the VDSW regime.  This excellent agreement for the TDSW parameters validates the assumptions (\ref{e:ui}) and 
(\ref{e:vrcm}) made above.

\section{Nematic Vacuum DSW}
\label{s:vacuum}

\begin{figure}
\centering
\includegraphics[width=0.6\textwidth,angle=270]{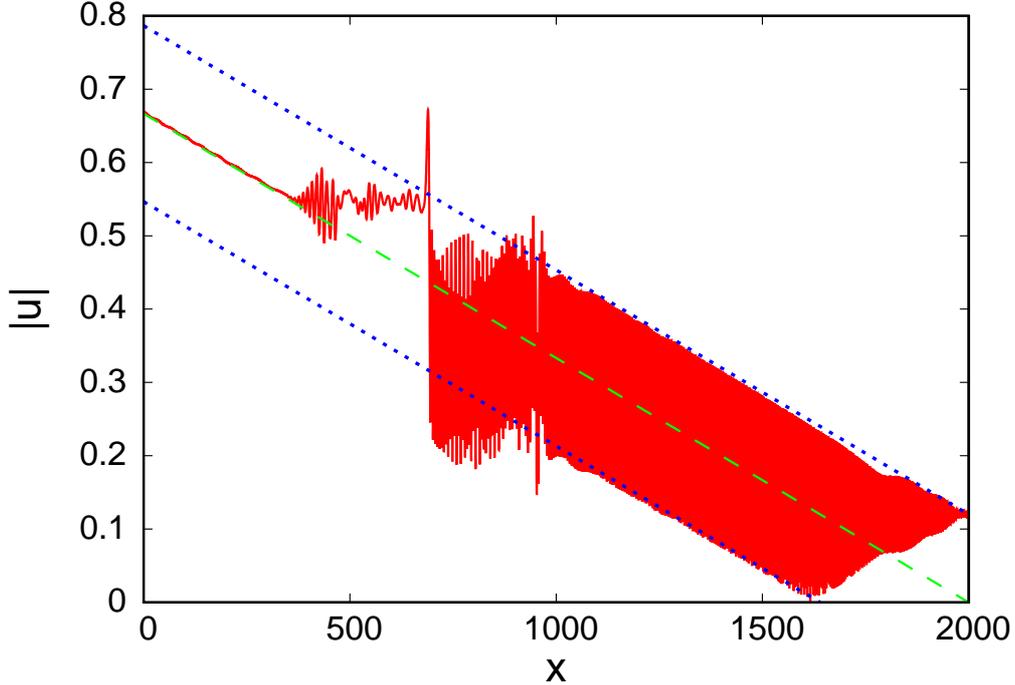}
   \caption{Numerical solution of the nematic equations (\ref{e:eeqn}) and (\ref{e:direqn}) for $|u|$:  red (solid) line;  
   expansion wave solution (\ref{e:midsolnu}):  green (dashed) line; expansion wave solution (\ref{e:midsolnu}) $\pm u_{+}$:  
   blue (dotted) line, $+u_{+}$ (upper), $-u_{+}$ (lower).  Here $u_{-} = 1.0$, $u_{+}=0.12$, $\nu =200$ and $q = 2$.} 
\label{f:vdswjust}
\end{figure}

As the level ahead $u_{+}$ decreases in the TDSW regime, the resonant wavetrain hits the vacuum point at which $\bar{u}_{r}-a_{r}=0$, 
at which point the DSW changes form \cite{gennady,moro}.  When this critical level is reached, the DSW changes to the vacuum 
DSW (VDSW) regime, with Figure \ref{f:types}(e) showing a typical VDSW solution.  Table \ref{t:regions} shows this critical 
value of $u_{+}$ for the vacuum point as given by the jump conditions (\ref{e:massjump})--(\ref{e:energyjump}) and by numerical 
solutions, with excellent agreement seen.  The partial DSW of the TDSW regime which brings the resonant wavetrain down to 
$u_{+}$ is now on a (linearly) varying mean.  As for the TDSW regime, there is a Whitham shock which links the resonant wavetrain 
to the intermediate level $u_{i}$, so that this will be determined by the same jump conditions as for the TDSW regime.  As for 
the TDSW regime, some assumptions based on numerical solutions will need to be made to derive the VDSW solution due to the 
lack of Whitham modulation equations in Riemann invariant form.  

Figure \ref{f:vdswjust} shows an expanded version of the VDSW solution of Figure \ref{f:types}(e).  On the same figure 
the expansion wave solution (\ref{e:midsolnu}) continued down to $u=0$ and this solution with $\pm u_{+}$ added are shown.  
It can be seen that the mean and envelopes of the transition wave bringing the resonant wavetrain down to $u_{+}$ from the 
intermediate shelf are well approximated by this expansion wave solution.  This type of wavetrain structure does not exist 
for the NLS DSW in the vacuum case as there is no resonance \cite{gennady}.  Without full Whitham modulation equations for 
the fully nonlinear nematic equations, there is no analytical method to justify this observed structure. This structure will 
be assumed here, with some justification based on weak shock theory. 


Figure \ref{f:vdswjust} shows that the resonant wavetrain and its leading edge have constant amplitudes, so that $a_{r} = u_{+}$ 
will be assumed, which is consistent with the wave envelopes of this figure.  The resonant wave amplitude is small, so that it 
decouples from the mean height variation in the modulation equations.  As the resonant wave amplitude is small, it will be 
ignored in the modulation equations (\ref{e:lambda1})--(\ref{e:lambda4}), so that the modulation equations (\ref{e:lambda3}) 
and (\ref{e:lambda4}) become the shallow water equations \cite{whitham}.  The jump conditions of Section \ref{s:tdsw} then 
become the shallow water equation jump conditions, for which the most convenient form is \cite{whitham}
\begin{equation}
 U_{\mbox{shock}} = \bar{v}_{r} \pm \sqrt{\frac{2}{q}} \left[ \frac{u_{i}^{2}\left( u_{i}^{2} 
 + \bar{u}_{r}^{2}\right)}{2\bar{u}_{r}^{2}} \right]^{1/2}, \quad
 v_{i} = \bar{v}_{r} \pm \sqrt{\frac{2}{q}} \frac{u_{i}^{2} - \bar{u}_{r}^{2}}{u_{i}^{2}} 
 \left[ \frac{u_{i}^{2}\left( u_{i}^{2} + \bar{u}_{r}^{2}\right)}{2\bar{u}_{r}^{2}} \right]^{1/2}. 
 \label{e:shallow12}
 \end{equation}
The sign choice for the shallow water equations (and the compressible gas equations) is the $+$ sign.  However, with this 
choice, in the limit $u_{+} \to 0$, the shock velocity approaches the front velocity of the dam break solution $2u_{-}\sqrt{2/q}$, 
and $u_{i} \to 0$.  This does not accord with numerical solutions, which show that $u_{i} \to u_{-}/2$ in this limit, which
is the Riemann invariant value (\ref{e:ui}).  The correct behaviour is obtained with the $-$ sign choice in the shallow water 
jump conditions (\ref{e:shallow12}).  Figure \ref{f:vdswjust} also shows that the Whitham shock is weak, the jump is small.  
Expanding the jump conditions (\ref{e:shallow12}) for small jump height, $|u_{i} - \bar{u}_{r}|$ small, gives
\begin{eqnarray}
 U_{\mbox{shock}} & = & \bar{v}_{r} - \sqrt{\frac{2}{q}} \bar{u}_{r} \left[ 1 + \frac{3}{4} \frac{u_{i}^{2} 
 - \bar{u}_{r}^{2}}{\bar{u}_{r}^{2}} 
 - \frac{9}{32} \frac{\left( u_{i}^{2} - \bar{u}_{r}^{2}\right)^{2}}{\bar{u}_{r}^{4}} + \ldots \right], \label{e:jump1small} \\
 v_{i} & = & \bar{v}_{r} - \sqrt{\frac{2}{q}} \frac{u_{i}^{2} - \bar{u}_{r}^{2}}{\bar{u}_{r}} \left[ 1 - \frac{u_{i}^{2} 
 - \bar{u}_{r}^{2}}{4\bar{u}_{r}^{2}} + \ldots \right] . \label{e:jump2small}
\end{eqnarray}

We see from the upper expansion wave envelope in Figure \ref{f:vdswjust} and the expansion wave solution (\ref{e:midsolnu}) that
\begin{equation}
 u_{i} = \frac{\sqrt{q}}{3\sqrt{2}} \left( \frac{2\sqrt{2}}{\sqrt{q}} u_{-} - U_{\mbox{shock}} \right) + u_{+}, 
 \label{e:uiexpand}
\end{equation}
so that 
\begin{equation}
 U_{\mbox{shock}} = 2\sqrt{\frac{2}{q}} u_{-} + 3\sqrt{\frac{2}{q}} \left( u_{+} - u_{i} \right).
 \label{e:ushockexpand}
\end{equation}
As for the TDSW solution, the shallow water Riemann variables (\ref{e:lambda3}) and (\ref{e:lambda4}) will be used to 
propagate the solution.  Furthermore, as for the TDSW regime, the amplitude corrections to these Riemann variables will 
be neglected, so that they become the Riemann invariants (\ref{e:cp}) and (\ref{e:cm}).  As the shock is weak, at leading 
order the Riemann invariant $R_{+}$ (\ref{e:cp}) is conserved through the shock, so that at first order, that is shock 
strength zero,
\begin{equation}
 2\sqrt{\frac{2}{q}} u_{-} = \bar{v}_{r} + 2\sqrt{\frac{2}{q}} \bar{u}_{r}.
 \label{e:cpshock}
\end{equation}
A measure of the shock strength is $u_{+}$ as it is small.  We then seek a weak shock correction to this Riemann invariant as
\begin{equation}
 \bar{v}_{r} + 2\sqrt{\frac{2}{q}} \bar{u}_{r} = 2\sqrt{\frac{2}{q}} \left( u_{-} + \delta_{1}u_{+} + \delta_{2} u_{+}^{2} 
 + \ldots \right).
 \label{e:cpperturb}
\end{equation}

Substituting the Riemann invariant correction (\ref{e:cpperturb}) and the assumed upper envelope expression 
(\ref{e:ushockexpand}) into the weak shock jump conditions (\ref{e:jump1small}) and (\ref{e:jump2small}) and solving as a 
series in small $u_{+}$, we find $\delta_{1} = \delta_{2} = 0$ and
\begin{equation}
 u_{i} = \bar{u}_{r} + 2u_{+} - \frac{u_{+}^{2}}{\bar{u}_{r}}.
 \label{e:uivdsw}
\end{equation}
To $O(u_{+}^{2})$ there is then no correction to the shallow water Riemann invariant.  Equation (\ref{e:ushockexpand}) then 
gives the Whitham shock velocity.  As the Riemann invariant $R_{+}$ is conserved to $O(u_{+}^{2})$ through the weak Whitham shock, 
the intermediate level result (\ref{e:ui}) holds to $O(u_{+}^{2})$, as confirmed from the intermediate level comparison 
of Figure \ref{f:intermediate}.  Using this intermediate level expression in the shock result (\ref{e:uivdsw}) gives that 
\begin{equation}
 \bar{u}_{r} = 
\frac{1}{4}u_{-} -\frac{3}{4}u_{+} + \frac{1}{4}\sqrt{ u_{-}^{2} - 6u_{+}u_{-} + 25u_{+}^{2} }
 = \frac{1}{2}u_{-} - \frac{3}{2}u_{+} + \frac{2u_{+}^{2}}{u_{-}} + \ldots
 \label{e:urvdsw}
\end{equation}
Figure \ref{f:shock} shows comparisons between numerical solutions for the Whitham shock velocity with the result 
(\ref{e:ushockexpand}) in the VDSW regime.  It can be seen that there is excellent agreement, except near the transition 
to the TDSW regime where the DSW form changes.  
As the transition to the TDSW regime is approached, the amplitude of the resonant wavetrain increases, so that its neglect 
in the Riemann variables becomes less valid.

Figure \ref{f:resonantwave}(a) gives comparisons for the height $H_{r} = a_{r} + \bar{u}_{r} = u_{+} + \bar{u}_{r}$ of the resonant 
wavetrain with numerical values.  Again, there is good agreement, except near the transition to the TDSW regime, again because 
the neglect of the resonant wave amplitude correction to the Riemann variables becomes less accurate.  The resonant wavetrain 
mean $\bar{u}_{r}$ comparison of Figure \ref{f:resonantwave}(b) is similar, with excellent comparison.
Figure \ref{f:intermediate} shows an excellent comparison between the numerical mean level $u_{i}$ in the VDSW regime and 
the VDSW result (\ref{e:uivdsw}), as expected as the Riemann invariant $R_{+}$ is conserved to $O(u_{+}^{2})$ through the weak 
Whitham shock.  Finally, Figure \ref{f:resonantwave}(c) shows comparisons for the resonant wavenumber $k_{r}$.  As for the CDSW 
regime, the agreement is poor, except near the boundary with the TDSW regime where the resonant wavetrain restabilises due to its 
increased amplitude.  As stated, it was found that the Riemann invariant $R_{+}$ (\ref{e:cp}) is conserved to $O(u_{+}^{2})$ 
through the Whitham shock, which means that the intermediate level $u_{i}$ is given by the mean value (\ref{e:ui}) to 
this order.  It can be seen from Figure \ref{f:intermediate} that there is a slight deviation in $u_{i}$ from the mean value 
in the VDSW regime which grows as $u_{+}$ decreases.  Presumably, this deviation is due to higher order corrections in the 
Riemann invariant expansion (\ref{e:cpperturb}).  As this correction is small, it is not considered here.  

\begin{figure}
\centering
 \includegraphics[width=0.3\textwidth,angle=270]{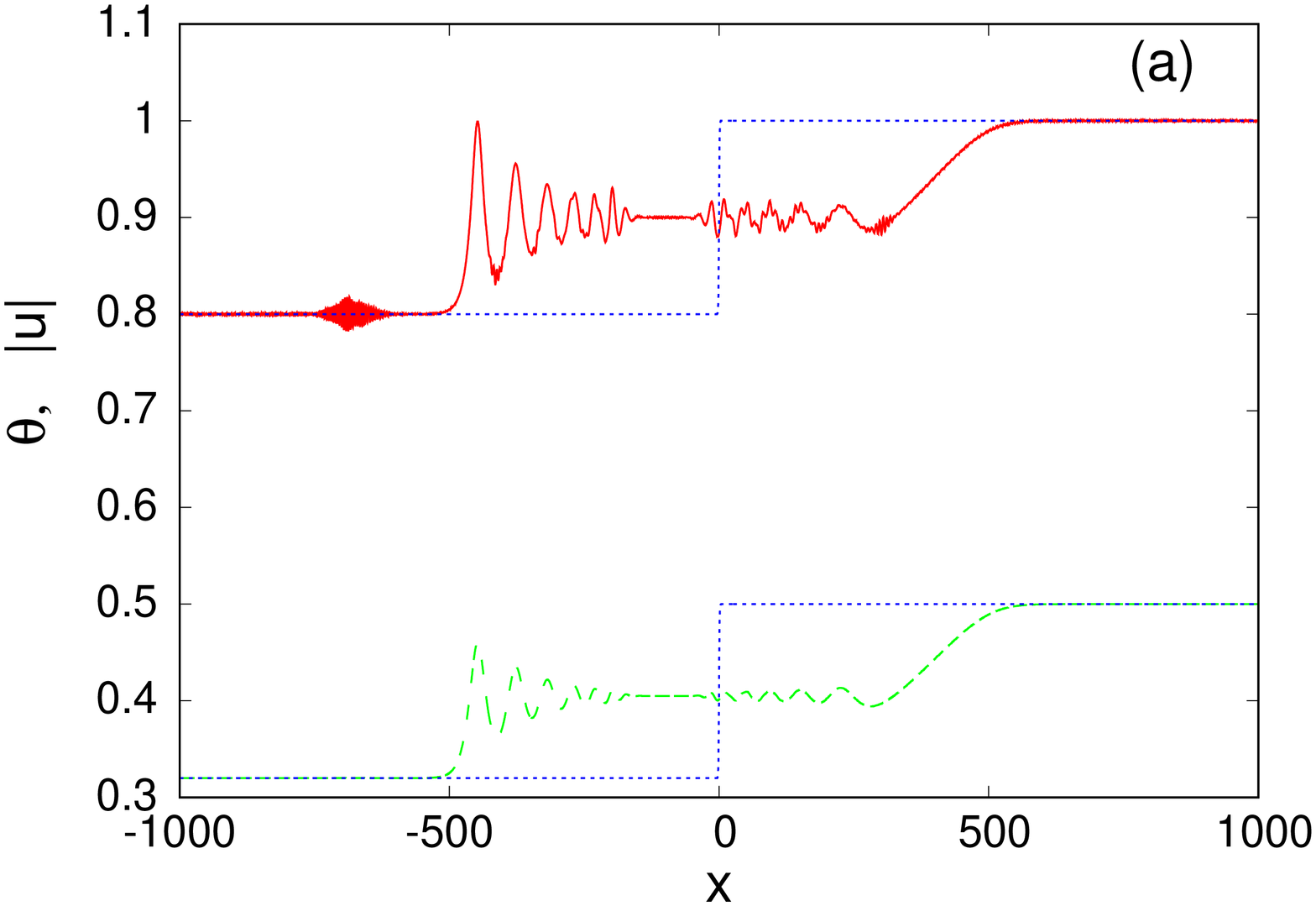}
 \includegraphics[width=0.3\textwidth,angle=270]{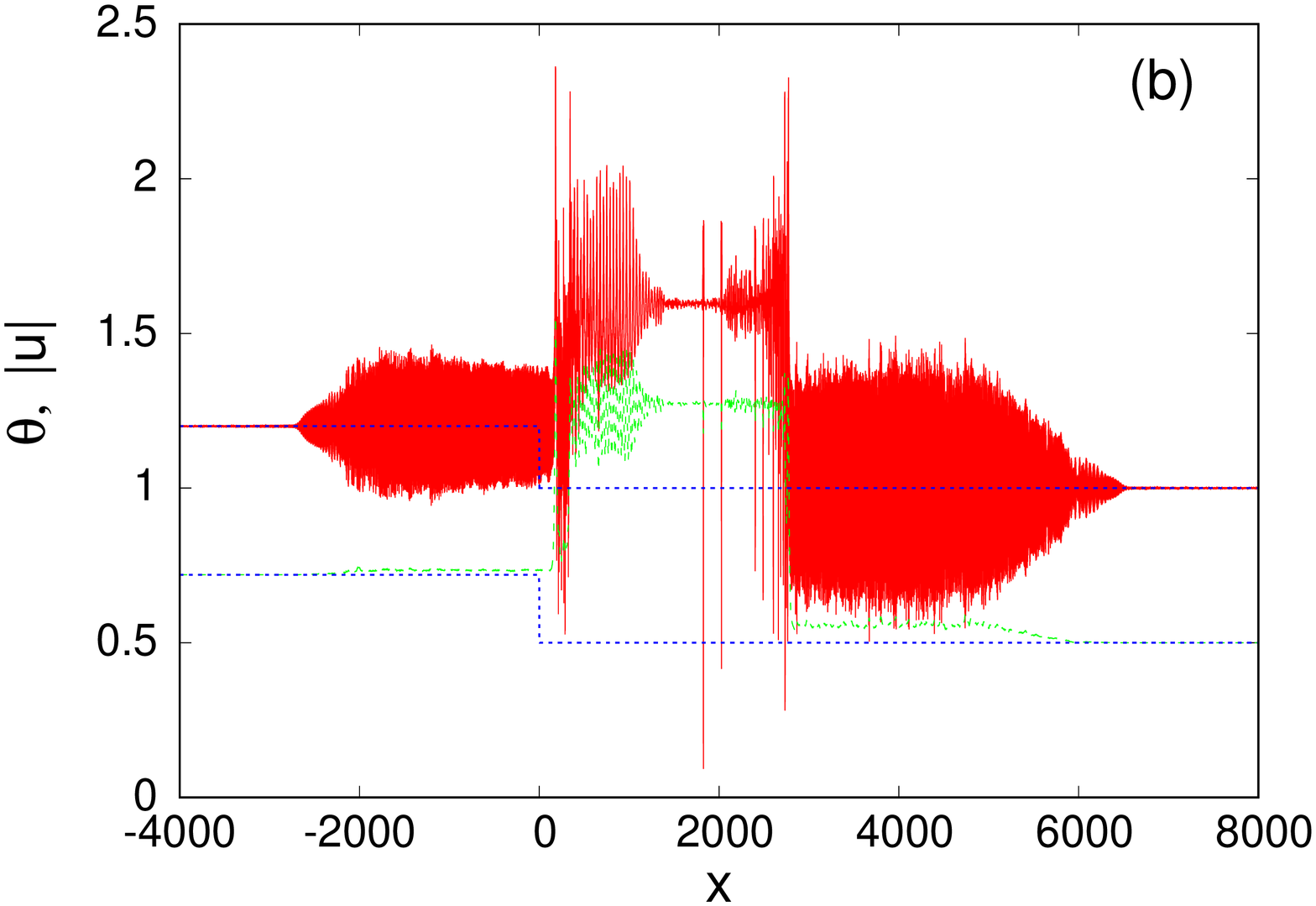}

 \includegraphics[width=0.3\textwidth,angle=270]{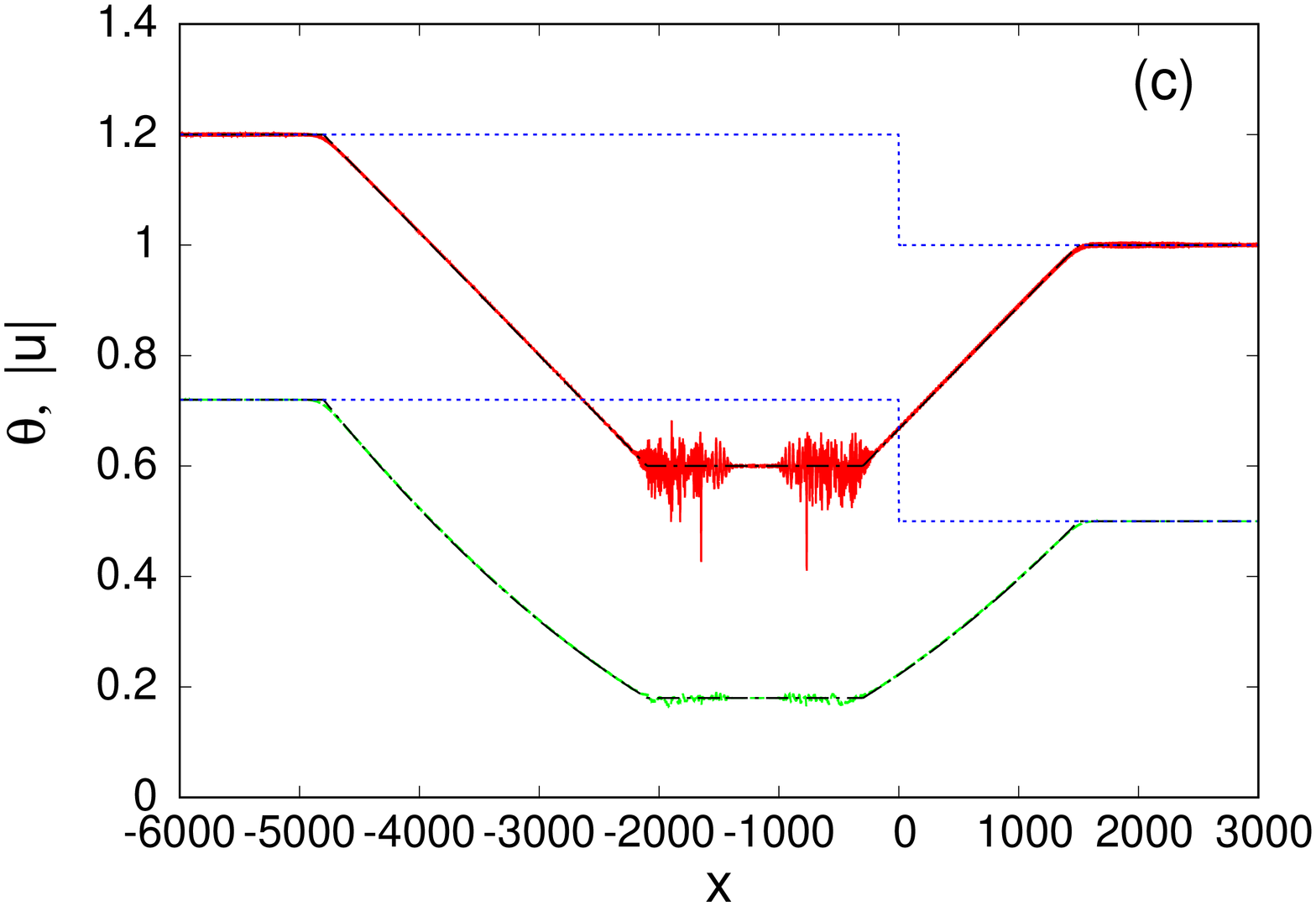}
 \includegraphics[width=0.3\textwidth,angle=270]{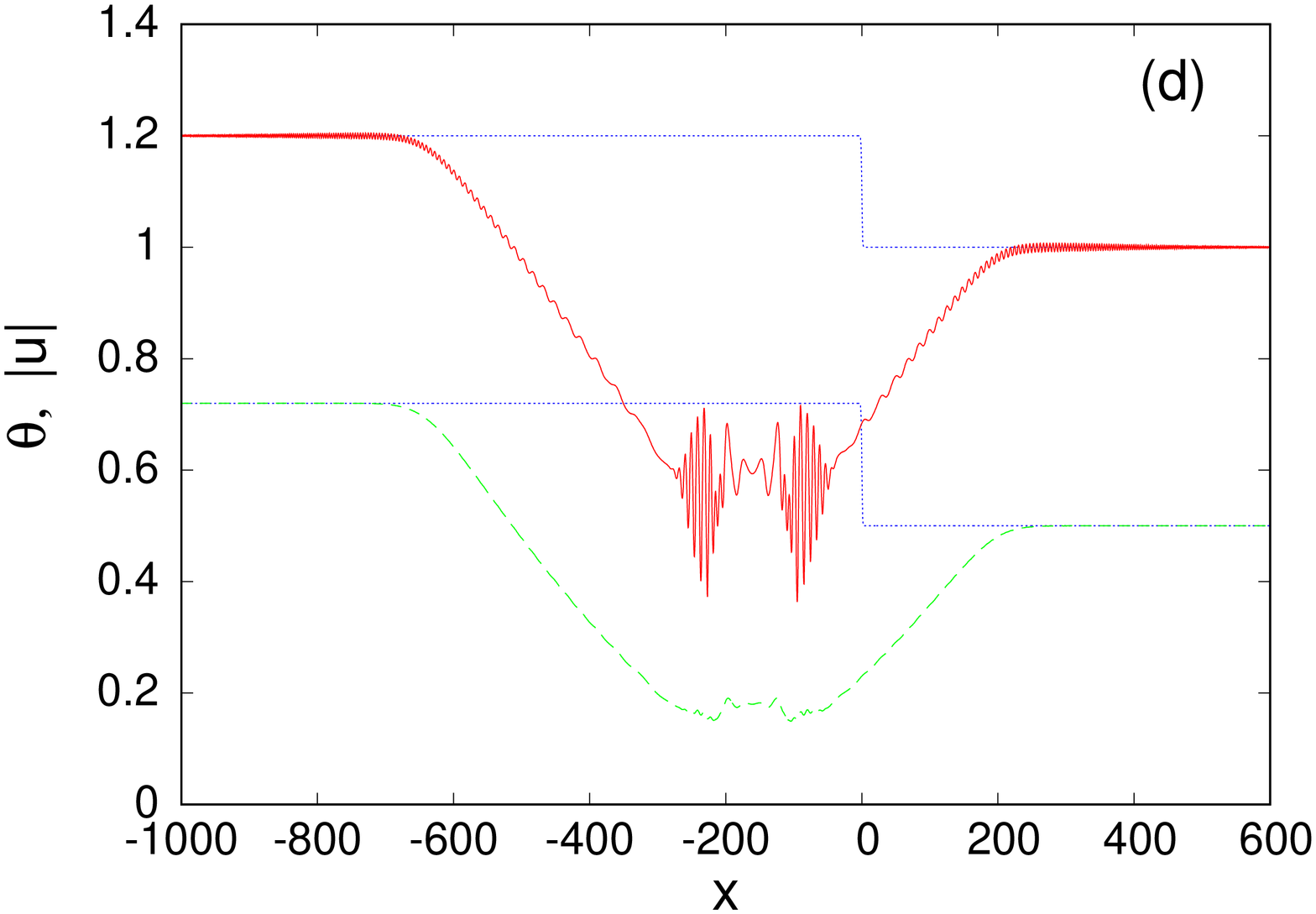}
 
 \includegraphics[width=0.3\textwidth,angle=270]{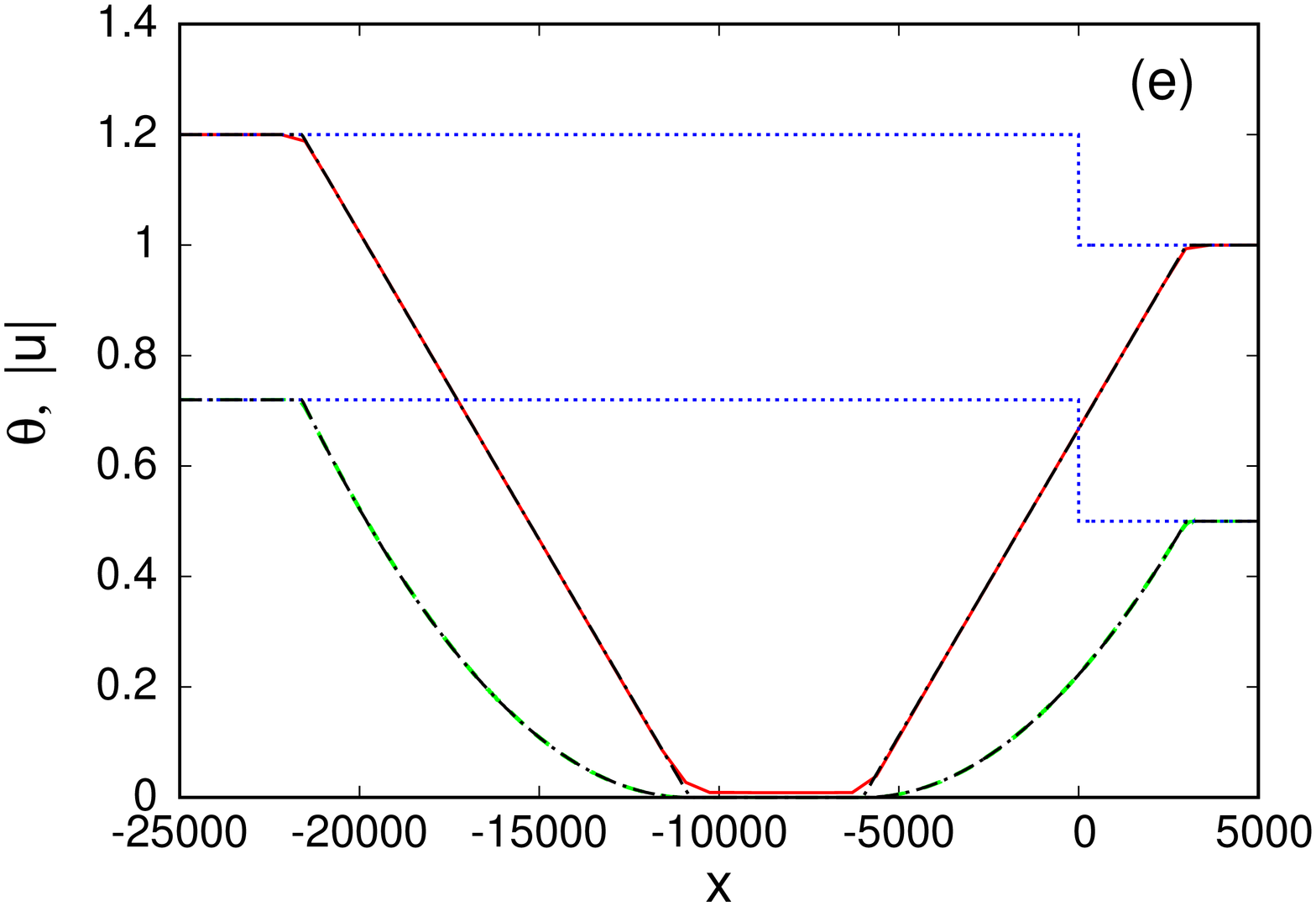}
 \includegraphics[width=0.3\textwidth,angle=270]{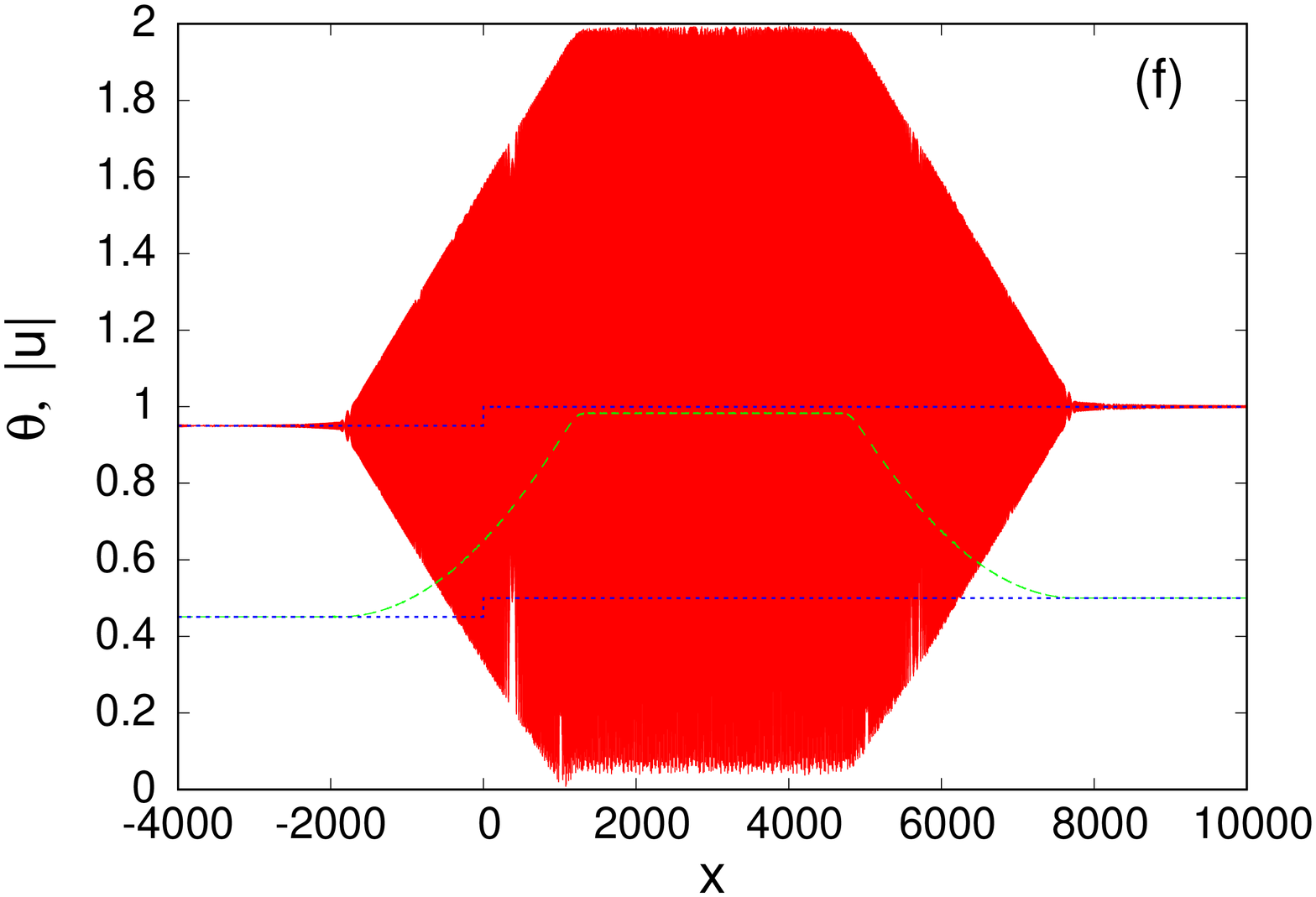}
\caption{Numerical solutions of the nematic equations (\ref{e:eeqn}) and (\ref{e:direqn}) for
the initial condition (\ref{e:ic}).  Red (solid) lines:  $|u|$; green (dashed)
lines $\theta$; blue (dotted) lines:  $|u|$ at $z=0$ (upper) and $\theta$ at $z=0$ (lower).
(a) $u_{-} = 0.8$, $u_{+} = 1.0$, $v_{-} = 0$, $v_{+} = 0$, $z=500$
(b) $u_{-} = 1.2$, $u_{+} = 1.0$, $v_{-} = 2$, $v_{+} = 0$, $z=1500$, 
(c) $u_{-} = 1.2$, $u_{+} = 1.0$, $v_{-} = -2.0$, $v_{+} = 0$, $z=1500$, black (dash-dot) line:  expansion
wave solution (\ref{e:phaseexpand}), 
(d) $u_{-} = 1.2$, $u_{+} = 1.0$, $v_{-} = -2.0$, $v_{+} = 0$, $z=200$, 
(e) $u_{-} = 1.2$, $u_{+} = 1.0$, $v_{-} = -6.0$, $v_{+} = 0.0$, $z=1500$, black (dash-dot) line:  expansion
wave solution (\ref{e:phaseexpand}), 
(f) $u_{-} = 0.95$, $u_{+} = 1.0$, $v_{-} = 6.0$, $v_{+} = 0$, $z=1000$.  Here $\nu = 200$ and $q =2$.} 
\label{f:phasetypes}
\end{figure}

\section{Dependence on Initial Wavenumber}

The analysis and comparisons of the previous sections were for initial jumps (\ref{e:ic}) with zero wavenumber 
$v_{-} = v_{+} = 0$.  The effect of initial wavenumber on the development of DSWs for the defocusing NLS equation has 
been investigated and it was found that solutions consist of a combination of DSWs and (non-dispersive) expansion waves 
\cite{gennady}.  In detail, there are six distinct types of solutions for the defocusing NLS equation.  The situation
for the nematic equations (\ref{e:eeqn}) and (\ref{e:direqn}) is more complicated, as can be seen from the DSW types 
illustrated in Figure \ref{f:types} for $v_{-} = v_{+} = 0$.  There are multiple DSW types, depending on the height of the 
initial jump, so that for each defocusing NLS solution with a DSW, there are five such solutions for the nematic equations.  
To finish the analysis of DSW solutions for the nematic equations, the broad extra five types of solution for the nematic 
equations will be briefly discussed, with the solutions for each of these types briefly outlined as the details are similar 
to those derived in the previous sections for $v_{-} = v_{+} = 0$.  The new general types of solutions, consisting of
various combinations of DSWs and expansion waves (solutions of the shallow water equations (\ref{e:cp}) and
(\ref{e:cm})) are illustrated in Figure \ref{f:phasetypes}.  

The simplest case is the reversal of the orientation of the initial jump, $u_{-} < u_{+}$, with $u_{-} < u_{i} < u_{+}$, 
as shown in Figure \ref{f:phasetypes}(a).  As the nematic equations are bidirectional, the solutions 
for this case are just the solutions of the previous sections, but reversed in direction.  
 
The mean height (\ref{e:ui}), valid for the initial condition (\ref{e:ic}) with no wavenumber jump, can be extended to
include such a jump, resulting in
\begin{equation}
 u_{i} = \frac{1}{2} \left( u_{-} + u_{+} \right) + \frac{1}{4} \frac{\sqrt{q}}{\sqrt{2}} \left( v_{-} - v_{+} \right),
 \label{e:uiv}
\end{equation}
which is obtained by propagating the rear conditions $u_{-},v_{-}$ on the characteristics $C_{+}$ (\ref{e:cp}) and the 
front conditions $u_{+},v_{+}$ on the characteristics $C_{-}$ (\ref{e:cm}).  A more substantive example than that of 
Figure \ref{f:phasetypes}(a) is shown in Figure \ref{f:phasetypes}(b).  This example shows two DSWs, both crossover cases, 
preceded by large resonant wavetrains.  This case arises when $u_{i} > u_{-} > u_{+}$, which results in two DSWs.  These CDSWs 
and resonant wavetrains can be analysed using the approximate method of Section \ref{s:cross}.  

When $0 < u_{i} < u_{+}$, two expansion waves are generated due to the change in orientation of the jumps between the initial
states $u_{-}$ and $u_{+}$ and the intermediate level.  One expansion wave is a generalisation of (\ref{e:midsolnu}) with 
$v_{-}, v_{+} \ne 0$.  The other is the equivalent simple wave solution on the characteristic $C_{+}$ (\ref{e:cp}) with the 
Riemann invariant $R_{-}$ (\ref{e:cm}) constant through the expansion fan.  This results in the solution
\begin{equation}
|u| = \left\{ \begin{array}{cc}
                        u_{-}, & \frac{x}{z} < v_{-} -\frac{\sqrt{2}}{\sqrt{q}}u_{-} \\
                        \frac{\sqrt{q}}{3\sqrt{2}} \left[ v_{-} + \frac{2\sqrt{2}}{\sqrt{q}}u_{-} - \frac{x}{z} \right], &
v_{-} -\frac{\sqrt{2}}{\sqrt{q}} u_{-} \le \frac{x}{z} \le \frac{1}{4} \left( v_{-} + 3v_{+} \right) + 
\frac{1}{2} \frac{\sqrt{2}}{\sqrt{q}} \left( u_{-} - 3u_{+} \right), \\
u_{i} = \frac{1}{2}\left( u_{-} + u_{+} \right) + \frac{1}{4} \frac{\sqrt{q}}{\sqrt{2}} \left( v_{-} - v_{+} \right), 
& \frac{1}{4} \left( v_{-} + 3v_{+} \right) + \frac{1}{2} \frac{\sqrt{2}}{\sqrt{q}} \left( u_{-} - 3u_{+} \right) 
< \frac{x}{z} \\
& < \frac{1}{4} \left( 3v_{-} + v_{+} \right) + \frac{1}{2} \frac{\sqrt{2}}{\sqrt{q}} \left( 3u_{-} - u_{+} 
\right), \\
\frac{\sqrt{q}}{3\sqrt{2}} \left[ \frac{2\sqrt{2}}{\sqrt{q}}u_{+} - v_{+} + \frac{x}{z} \right], & 
\frac{1}{4} \left( 3v_{-} + v_{+} \right) + \frac{1}{2} \frac{\sqrt{2}}{\sqrt{q}} \left( 3u_{-} - u_{+} 
\right) \le \frac{x}{z} \le v_{+} + \frac{\sqrt{2}}{\sqrt{q}} u_{+}, \\
u_{+}, & v_{+} + \frac{\sqrt{2}}{\sqrt{q}} u_{+} < \frac{x}{z} 
\end{array}
\right.
\label{e:phaseexpand}
\end{equation}
and
\begin{equation}
v = \left\{ \begin{array}{cc}
                        v_{-}, & \frac{x}{z} < v_{-} -\frac{\sqrt{2}}{\sqrt{q}}u_{-} \\
                        \frac{1}{3}v_{-} + \frac{2}{3}\frac{\sqrt{2}}{\sqrt{q}} u_{-} + \frac{2}{3}\frac{x}{z}, &
v_{-} -\frac{\sqrt{2}}{\sqrt{q}} u_{-} \le \frac{x}{z} \le \frac{1}{4} \left( v_{-} + 3v_{+} \right) + 
\frac{1}{2} \frac{\sqrt{2}}{\sqrt{q}} \left( u_{-} - 3u_{+} \right), \\
v_{i} = \frac{1}{2} \left( v_{-} + v_{+} \right) + \frac{\sqrt{2}}{\sqrt{q}} \left( u_{-} - u_{+}\right), 
& \frac{1}{4} \left( v_{-} + 3v_{+} \right) + \frac{1}{2} \frac{\sqrt{2}}{\sqrt{q}} \left( u_{-} - 3u_{+} \right) 
< \frac{x}{z} \\
& < \frac{1}{4} \left( 3v_{-} + v_{+} \right) + \frac{1}{2} \frac{\sqrt{2}}{\sqrt{q}} \left( 3u_{-} - u_{+} 
\right), \\
\frac{1}{3}v_{+} - \frac{2}{3}\frac{\sqrt{2}}{\sqrt{q}} u_{+} + \frac{2}{3} \frac{x}{z}, & 
\frac{1}{4} \left( 3v_{-} + v_{+} \right) + \frac{1}{2} \frac{\sqrt{2}}{\sqrt{q}} \left( 3u_{-} - u_{+} 
\right) \le \frac{x}{z} \le v_{+} + \frac{\sqrt{2}}{\sqrt{q}} u_{+}, \\
v_{+}, & v_{+} + \frac{\sqrt{2}}{\sqrt{q}} u_{+} < \frac{x}{z} ,
\end{array}
\right.
\label{e:phaseexpandv}
\end{equation}
which consists of two expansion fans either side of the intermediate level $u_{i}$.  The corresponding director solution is 
given by $\theta = |u|^{2}/q$.  This expansion wave solution is compared with a numerical solution in Figure 
\ref{f:phasetypes}(c), with near perfect agreement seen.  The main disagreement is the high frequency dispersive waves at the 
junctions of the expansion fans and the intermediate level $u_{i}$, which are not captured by the non-dispersive expansion 
waves.  The origin of these dispersive waves can be seen from Figure \ref{f:phasetypes}(d), which is the same solution as in 
Figure \ref{f:phasetypes}(c), but at a lower value $z=200$, rather than $z=1500$.  As the initial condition pushes down to 
create the intermediate level $u_{i}$, two small jumps in the centre of the intermediate level are created, which generate 
small DSWs.  As the DSWs are resonant, resonant diffractive waves are also created, which is the origin of the diffractive waves 
in Figure \ref{f:phasetypes}(c).

This expansion fan solution holds until the vacuum case is reached, $u_{i} \le 0$, so that (\ref{e:phaseexpand}) becomes 
invalid.  However, the expansion wave solution (\ref{e:phaseexpand}) is valid if in the sections for which $|u|$ is negative,
the solution is replaced by $|u|=0$.  Such a comparison is shown in Figure \ref{f:phasetypes}(e) and it can be seen that the 
comparison is excellent, as for that of Figure \ref{f:phasetypes}(c).

The final generic DSW solution is that illustrated in Figure \ref{f:phasetypes}(f).  This solution
occurs when the two DSWs of Figure \ref{f:phasetypes}(b) collide and interact, as for the equivalent case for the NLS equation 
\cite{bikbaev}.  The solution for this regime is then a two phase wavetrain, which is difficult to analyse without the full 
Whitham modulation equations.  The particular solution shown in Figure \ref{f:phasetypes}(f) is two interacting TDSWs.  There 
are two partial DSWs on either side of the central wavetrain, which is a (stable) uniform resonant wavetrain.  In essence,
the interaction has destroyed the intermediate level so that the two individual TDSWs can join.  It is difficult to give the 
range of validity in $(u_{+},u_{-},v_{+},v_{-})$ for this case as the borderline occurs when the tails of the two DSWs 
interact, noting that the velocities of the trailing edges of the DSWs critically depend on which regimes they lie in.

\section{Conclusions}

Dispersive shock wave (DSW) solutions of the nematic equations (\ref{e:eeqn}) and (\ref{e:direqn}) have been found in 
the high nonlocality limit $\nu \gg 1$, which is the experimentally relevant limit \cite{PR,Wiley}.  It was found that there 
are six distinct DSW types, see Figure \ref{f:types}.  In contrast to previous work \cite{nembore,nemgennady} the solutions 
in these six regimes were found based on different asymptotic and approximate techniques appropriate for each regime.  The 
work of \cite{nembore} assumed that the DSW was of KdV type in all regimes, except the dam break case, while that of 
\cite{nemgennady} assumed that the DSW solution was determined by a gas dynamics type shock, except in the dam break case.  
The present work shows that the nematic DSW is more complicated than this, with the PDSW and TDSW regimes consisting of 
(perturbed) KdV-type DSWs and the TDSW and VDSW regimes being determined by Whitham shocks in the appropriate Whitham modulation 
equations.  The CDSW regime is a transition between these two broad types.  One novel feature of the present work is the use of 
Whitham shocks for Whitham modulation equations, as pioneered by \cite{patjump}, validating ideas of Whitham 
\cite{whitham,modproc} when he first developed modulation theory.  The nematic DSW shows a wider range of behaviours and 
solution types than the equivalent resonant Kawahara DSW \cite{patkdv,pat}.  The DSW solutions for all six regimes show
excellent agreement with numerical solutions for the DSW itself (when it exists), the resonant wavetrain (when it exists) and 
the intermediate level linking the backwards propagating expansion wave with the DSW or Whitham shock.  The only exception is for 
the wavenumber of the resonant wavetrain when this wavetrain is unstable, as would be expected.

As stated, the present work obtained solutions for the nematic DSW in the highly nonlocal limit $\nu \gg 1$.  As can be 
seen from Figures \ref{f:types}(a) and \ref{f:nu10} the form of the DSW is highly dependent on the value of $\nu$.  Indeed, 
the nematic equations (\ref{e:eeqn}) and (\ref{e:direqn}) reduce to the defocusing NLS equation for $\nu = 0$ and a perturbed 
defocusing NLS equation for $\nu$ small, for which there is no resonance.  It is then of interest to study the transition from 
the highly nonlocal case to the local case with $\nu$ small.  In this regard, the DSW changes from resonant to non-resonant, as 
indicated by the change of sign of the third derivative of the small deviation KdV reduction  (\ref{e:kdv5nem}) of the nematic 
equations.  

Most of the solutions for the DSW types derived here have been based to a greater or lesser degree on the Whitham 
modulation equations for the periodic wave solution of the nematic equations.  Unfortunately, this periodic wave solution 
is not known in general and a weakly nonlinear Stokes' wave approximation was used.  While this generally gave satisfactory 
results, the Whitham modulation equations based on it gave incorrect modulational stability when the wave amplitude was not 
small.  Whether these weakly nonlinear modulation equations can be improved on is not clear.

As stated in Section \ref{s:types}, the nematic system (\ref{e:eeqn}) and (\ref{e:direqn}) is general and applies to other 
nonlinear optical media.  In particular, for optical thermal media $q=0$ in the director equation (\ref{e:direqn}) is a 
physical limit, in which case $\theta$ is the temperature of the medium.  In this limit, $|u|$ constant ceases to be a valid 
solution of the system, so that the step initial condition (\ref{e:ic}) does not result in an expansion wave and DSW being 
generated between the initial levels $u_{-}$ and $u_{+}$.  The form of a DSW for $q=0$ is an open question.

In summary, while the form of the DSW in the highly nonlocal limit has been largely resolved, there are still many open 
questions in regard to DSW solutions of the nematic system for general nonlocality and other parameter values.

Saleh Baqer thanks Kuwait University for a scholarship to undertake his Ph.D. at the University of Edinburgh.  The authors thank 
C\^ome Houdeville for pointing out errors in the manuscript.

\end{document}